\documentclass[twocolumn]{aastex63}

\usepackage{lineno}

\usepackage{graphicx}
\usepackage{graphics}
\usepackage{amssymb}
\usepackage{bm}
\usepackage{times}
\usepackage[flushleft]{threeparttable}
\usepackage{booktabs}
\usepackage{footnote}
\usepackage{multirow}

\usepackage{rotating}
\usepackage{color}
\usepackage{soul}

\newcommand{\kms}{km~s$^{-1}$}

\newcommand{\arcs}{\ensuremath{^{\prime\prime}}}
\newcommand{\arcm}{\ensuremath{^{\prime}}}
\newcommand{\hicm}{H{\sc i} 21 cm}
\newcommand{\hi}{H{\sc i}}
\definecolor{cerulean}{rgb}{0.0, 0.48, 0.65}
\definecolor{red}{rgb}{1.0, 0.0, 0.0}

\definecolor{green}{rgb}{0.0, 0.6, 0.6}

\shorttitle{Dark and bright galaxies on an unusual  filament}
\shortauthors{Arabsalmani et al.}
\begin{document}


\title
{
Pearls on a string: Dark and bright galaxies on a strikingly straight and narrow  filament
}

\correspondingauthor{Maryam Arabsalmani, Vera Rubin Fellow}
\email{maryam.arabsalmani@origins-cluster.de}

\author{M. Arabsalmani}, 
\affiliation{Excellence Cluster ORIGINS, Boltzmannstra{\ss}e 2, 85748 Garching, Germany}
\affiliation{Ludwig-Maximilians-Universit\"at, Schellingstra{\ss}e 4, 80799 M\"unchen, Germany}

\author{S. Roychowdhury}
\affiliation{Ludwig-Maximilians-Universit\"at, Schellingstra{\ss}e 4, 80799 M\"unchen, Germany}
\affiliation{University Observatory Munich (USM), Scheinerstra{\ss}e 1, 81679 M\"unchen, Germany}

\author{B. Schneider}
\affiliation{MIT Kavli Institute for Astrophysics and Space Research, 70 Vassar Street, Cambridge, MA 02139, USA}

\author{V. Springel}
\affiliation{Max Planck Institute for Astrophysics, Karl-Schwarzschild-Stra{\ss}e 1, D-85748 Garching, Germany}

\author{E. Le Floc'h}
\affiliation{Universit\'e Paris-Saclay, Universit\'e Paris Cit\'e, CEA, CNRS, AIM, 91191, Gif-sur-Yvette, France}

\author{F. Bournaud}
\affiliation{Universit\'e Paris-Saclay, Universit\'e Paris Cit\'e, CEA, CNRS, AIM, 91191, Gif-sur-Yvette, France}

\author{A. Burkert}
\affiliation{University Observatory Munich (USM), Scheinerstra{\ss}e 1, 81679 M\"unchen, Germany}
\affiliation{Max-Planck-Institut f\"ur extraterrestrische Physik (MPE), Giessenbachstr. 1, 85748 Garching, Germany}

\author{J. Cuillandre}
\affiliation{Universit\'e Paris-Saclay, Universit\'e Paris Cit\'e, CEA, CNRS, AIM, 91191, Gif-sur-Yvette, France}

\author{P. Duc}
\affiliation{Universit\'e de Strasbourg, CNRS, Observatoire astronomique de Strasbourg, UMR 7550, F-67000 Strasbourg, France}

\author{E. Emsellem}
\affiliation{European Southern Observatory, Karl-Schwarzschild-Stra{\ss}e 2, 85748 Garching, Germany}
\affiliation{Univ. Lyon, Univ. Lyon1, ENS de Lyon, CNRS, Centre de Recherche Astrophysique de Lyon, UMR5574, 69230 Saint-Genis-Laval, France}

\author{D. Gal\'arraga-Espinosa}
\affiliation{Max Planck Institute for Astrophysics, Karl-Schwarzschild-Stra{\ss}e 1, D-85748 Garching, Germany}

\author{E. Pian}
\affiliation{INAF, Astrophysics and Space Science Observatory, via P. Gobetti 101, 40129 Bologna, Italy}
\affiliation{Max Planck Institute for Astrophysics, Karl-Schwarzschild-Stra{\ss}e 1, D-85748 Garching, Germany}

\author{F. Renaud}
\affiliation{Universit\'e de Strasbourg, CNRS, Observatoire astronomique de Strasbourg, UMR 7550, F-67000 Strasbourg, France}
\affiliation{University of Strasbourg Institute for Advanced Study, 5 all\'ee du G\'en\'eral Rouvillois, F-67083 Strasbourg, France}

\author{M. Zwaan}
\affiliation{European Southern Observatory, Karl-Schwarzschild-Stra{\ss}e 2, 85748 Garching, Germany}


\begin{abstract}
We identify a chain of galaxies along an almost straight line in the nearby Universe with a  projected length of $\sim$ 5 Mpc. The galaxies are  distributed within projected distances of only 7 to 105 kpc from the axis of the identified filament. They  have redshifts in a very small range of $z=0.0361-0.0370$ so that their radial velocities are consistent with galaxy proper motions. The filament galaxies are mainly star-forming and have stellar masses in a range of $\rm 10^{9.1}-10^{10.7}\,M_{\odot}$. We search for systems with similar geometrical properties in the full-sky mock galaxy catalogue of the MillenniumTNG simulations and find that although such straight filaments are unusual and rare, they are predicted by $\Lambda$CDM simulations  ($4\%$ incidence). 
We study the cold \hi\ gas in a 1.3 Mpc section of the filament through \hicm\ emission line observations and detect eleven \hi\ sources,  many more than expected from the  \hi\ mass function in a similar volume. They have \hi\  masses $\rm 10^{8.5}-10^{9.5}\,M_{\odot}$ and are mostly  within $\sim$ 120 kpc projected distance from the filament axis. None of these \hi\ sources has a  confirmed optical counterpart.  
Their  darkness  together with their  large \hicm\ line-widths indicate that they contain gas that might not yet be virialized. 
These clouds must be marking the peaks of the dark matter and \hi\ distributions over large scales within the filament. 
The presence of such gas clouds around the filament spines is  predicted by simulations, but this is the first time that the existence of such clouds in a filament is observationally confirmed.

\end{abstract}




\section{Introduction} 
\label{sec:int}

The dark and baryonic matter in the Universe  is predicted to be distributed over an intriguing  pattern, known as the cosmic web \citep[][]{Zeldovich1970-1970A&A.....5...84Z, Bond1996-1996Natur.380..603B, Springel05-2005Natur.435..629S, Springel06-2006Natur.440.1137S}.  
This multiscale and highly interconnected  network consists of   knots, filaments, sheets, and voids, with dense knots at the intersection of filaments,  and filaments at the intersection of sheets.  And these  surround nearly empty void regions.
The filaments, connecting massive galaxy clusters located at the knots, contain the largest fraction ($\sim 50\%$) of mass in the Universe \citep[][]{Cen06-2006ApJ...650..560C,  Aragon-calvo2010-2010MNRAS.408.2163A, Cautun14-2014MNRAS.441.2923C}.     
High resolution simulations show that filaments are sparsely inhabited  by haloes, and that galaxies make up only a few percent of the baryonic  mass in the filaments \citep[][]{Cautun14-2014MNRAS.441.2923C, Martizzi19-2019MNRAS.486.3766M}. The dominating component of baryons in filaments  
is predicted  to be the  low density, diffuse Inter Galactic Medium (IGM), and mainly at temperatures $10^5\ {\rm K} \lesssim T \lesssim 10^7\ \rm K$ \citep[e.g.,][]{Cen06-2006ApJ...650..560C, Martizzi19-2019MNRAS.486.3766M}. 
The low densities   of the filaments  suggest  that their  identifications in observations  are quite challenging.


Only a handful of  inter cluster filaments have been identified through the observations  of X-ray emission or Sunyaev--Zel'dovich signal from the warm/hot phase  between  pairs of close clusters or inside superclusters in the nearby Universe \citep[][]{Scharf00-2000ApJ...528L..73S,  Eckert15-2015Natur.528..105E, Alvarez18-2018ApJ...858...44A, Tanimura19-2019MNRAS.483..223T, Tanimura20-2020A&A...643L...2T, Tanimura20-2020A&A...637A..41T}. 
Such studies (mainly done through stacking) have succeeded in measuring the density and temperature of the densest parts of the warm/hot gas in the filaments around clusters.  
In terms of finding filaments the primary  approach though  has been  structure-finding algorithms 
based on the  distribution of galaxies and galaxy clusters in large sky area databases \citep[]{Sousbie08-2008ApJ...672L...1S, Jasche10-2010MNRAS.409..355J, Tempel14-2014MNRAS.438.3465T, Malavasi20-2020A&A...642A..19M, Santiago-Bautista20-2020A&A...637A..31S} which has led to the identifications of a large number of filaments in the nearby Universe and their geometrical properties such as length and radius \citep[see also][]{Kim16-2016ApJ...833..207K}.  
In some cases  filaments have been identified through  weak lensing techniques \citep[at $z\sim 0.5$][]{Jorg12-2012Natur.487..202D, Jauzac12-2012MNRAS.426.3369J, Higuchi15-2015arXiv150306373H, Martinet16-2016A&A...590A..69M, Kondo20-2020MNRAS.495.3695K}. 
At higher redshifts ($z>2$) 
detection of rest-frame ultraviolet Lyman-$\alpha$ radiation  through absorption has resulted in the identification of a few filaments \citetext{\citealp[e.g.,][]{Cantalupo14-2014Natur.506...63C, Martin15-2015Natur.524..192M, Umehata19-2019Sci...366...97U}, \citealp[see also][]{Gallego18-2018MNRAS.475.3854G} for stacking analysis}. 
These studies have obtained indirect estimates of cold gas mass ($T \sim 10^4\ \rm K$) in the observed sections of the filaments and in some cases have found it to be significantly larger than what is predicted by simulations \citep[e.g.,][]{Cantalupo14-2014Natur.506...63C}.  

Theoretical studies  suggest that streaming of cold gas along the  filaments  plays a prominent role in the formation and growth of galaxies by providing the required gas for star formation \citep[][]{Bournaud05-2005A&A...438..507B, Dekel09-2009Natur.457..451D, Fumagalli11-2011MNRAS.418.1796F}. 
This gas component not only is expected to have very low densities,  
it is also predicted to comprise only a small fraction (a few percent) of the baryonic mass in the filaments  \citep[][]{Popping09-2009A&A...504...15P, Martizzi19-2019MNRAS.486.3766M, Galarraga-Espinosa21-2021A&A...649A.117G}. 
There have been attempts to detect the diffuse cold gas in the nearby filaments through \hicm\ emission line observations \citep[e.g.,][]{Popping11-2011A&A...533A.122P, Poppoing11-2011A&A...527A..90P, Popping11-2011A&A...528A..28P}, but none has been successful  to date. 


In this letter we report the identification of a strikingly narrow filament of galaxies in the nearby universe. 
{We chanced upon this filament while studying the environment of the galaxy 2MASX J11093966-1235116 \citep[the host galaxy of Gamma Ray Burst, GRB,  171205A at $z=0.037$, see][]{Arabsalmani22-2022AJ....164...69A}. We  noticed the alignment of a number of \hi\ sources that we had detected in the  field of the galaxy along a direction with a position angle of $\sim 45^\circ$ with respect to North.  We then inspected the  Pan-STARRS optical images  of the  field of the galaxy in large scales  and found a number of galaxies aligned in the same direction, covering about 2$^\circ$ in the sky. 
The tantalizing possibility that these  were  marking  a cosmic filament prompted us to further observe the field and investigate the properties of the identified  galaxies and the \hi\ sources, both in radio and optical wavelengths.  The details of the observations used in this study are presented in Section \ref{sec:obs}. We describe and discuss  our findings in Sections \ref{sec:res} and \ref{sec:dis} and summarise them in Section \ref{sec:sum}.
} 
Throughout this paper we use a standard flat $\rm\Lambda$CDM model with $H_0=69.6\,\rm km\,s^{-1}\,Mpc^{-1}$ and $\Omega_m=0.286$.


\section{Observations and data analysis}
\label{sec:obs}

\subsection{JVLA data and \hi\ source identification}
\label{sec:jvla}
{Observations with the  L-band receivers of the Karl J. Jansky Very Large Array (JVLA) in B and C configurations were used  to map  the \hicm\ emission in a section of the identified filament.}
The observations in the B-configuration, originally as  part of a program targeting the \hi\ properties of  the galaxy 2MASX J11093966-1235116, were   carried out in March-2019 and May-2019 for a total time of $\sim$ 10.5 hours (proposal ID: VLA/2018-07-058; PI: Arabsalmani). The JVLA primary beam 
was  centred at the position of 2MASX J11093966-1235116.   With a half-power primary beam width (HPBW) of $\sim$ 30\arcm\ at the observed frequency  in the L band, a 1.3 Mpc long section of the filament was covered within the HPBW (marked in  the right panel of Figure \ref{fig:field}). 

Follow up C-configuration observations were carried out in October-2022 for a total time of $\sim$ 17.5 hours (proposal ID: VLA/2022-00-109; PI: Arabsalmani) {in order to study the same section of the identified filament}. For both configurations, we used the JVLA Software Backend with 16 MHz bandwidth, centred on $\sim$ 1.368 GHz, sub-divided into 4096 channels, yielding a velocity resolution of $\sim$ 0.9 km~s$^{-1}$ and a total velocity coverage of $\sim 3500$~km~s$^{-1}$. 
The bright calibrator 3C286 and the secondary calibrator J1130-1449 were observed to calibrate the flux, time dependent part of the gain, and the system bandpass.

Following  the procedure described in \citet[][]{Arabsalmani22-2022AJ....164...69A}, ``Classic'' {\sc aips}   \citep[][]{Greisen03-2003ASSL..285..109G}  was used to calibrate, self-calibrate, subtract the continuum from the calibrated visibilities for  each day's data, and create cumulative visibility datasets separately for B and C configurations (see Appendix \ref{app:data} for more details). We also combined the residual visibilities from the full 5 days in order to create a combined B+C configuration dataset. 
The three sets of combined visibilities (B-configuration, C-configuration, and B+C-configuration) were Fourier transformed to produce spectral cubes using the task {\sc imagr}. 
For each configuration, we created two data cubes, with optimized velocity resolutions of $\sim 34$ \kms\ and  $\sim 50$ \kms\ to improve the statistical significance of the detected H{\sc i} 21\,cm emission in independent velocity channels while still having sufficient resolution to accurately trace the velocity field of any detected source. 
In order to optimize between the signal-to-noise ratio of the detections and the spatial resolution,  robust factors of 0.5, $-$0.5, and 0.5  were  used for the B-configuration, C-configuration, and B+C-configuration visibilities respectively, when creating the cubes, resulting in angular resolutions of 7.5\arcs $\times$ 6.4\arcs, 20.3\arcs $\times$ 17.0\arcs, and 10.5\arcs $\times$ 9.1\arcs, respectively. 
The properties of the produced cubes are provided in Table \ref{tab:cubes} in Appendix \ref{app:data}. 
We also applied the primary beam correction using the task {\sc pbcor}. 

Identifications of \hi\ sources were carried out in two steps: (i) we first did a blind search using the Source Finding Application (SoFiA) software (see Appendix \ref{app:data} for more details), (ii) next for each of the sources identified by SoFiA, we created two separate spectra from the 35 and 50 \kms\ data cubes. The spectra were extracted from the cubes in the regions defined based on the moment 0 maps created by SoFiA. We only kept those  detected \hi\ sources  which have signal-to-noise ratios (SNRs) of 5 and above in both 35 and 50 \kms\ cubes. 
We measured the \hi\ integrated flux density  by integrating over the adjacent channels with fluxes above the rms noise, and converted this to  a \hi\ mass. We  also fitted a Gaussian function to the \hicm\ emission line to measure the redshift and the FWHM of the emission.

\subsection{CFHT Data and optical source identification}
\label{sec:cfht} 

We obtained deep multiband (u, g, r, and i) images of the same portion of the filament described in Section \ref{sec:jvla} with the MegaCam camera installed on the CFHT  on 16-04-2022 (proposal ID: 22ad91; PI: Le Floc'h). As described in \citet[][]{Duc15-2015MNRAS.446..120D}, the images were obtained by  (a) acquiring consecutively seven individual images with large offsets between them, (b) building a master sky median-stacking the individual images,  and (c) subtracting a smoothed and rescaled version of the master sky from each individual image before stacking them. The final stacked images consist of $4 \times 7$ frames with a total exposure time of 82 min for the u band, 42 min for the g and r bands, and 27 min for the i band. The observations were obtained with an average seeing of 1.1 arcsec, under photometric conditions. Images were processed by the ELIXIR - LSB pipeline. We estimate the upper limits of  29.5, 29.5, 29, and 28 $\rm mag\,arcsec^{-2}$ for the local  surface brightnesses in the u, g, r, and i bands respectively.

To detect the possible optical counterparts of the \hi\ sources, we ran the SExtractor software \citep[][]{Bertin96-1996A&AS..117..393B} in ``double image" mode, using the deepest CFHT r-band image as the detection reference for all bands. This mode also ensures that object magnitudes were measured consistently for all images (e.g. aperture positions and sizes). 
We used   {SExtractor MAG{\_}AUTO parameter}  to measure  the total flux of the detected sources in each image. 
To ensure detection of very faint optical sources, the detection threshold of the SExtractor was set to 1$\sigma$ and spurious sources were {removed manually after a visual inspection.} In case of non-detection in a band, the image depth obtained with small blank apertures around the source position was used as an upper limit.

\subsection{Ancillary Data}
\label{sec:Adata}

We  used the SIMBAD Astronomical Database for creating a list of  galaxies identified in optical surveys in the field around  2MASX J11093966-1235116.       
For the eight galaxies listed in Table \ref{tab:g8} we used all the publicly available photometries (based on measurements of total fluxes) from 2MASS, GALEX, Pan-STAARS, and WISE surveys.  
For the detected optical source candidates in our CFHT images, we used  the publicly available photometries  from the  Legacy and WISE Surveys \citep[][]{Dey19-2019AJ....157..168D}. We cross-matched the magnitudes obtained from the CFHT images with those of the Legacy Survey  and found them to be consistent.


\begin{figure*}
\centering
\includegraphics[width=1 \textwidth]{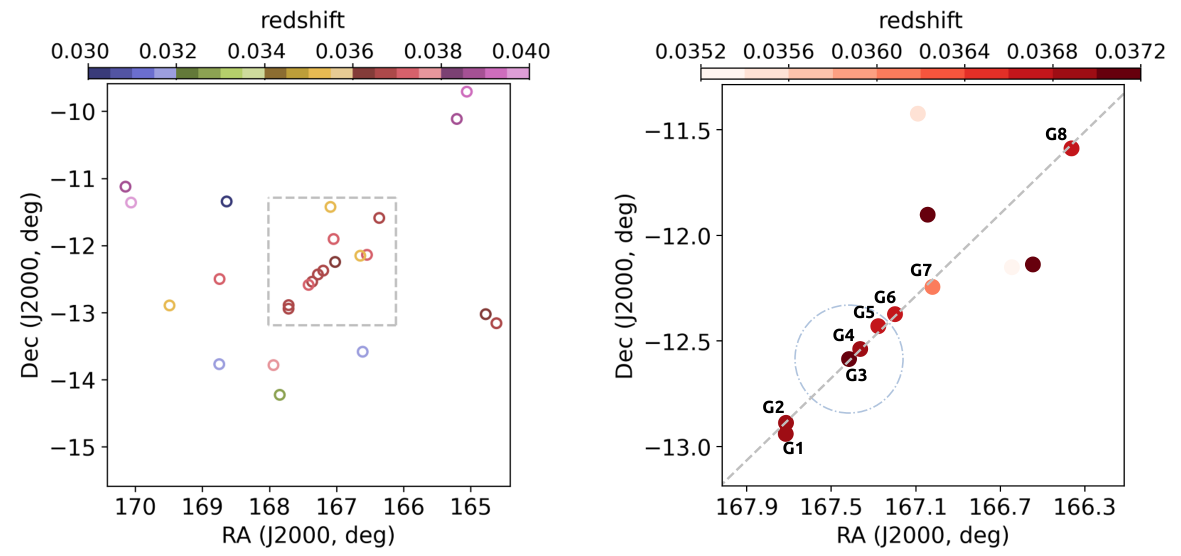}
\caption{ 
\textbf {Left:} $6^{\circ}\times 6^{\circ}$ image, centred at the position of 2MASX J11093966-1235116 (G3), showing galaxies with measured spectroscopic redshift listed in SIMBAD. We have also added LEDA 951348 (G4) with its redshift measured from the \hicm\ emission line. The gray-dashed box marks the frame of the right panel. 
\textbf{Right:} $2^{\circ}\times 2^{\circ}$ image, showing the eight galaxies (G1-G8)  along  the narrow filament (see Table \ref{tab:g8} for galaxy properties). The gray-dashed line shows the best fitted straight line to the positions of the eight galaxies. 
In both panels the color of the points represents the redshift values of the galaxies following the representative color-bars above the panels. 
The dot-dashed-blue  circle (centred at the position of 2MASX J11093966-1235116, G3) marks the JVLA HPBW for the \hicm\ emission line observations we present in this letter.  
\label{fig:field}}
\end{figure*}

\section{Results}
\label{sec:res}

\subsection{A  narrow filament of galaxies}
\label{sec:filament}
The left panel of Figure \ref{fig:field} shows the distribution of galaxies with known spectroscopic redshifts within a  $6^{\circ}\times 6^{\circ}$ image, centred on the galaxy 2MASX J11093966-1235116 at $z=0.037$. One can immediately distinguish the assembly  of eight  galaxies (hereafter as G1 to G8) along an almost straight  line at $z\sim 0.037$. The right panel of the figure shows  the eight galaxies G1--G8 lined up, creating a very straight galaxy chain.  
The redshifts of all galaxies but G4 are   measured  from optical spectroscopic data and  are taken from   \citet[][]{Jones04-2004MNRAS.355..747J, Jones09-2009MNRAS.399..683J} and \citet[][]{Shectman96-1996ApJ...470..172S}. For G4 (LEDA 951348) the redshift is derived from our \hicm\ emission line observations \citep[][]{Arabsalmani22-2022AJ....164...69A}. All the redshift measurements are quite accurate (see Table \ref{tab:g8}). These are in a very small range between $z=0.03609$ and $z=0.03702$, with a median redshift of {$z_{\rm med}=0.03683$} {same as the average redshift}. 

\begin{figure*}
\centering
\begin{tabular}{cc} 
\includegraphics[width=0.45 \textwidth]{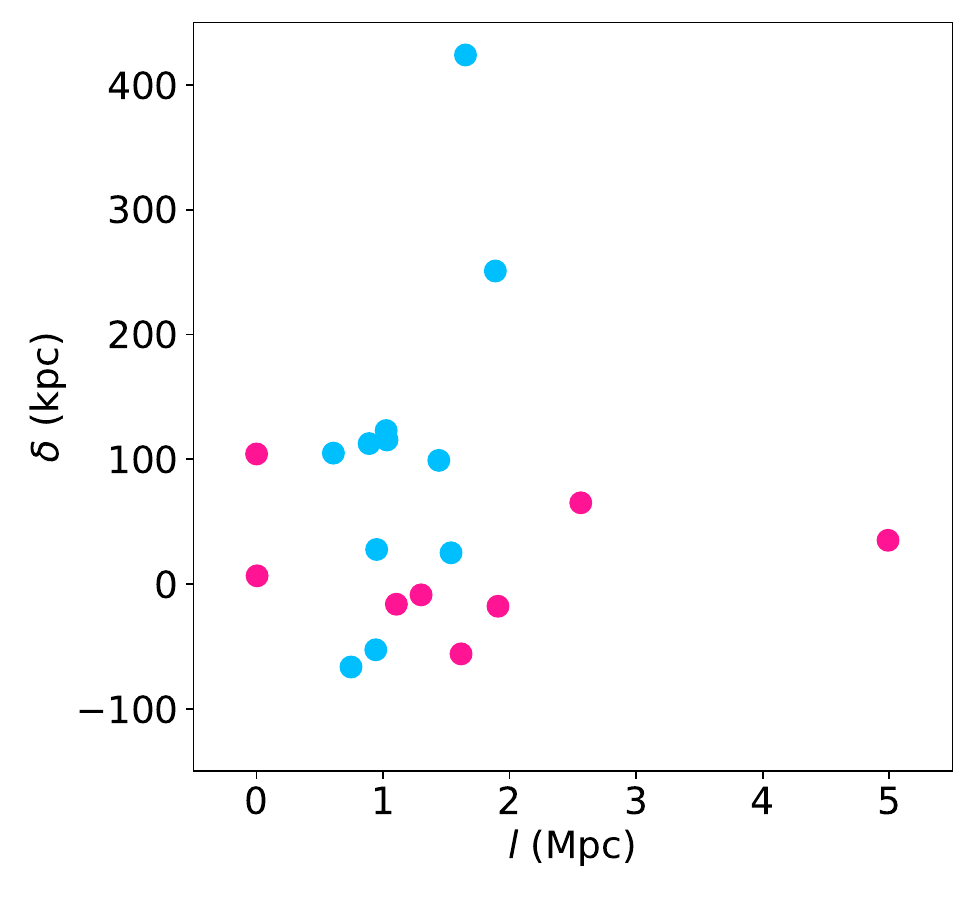}&\hskip 5 mm
\includegraphics[width=0.45 \textwidth]{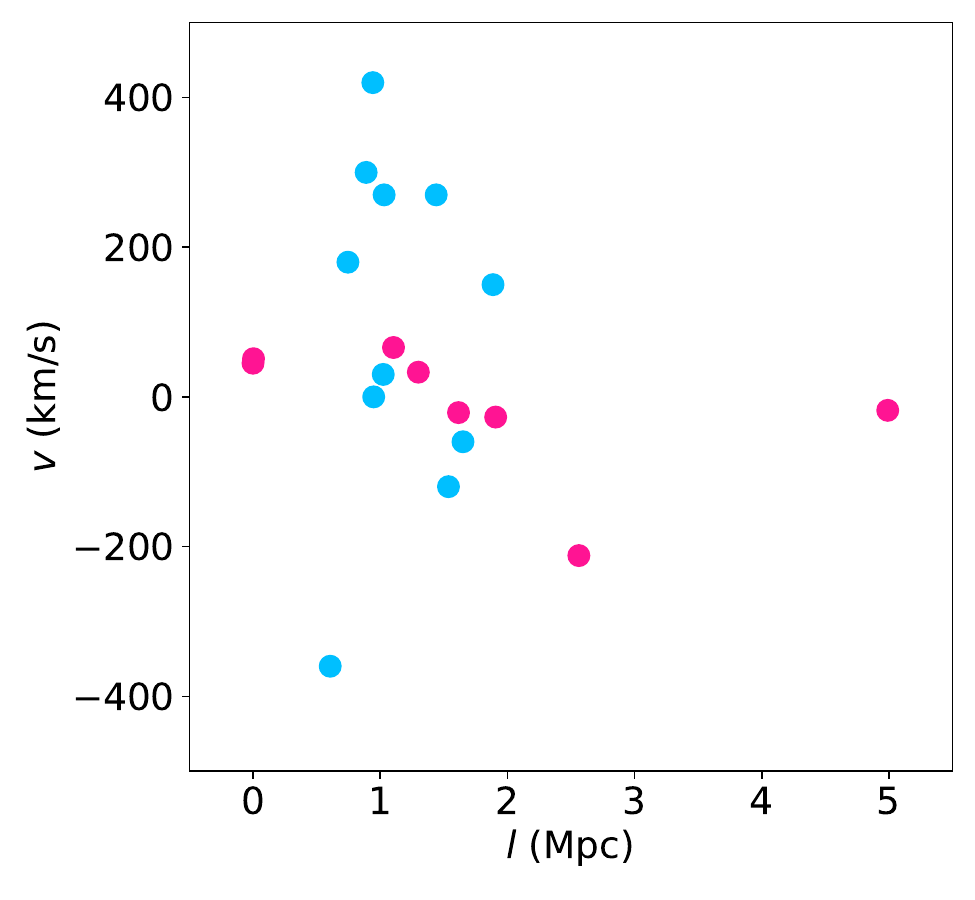}
\end{tabular}
\caption{ 
\textbf {Left:}  The  perpendicular {(on-sky)} distance  to  the filament axis, $\delta$, versus the distance   along the filament axis from a reference point on the line, $l$. {We have chosen G1 as  the reference point which is located at $l=0$} (hence G8 is located at $l=4.99$ Mpc, the length of the filament). The filament axis is presented with the dashed line  in the right panel of Figure \ref{fig:field}. 
\textbf{Right:}  The radial velocity with respect to $z_{\rm med}=0.0368$ ({the median, and also the average,} redshift of the eight galaxies on the filament), $v$, versus $l$. The pink and blue points represent  the eight galaxies and the eleven \hi\ sources, respectively, in both panels. 
\label{fig:distance}}
\end{figure*}

In order to investigate the structure of the galaxy chain  in the perpendicular direction to the sky, we use the redshift measurements of G1--G8 and obtain their relative radial velocities with respect to  {$z_{\rm med}=0.0368$}. These velocities, presented in the right panel of Figure \ref{fig:distance}, vary   from  -212 \kms\ to 66 \kms, similar to the velocities expected from the  proper motion of galaxies. This implies that G1--G8 are creating a very straight filament on the plane of the sky. 
One cannot fully discount the possibility that the slight differences in the redshifts of G1--G8 are due to their different  distances (and not their proper motions), ie., G1--G8   are distributed over a 2-dimensional plane (sheet) rather than being along a filament on the plane of the sky. Investigating this possibility requires independent distance measurements to the galaxies, which are exceedingly difficult, if at all possible given the distance to the sources and their properties and magnitudes \citep[see for e.g.,][]{Steer17-2017AJ....153...37S}. But the even distribution of the radial velocities of G1-G8 with respect to $z_{\rm med}$  (right panel of Figure \ref{fig:distance}) indicates that they are likely on the plane of the sky and hence belong to a filament.

Even more striking is the narrowness of the alignment on the plane of the sky. We fit a straight line to the positions of G1--G8 on the sky. This is shown with the dashed line in the right panel of Figure \ref{fig:field}.  The eight galaxies reside  very close to this line (hereafter filament axis), with projected (on the sky) distances varying in a  small  range between 7 kpc to 104 kpc. This is very small compared to the projected length of the filament which is about 5 Mpc. The left panel of Figure \ref{fig:distance} shows the distribution of the separation  from filament axis, $\delta$,   versus the distance from a reference point along the line, $l$,  both projected on the sky. We have chosen this reference point to be the projected position of G1 on the line, so that $l$ is 0 and 4.99 Mpc for G1 and G8 respectively.

In principle a chance alignment of galaxies in the plane of the sky can result in the filament we have identified. We cannot precisely determine the probability of such a chance alignment   given that the identification of the filament  was serendipitous and  emerged   while  studying the field of a nearby GRB (a single field).  However, the probability should be insignificant since the discovery was not  based on a systematic search.  
Future studies are planned in order to address the probability of detecting such aligned systems by examining large data sets.

We note that there are several more objects, identified as galaxies in large surveys, whose on-sky locations are coincident with the filament. But these galaxies do not have spectroscopic redshift measurements (possibly for being too faint) and hence their association with the filament is not clear. We do not include these galaxies in this study, but further spectroscopic observations are planned for these objects in order to explore their association with the filament.  

\begin{table*}
\begin{center}
\caption{The properties of the eight galaxies  that reside along a narrow filament shown in  the right panel of Figure \ref{fig:field}.}
\label{tab:g8}
\begin{tabular}{llcccccc}
\hline
ID & Name as in optical surveys & RA(J2000) & Dec(J2000) & redshift$^*$ &  $\rm log_{10}(M_{*}/M_{\odot})$  &  $\rm log_{10}(SFR/M_{\odot}\,yr^{-1})$  & E(B-V)  \\
\hline
G1 & LEDA 114626 & 11:10:51.48 & -12:56:27.3 & 0.03695$\pm$0.00015 &  9.75$\pm$0.03 & -1.90$\pm$0.07 & 0.1 \\
G2 & 2MASX J11105122-1253193 & 11:10:51.22 & -12:53:19.1 & 0.03697$\pm$0.00015 &  10.70$\pm$0.03  &    -1.70$\pm$0.07 & 0.1 \\
G3 & 2MASX J11093966-1235116  & 11:09:39.66 & -12:35:11.6 & 0.03702$\pm$0.00015 &  10.15$\pm$0.03  &    0.10$\pm$0.07 & 0.2 \\
G4 & LEDA 951348 & 11:09:26.97 & -12:32:20.6 & 0.03691$\pm$0.00012 & 9.10$\pm$0.03  &    -0.80$\pm$0.07 & 0 \\
G5 & LEDA 104056 & 11:09:06.43 & -12:25:51.6 & 0.03673$\pm$0.00022 &  9.80$\pm$0.03  &    -0.70$\pm$0.07 & 0.1 \\
G6 & 2MASX J11084744-1222264 & 11:08:47.44 & -12:22:26.6 & 0.03671$\pm$0.00015 &  10.45$\pm$0.03  &    0.10$\pm$0.07 & 0.2 \\
G7 & 2MASX J11080494-1214422 & 11:08:04.93 & -12:14:42.2 & 0.03609$\pm$0.00015 & 10.15$\pm$0.08  &    -0.15$\pm$0.18 & 0.2 \\
G8 & 2MASX J11052703-1135216 & 11:05:27.05 & -11:35:21.4 & 0.03674$\pm$0.00015 & 10.20$\pm$0.03  &    -1.30$\pm$0.07 & 0 \\
\hline
\end{tabular}   
\end{center}
\flushleft{
The last three columns list the stellar mass, star formation rate, and E(B-V) of the galaxies obtained from the SED modeling (see Figure \ref{fig:seds-galaxies} for the  best fitted SEDs). 
G1, G2 and G8, the three galaxies  at the two ends of the filament, appear to be passive galaxies while the  remaining five galaxies are all actively making stars and are  amongst the Main-Sequence galaxies.
\\
$^*$ The redshifts of G2, G3, G6, G7, and G8 are taken from  the Final Release of 6dFGS \citep[][]{Jones04-2004MNRAS.355..747J, Jones09-2009MNRAS.399..683J}. 
The redshifts of G1 and  G5  are taken from The Las Campanas Redshift Survey \citep[][]{Shectman96-1996ApJ...470..172S}. For  G4 (LEDA 951348), the redshift is obtained from the \hicm\ emission line \citep[][]{Arabsalmani22-2022AJ....164...69A}. 
}
\end{table*}


We model    the  spectral energy distribution (SED) of G1--G8  using {\textit{LePhare}} \citep[][]{1999MNRAS.310..540A, 2006A&A...457..841I}  through  {\textit{GAZPAR}} (Galaxy redshifts and physical parameters)  community service  in order to obtain  their stellar properties such as stellar mass and star formation rate (SFR) values. These properties are  listed in Table \ref{tab:g8} (the fitted SEDs are provided in Appendix \ref{app:opt}). The galaxy stellar masses are in a range of $\rm 10^{9.1}-10^{10.7}\,M_{\odot}$. 
G1, G2 and G8, the three galaxies  at the two ends of the filament, appear to be passive galaxies based on their location on the $\rm M_*-SFR$ plane \citep[][]{Brinchmann04-2004MNRAS.351.1151B}. The color-color diagram analysis \citep[][]{Arnouts13-2013A&A...558A..67A} confirms that G2 is   indeed a passive galaxy, but suggests  that G8  is  transiting from  star-forming to passive. The limited  photometries for G1 does not allow studying its position in the   color-color diagram.   
The remaining five galaxies all lie within the star-forming Main Sequence galaxies on the $\rm M_*-SFR$ plane, and are also categorised as star-forming galaxies  based on the galaxy color-color diagram. 
Using the \hicm\ emission line observations in the 1.3 Mpc long  section of the filament (see Section \ref{sec:jvla}), we  measure the \hi\ masses for  G3, G4, and G5, listed in Table \ref{tab:HI}.   We find G3 and G4 to  be  gas-rich galaxies  with   \hi\ masses above average for their stellar masses \citep[see][for $\rm M_*-M_{HI}$ relation of nearby galaxies]{Catenilla18-2018MNRAS.476..875C}. The $\rm M_{HI}/M_{*}$ ratio of G5 though is towards the lower side of the $\rm M_*-M_{HI}$ relation.

\begin{figure*}
\centering 
\includegraphics[width=0.8 \textwidth]{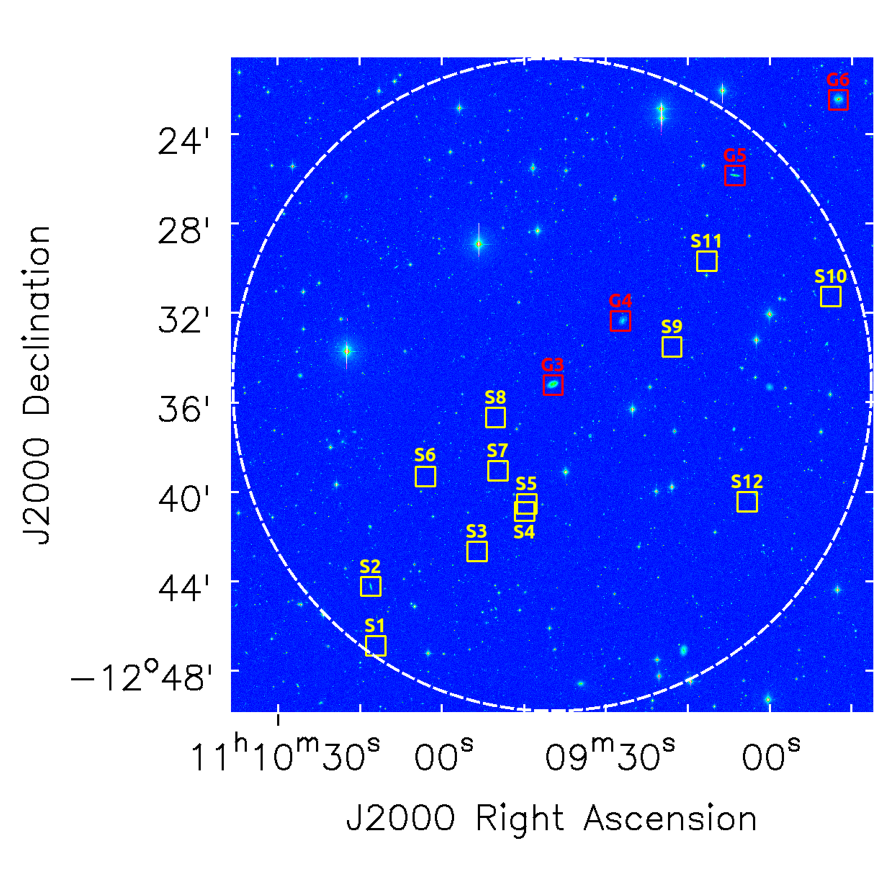}
\caption{ 
The r-band CFHT image, with the  squares marking  the positions of the \hi\ sources detected within the JVLA HPBW (listed in Table \ref{tab:HI}).    G3-G6 are marked with  red squares. The HPBW of JVLA is marked with a white-dashed circle.
\label{fig:HI}}
\end{figure*}

\subsection{Dark \hi\ clouds on the filament}
\label{sec:HI}
The  HPBW of JVLA in our \hicm\ emission line observations, marked with a dot-dashed-blue circle in the  right panel of Figure \ref{fig:field},  covers  a 1.3 Mpc long portion of the filament centred at G3. We restrict {mapping the \hicm\ emission line}  to within the JVLA HPBW in order to only detect \hi\ emission with high sensitivity. 
We identify fifteen  \hi\ sources in this region, all   marked on the  r-band CFHT image presented in Figure \ref{fig:HI}. The \hi\ properties of these sources, including the  (spectroscopic) redshift measurements   obtained from their \hicm\ emission lines, are listed in Table \ref{tab:HI}.  The \hi\ masses are within a  range of $\rm 10^{8.5}-10^{9.5}\,M_{*}$ and the \hicm\ line-widths vary in a range of $\rm 110-420$ \kms. 

Three of the detected sources are associated with G3, G4, and G5. Note that we do not detect the \hicm\ emission line for G6 as expected since  this galaxy falls out of the JVLA HPBW. 
One of the detected \hi\ sources, S2,  is spatially coincident with a spiral galaxy with no measured {spectroscopic} redshift in optical surveys. This galaxy resides at a projected distance of 12 kpc from the filament axis (the dashed line in the right panel of Figure \ref{fig:field}). We obtain a redshift of $z=0.0341$ from the \hicm\ emission line of this source, consistent with the  photometric redshift of $z < 0.2$  that we obtain from modeling the SED of the galaxy.
However, we detect signatures of systematics, appearing like  absorption features over a velocity width of $\sim$ 1000 \kms\  in the \hicm\ spectrum of this source (see Figures \ref{fig:specs} and \ref{fig:specs-50} in Appendix \ref{app:data}). We examine these features in the intensity map of the source   and find them not to be present. In addition, no continuum emission in our JVLA observations  is detected at the location of S2. We therefore conclude  that the apparent features in the spectrum are due to systematics of unknown origin.  This could have affected the redshift measured from the \hicm\ emission of this source. We therefore exclude this source from any further analysis.  

The remaining eleven sources are distributed along the filament, with projected distances of 25-424 kpc from the filament axis.  They are marked with blue circles in Figure \ref{fig:distance}. Nine out of the eleven  \hi\ sources are located within 123 kpc from the filament axis, similar to the $\delta$ range for G1--G8. Only two sources, S10 and S12,  are located at slightly larger  distances of $\delta=$ 251 kpc and 424 kpc respectively. 
We also derive the radial velocities of these sources with respect to $z_{\rm med}=0.0368$. These are within a range between -360 \kms\ and 420 \kms, and are shown in the right panel of Figure \ref{fig:distance}. The radial velocities of the \hi\ sources are somewhat larger than those  of G1--G8.

We search for the  optical counterparts of the eleven  \hi\ sources  using our deep, multiband CFHT images (see Section \ref{sec:cfht} for the magnitude limits). 
For each \hi\ source, we identify all the detected optical sources within the extent of the \hi\ emission. This extent  is   defined by the 1$\sigma$ contour of the total \hi\ emission. For each of the identified optical sources we  model the SED  with {\textit{LePhare}} (through {\textit{GAZPAR}}) and obtain the probability distribution of the photometric redshift ($z_{phot}$) based on modelling its SED. 
Most of the identified  optical  sources are very unlikely to have  a $z_{phot} \lesssim 0.1$  (with probabilities $<$ 1\%; see Appendix \ref{app:opt} for more details). 

To very  conservatively  estimate the lower limit of $\rm M_{HI}/M_*$ for each \hi\ source, we assume that those   optical counterpart candidates with $\gtrsim 1\%$ probability of having $z_{phot} \lesssim 0.1$ are  associated with the \hi\ cloud. These are marked with yellow circles in the middle panel  of Figure \ref{fig:maps}).   Assuming a redshift of {$z=0.0368$ (the median/average redshift of G1--G8)} for these sources, we  model their  SEDs and estimate their stellar masses.  By adding the stellar masses of the associated optical sources for each \hi\ cloud, we  obtain the corresponding $\rm M_{HI}/M_*$  for the \hi\ cloud.  The estimated $\rm M_{HI}/M_*$ values are    listed in Table \ref{tab:HI}. These  are within a range of $\sim 30-15500$, with a median of $\sim 900$. 
Note that the actual \hi\ masses for our detected \hi\ sources should be larger than what we have estimated given that the performed JVLA observations are not sensitive to low column densities (N(\hi)$\rm \leq$ a few times $\rm 10^{20}\, cm^{-2}$, depending on the array configuration), making the  $\rm M_{HI}/M_*$ estimates even more conservative. This  implies   that the \hi\ sources that we have detected in the filament are quite dim/dark.

\begin{table*}
\begin{center}
\caption{The detected \hi\ sources marked in  Figure \ref{fig:HI} and their  properties obtained from their \hicm\ emission lines.}
\label{tab:HI}
\begin{tabular}{lcccccccccc}
\hline
ID & $\rm z_{HI}$ & $\rm \Delta v_{0.037}$ & FWHM$_{\rm HI}$ & $\int SdV$	&$\rm log_{10}(M_{HI}/M_{\odot})$   & $\rm M_{HI}/M_{*}$ &  N(\hi)   & S/N$_{\rm 35}$ & S/N$_{\rm 50}$ & Array\\
   &              &  (\kms)                & (\kms)          & (mJy \kms)   &                                   &                    & (cm$^{-2}$) &                &                &      \\
\hline
G3          &  0.0370  &  3    &  317$\pm$58  &  501$\pm$47  &  9.50$\pm$0.04  &  0.2	& 6.1 $\times 10^{20}$ 	&  11  &  10  &  B   \\
G4          &  0.0369  &  -27  &  117$\pm$18  &  485$\pm$53  &  9.49$\pm$0.05  &  2.5	& 6.2 $\times 10^{20}$	&   9  &   9  &  B  \\
G5          &  0.0367  &  -92  &  170$\pm$37  &  83$\pm$14   &  8.72$\pm$0.07  &  0.1	& 2.8 $\times 10^{20}$	&   6  &   5  &  B+C  \\
\hline                                                          
      	    &          &       &              &                 &                 & $\rm (M_{HI}/M_{*})_{LL}$\\ 
\hline  
S1          &  0.0356  &  -428 &  127$\pm$34  &  198$\pm$37  &  9.10$\pm$0.08  &  89	& 2.8 $\times 10^{20}$	&   5  &   5  &  B+C  \\                                                       
S2$^*$      &  0.0341  &  -875 &  132$\pm$45  &  139$\pm$29  &  8.95$\pm$0.08  &  1	& 2.7 $\times 10^{20}$	&   6  &   5  &  B+C  \\
S3          &  0.0378  &  245  &  209$\pm$59  &  102$\pm$20  &  8.81$\pm$0.08  &  2570	& 6.6 $\times 10^{20}$	&   5  &   6  &  B  \\
S4          &  0.0369  &  -21  &  112$\pm$26  &  156$\pm$25  &  9.00$\pm$0.07  &  10000& 5.9 $\times 10^{20}$	&   6  &   6  &  B  \\
S5          &  0.0377  &  199  &  177$\pm$56  &  133$\pm$24  &  8.93$\pm$0.07  &  13/5370$^{**}$	& 6.7 $\times 10^{20}$	&   6  &   5  &  B  \\
S6          &  0.0374  &  108  &  178$\pm$45  &  198$\pm$35  &  9.10$\pm$0.08  &  1000	& 6.4 $\times 10^{20}$	&   6  &   7  &  B  \\
S7          &  0.0368  &  -46  &  229$\pm$58  &  50$\pm$9    &  8.50$\pm$0.07  &  316	& 1.6 $\times 10^{20}$	&   6  &   5  &  B+C  \\
S8          &  0.0382  &  369  &  304$\pm$105 &  45$\pm$9    &  8.46$\pm$0.09  &  912	& 2.1 $\times 10^{20}$	&   5  &   7  &  B+C  \\
S9          &  0.0377  &  216  &  126$\pm$33  &  110$\pm$18  &  8.85$\pm$0.07  &  224	& 1.8 $\times 10^{20}$	&   6  &   5  &  B+C  \\
S10         &  0.0373  &  83   &  424$\pm$91  &  386$\pm$60  &  9.39$\pm$0.06  &  15488& 8.1 $\times 10^{20}$	&   6  &   5  &  B  \\
S11         &  0.0364  &  -190 &  134$\pm$29  &  199$\pm$36  &  9.10$\pm$0.08  &  96	& 7.8 $\times 10^{20}$	&   6  &   6  &  B  \\
S12         &  0.0366  &  -124 &  203$\pm$42  &  203$\pm$31  &  9.11$\pm$0.06  &  28  	& 9.3 $\times 10^{19}$	&   7  &   7  &  C\\
\hline
\end{tabular}   
\end{center}
\flushleft{
Columns 2 to 6 are, respectively,  redshift obtained from the \hicm\ emission line, velocity offset of the  \hicm\ emission  with respect to $z=0.037$, Full-Width-at-Half-Maximum of the \hicm\ emission line, integrated flux density of the \hicm\ emission line, and \hi\ mass. Column 7 is the ration of the \hi\ mass to stellar mass for G3-G5, and   the lower limit on $\rm M_{HI}/M_{*}$ for the rest of the \hi\ sources assuming that all the optical sources listed in Table \ref{tab:optical} in Section \ref{app:opt} are associated with the \hi\ sources. Column 8 is the \hi\ column density equivalent to the $3\sigma$ level of emission in a single channel with a velocity width of 34 \kms\ (see Figure \ref{fig:maps} for the \hi\ contours).  Columns 9 to 11 are the signal to noise ratios of the \hicm\ emission line detected in cubes with channel widths of 35 \kms\ and 50 \kms, and the array configuration used for detecting the \hi\ source and obtaining the \hi\ spectrum. \\
$^*$  The \hi\ spectrum of this source is affected by systematics of  unknown origin (see Section \ref{sec:HI} for details).  We therefore exclude this source from our analysis.   \\
$^{**}$ For this source we have provided the $\rm M_{HI}/M_{*}$ ratio  with including/excluding the optical source that is outside the \hi\ emission (marked by e in Figure \ref{fig:maps}).
}
\end{table*}

\section{Discussion}
\label{sec:dis}

\subsection{The peculiarity of the filament}
Following the identifications of large numbers of filaments   in the nearby Universe, mainly through structure-finding algorithms, the geometrical properties of  filaments   have been studied in details.  
Filament lengths  are reported to  vary between a few Mpc up to $\sim$ 150 Mpc, with typical lengths of a few tens of Mpc \citep[e.g.,][]{Tempel14-2014MNRAS.438.3465T, Santiago-Bautista20-2020A&A...637A..31S, Bonjean20-2020A&A...638A..75B}. 
The radius of the nearby filaments are  found to be   around or more than a Mpc, consistent with predictions from simulations \citep[e.g.,][]{Cautun14-2014MNRAS.441.2923C}. 
Filament galaxies are typically at distances $>$ 100 kpc from the filament spine \citep[see for e.g.,][]{Santiago-Bautista20-2020A&A...637A..31S, Bonjean20-2020A&A...638A..75B, Castignani22-2022A&A...657A...9C},   with typical distances of about a few Mpc in most cases. 
In contrast, the filament we present here  has all its identified  galaxies residing at distances of $\lesssim$ 100 kpc (mainly a few tens of kpc) from its axis, making  this system strikingly narrow (see the left panel of Figure \ref{fig:distance}). With a $\sim$ 5 Mpc length, this filament is amongst the shortest filaments that have been identified. 
\citet[][]{Galarraga-Espinosa20-2020A&A...641A.173G} examined the geometrical properties of the simulated filaments in TNG simulations and found that short filaments with lengths of a few Mpc  are puffier than long filaments.  Therefore, the short length of the presented filament here makes its narrowness even more unusual. 
In addition to its narrowness, the straightness of this filament is quite  intriguing.

Although the low number of the galaxies (eight) in the  narrow filament does not allow a statistical comparison with the properties of  large filament galaxy samples, we find the properties of the eight   galaxies  to  stand out. 
Properties of  filament galaxies have been extensively studied in order to explore  the  effect of filaments (predicted by simulations) on the formation and evolution of galaxies \citep[][]{Alpasla16-2016MNRAS.457.2287A, Martinez16-2016MNRAS.455..127M, Kleiner17-2017MNRAS.466.4692K, Kuutma17-2017A&A...600L...6K, Malavasi17-2017MNRAS.465.3817M,  Kraljic18-2018MNRAS.474..547K,  Laigle18-2018MNRAS.474.5437L, Sarron19-2019A&A...632A..49S, Bonjean20-2020A&A...638A..75B, Castignani22-2022A&A...657A...9C, Donnan22-2022NatAs...6..599D}. 
The focus of these investigations  has  been on how the properties of galaxies change   as one moves towards the spines of the filaments. Most of these studies    find an increase in the fraction of passive galaxies compared to star-forming galaxies when going closer to filament spines.  
At fixed stellar mass, \citet[][]{Laigle18-2018MNRAS.474.5437L}  showed  passive galaxies are  found closer to their filament than active star-forming galaxies. 
\citet[][]{Malavasi17-2017MNRAS.465.3817M} reported the presence of the most massive and quiescent galaxies  closer to the filament axis. 
\citet{Castignani22-2022A&A...657A...9C}  found the number of quenched galaxies at close distances to filaments to be significantly larger than that of active galaxies. 
Keeping in mind the  low number of the galaxies,  it is notable that five out of the eight   galaxies in this filament  are actively forming stars. This is in particular interesting given that the eight galaxies are very close to the filament axis ($7-104$ kpc). In  all the  studies mentioned above,  the filament galaxies are typically at distances $>$ 100 kpc from the filament spine, ie., the star-forming galaxies are at much larger distances from the filament spines compared to what we find here. 

It is notable that the star-forming galaxies in the filament are within its  mid-section, quite away from the two nodes. These star-forming galaxies are predicted to fall into the nodes of the filament where passive galaxies reside.  The concentration of the star-forming galaxies in the middle of the filament  might indicate that the filament is in its initial evolutionary stages.

In order to assess the rarity of the identified narrow  filament based on  theoretical predictions, we use   the filament catalogue of the TNG300-1 simulations at $z=0$ \citep[][]{Galarraga-Espinosa20-2020A&A...641A.173G} to search for filaments with similar galaxy and geometrical properties in the TNG catalogues.    
This catalogue is built up using the discrete persistent structure extractor (DisPerSe)  algorithm \citep[][]{Sousbie11-2011MNRAS.414..350S} which   operates  in three dimensions and detects filaments based on  the galaxy density field.
At first we select  filaments with lengths within a range of 4 to 6 Mpc,  containing 7 to 9 galaxies of stellar masses  $\rm 10^{9.1}-10^{10.7}\,M_{\odot}$ within 100 kpc from the filament spine. We do not put any constraint on the number of galaxies  that the filaments might contain  at distances larger than 100 kpc  from their  spines. We also allow selecting filaments that contain  any number of galaxies with stellar masses $\rm < 10^{9.1}\,M_{\odot}$ or $\rm > 10^{10.7}\,M_{\odot}$ within the 100 kpc from their  spines. 
With these constraints, we  find only 13 filaments  out of 10106 filaments in the filament catalogue, but more importantly, none is like the filament we have identified.

None of the selected  13 filaments is nearly as straight, and all of them are much thicker, with a few tens of  galaxies at distances of 100 kpc to 1 Mpc from the spine. 
Several ($>$10) of these galaxies have stellar masses of $\rm  10^{10}-10^{12}\,M_{\odot}$ -- well above the detection limit of large galaxy surveys in the field of the identified filament. The environment of the filament we present here is much emptier. There are only three galaxies detected  in large galaxy surveys (with $\rm \gtrsim 10^{10}\,M_{\odot}$) within the 1 Mpc distance of the filament axis (see  Figure \ref{fig:field}). Two of them (2MASX J11081033-1154122 and 2MASX J11061093-1208193), at distances of 630 kpc and 740 kpc  from the filament axis, are at $z\sim0.037$, the redshift  of the filament, and  the third galaxy (2MASX J11063472-1209059) is at $z\sim0.035$ ($\sim$ -500 \kms\ with respect to $z_{\rm med}=0.0368$). 
We also note that  in  11 out of the 13 selected filaments, all the  7--9 galaxies are  passive, and  in the remaining two filaments only  one or two of the 7--9 galaxies are star-forming.

The narrow and straight filament we have identified  seems  to have no match in the TNG catalogue mentioned above. This,  however,  could be a result of how the catalogue has been built up, i.e., using the  DisPerSe  method.  For instance,  in 10 out of the 13 selected  filaments in the catalogue, all the 7--9  galaxies are located in a $\sim$ 0.5 Mpc long section along the filament. This,  perhaps, is expected given that the DisPerSe  algorithm works based on identifying the peaks of the galaxy density.

\begin{figure}
\resizebox{8cm}{!}{\includegraphics{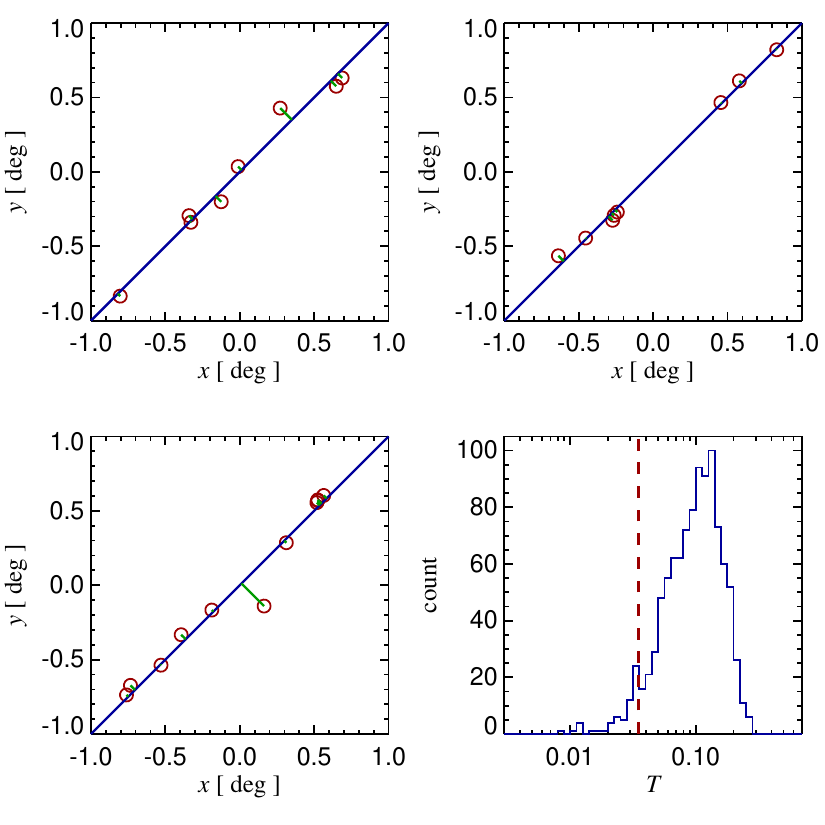}}
\caption{Three examples for $2\times 2$ degree fields in the full-sky
  galaxy lightcone of the MillenniumTNG simulation, selected to
  contain galaxies with similar masses and redshifts as observed, and
  to exhibit a similarly long and narrow filamentary distribution. We
  measure the thinness of the galaxy distribution through a quantity
  $T$ defined as the ratio between the averaged perpendicular
  distances (green) between galaxies and the ridge-line of their
  filament (blue), and the length of the extension of the galaxy
  distribution in the filament direction. The histogram in the bottom
  right panel shows the distribution of this quantity for the 1017
  fields we have with matching galaxy number density. For our observed
  galaxy distribution, the value of the thinness parameter is
  $T_{\rm obs} = 0.035$ (vertical dashed line). About 4\% of the
  fields have an equal or lower value for $T$, implying that the
  configuration we have discovered is rare in $\Lambda$CDM, but not
  exceptionally so. \label{FigThinness}}
\end{figure}

In order  to further  investigate   whether the existence of the  filament we have identified  challenges  $\Lambda$CDM and our understanding of  structure formation,  we perform a more comprehensive comparison with the simulations. 
To do so, we make use of the full-sky mock galaxy catalogue
computed by \citet{Barrera2023} as part of the MillenniumTNG
simulation project \citep{HernandezAguayo2023}, using a 740 Mpc box
with $4320^3$ simulation particles. The galaxy catalogue is produced
in a completely seamless fashion for the backwards lightcone of a
fiducial observer. We first select all galaxies in the redshift
interval $0.0360 < z < 0.0371$ and with stellar mass
$10^9\,M_\odot < M_\star < 10^{11}\,M_\odot$, closely matching the
characteristics of our filament galaxies. We then construct a set of
12288 observational fields of 2 degrees on a side based on pointings
that uniformly cover the sky with the help of a Healpix tessellation
of the unit sphere. To account for the centering onto the filament we
carry out in our observational analysis, we first identify all
galaxies in a spherical aperture with a radius of 2 degrees around
each of our pointings, and then adopt the center-of-mass of the
identified galaxies as new center. We then constrain to the galaxies
in a circular region of radius $\sqrt{2}$ degrees around the new
center, and compute a second order shape tensor for the
two-dimensional distribution of galaxies, allowing us to determine the
major and minor axes. We also turn the galaxy distribution such that
the major axis lies approximately diagonal, and then restrict to a
$2\times 2$ degree square-shaped field, just like in our observations
shown in the right panel of Fig~1. We only keep fields that have
between 8 to 12 galaxies in them, in order to closely mock up the
conditions we have in our observational field. This leaves 1017 fields
in our sample.

Finally, for the galaxies in each remaining field, we compute the
average distance $\left< D\right>$ of the galaxies to the major axis
-- which we identify as the filament spine -- and the maximum
extension of the galaxies along the filament direction, $L_{\rm
  max}$. We then use these two quantities to define a dimensionless
``thinness'' quantity $T = \left< D\right> / L_{\rm max}$ that
measures how narrow the galaxy distribution is. We can also measure
the value of $T$ for our observed filament, i.e.~the $8$ galaxies G1
to G8 plus the two galaxies that have somewhat larger distances from
this filament spine. This yields $T_{\rm obs} = 0.035$.

In the top two and bottom left panels of Figure~\ref{FigThinness}, we
show 3 examples of filamentary galaxy distributions found in this way,
all with $T$-values lower than $T_{\rm obs}$. They indeed look
comparably straight and narrow.  In the bottom right panel of
Figure~\ref{FigThinness} we examine this quantitatively in terms of
the distribution of $T$ over all our 1017 fields. The vertical dashed
red line marks the observed galaxy distribution. The observations are
in the tail of the distribution, confirming that the filament is
indeed unusual and rare. But it is not exceptionally uncommon. About
4\% of our fields have comparable or even smaller values for the
thinness parameter, so some fairly straight filaments can be found and
are expected in the $\Lambda$CDM galaxy distribution as well.

\subsection{Dark clouds: peaks of dark matter and \hi\ distributions}
In addition to the peculiarity of the filament in its geometrical properties, the identification of several  dark  \hi\ clouds in the filament is quite intriguing. 
The number of \hi\ sources we have detected in the field exceeds   the average number of \hi\ sources  that is typically detected in a similar volume. In order to estimate the excess of the \hi\ source count, we use the local \hi\ mass function derived  for large  \hi\ surveys.  
We  calculate the average number of \hi\ sources   (i) within  the \hi\ mass range of $\rm 10^{8.5}-10^{9.5}\,M_{\odot}$, the mass range of our detected \hi\ sources, and (ii) in a volume corresponding to a cylinder with a base as large as the HPBW of JVLA (marked with a white-dashed circle in Figure \ref{fig:HI}), and a  perpendicular distance of $d_l(z=0.0382)-d_l(z=0.0356)$. Here  $d_l(z)$ is the luminosity distance at redshift $z$, and  $z=0.0382$ and $z=0.0356$ are the maximum and minimum redshifts of our detected \hi\ sources, respectively.   Using the \hi\ mass function of the HIPASS Survey presented in \citet[][see their  Figure 1. and their best fitted parameters for a Schechter function]{Zwaan05-2005MNRAS.359L..30Z}, we find the average   number of \hi\ sources to be $0.16\pm0.04$. This is even lower ($0.11\pm0.01$) when we use the \hi\ mass function of the ALFALFA survey presented in \citet[][see their  Figure 2. and their best fitted parameters for a Schechter function]{Jones18-2018MNRAS.477....2J}. The number of the \hi\ sources that we have detected (see Table \ref{tab:HI} and Figure \ref{fig:HI}) are about 100 times   larger than that expected from the average counts in large \hi\ surveys.   

The \hi\ clouds we have detected in the filament  have larger radial velocities compared to the eight galaxies in the filament (see the right panel of Figure \ref{fig:distance}). This might imply that they are at larger distances (in 3D)  from the filament axis, and hence in regions with lower densities. On the other hand, the galaxies (that have  formed stars) are in higher density regions close to the filament axis.  
We also note that the typical distances between the \hi\ sources   (the eleven \hi\ clouds and the three galaxies with detected \hi\, presented in Figure \ref{fig:HI}) are $\sim$ 180 kpc.  
We do not have sufficient information to infer a robust origin (e.g., environment, filament's characteristics, Jean's scale length, etc.) for this preferred length scale.

Previous studies of \hi\ in filaments  resulted in the identification of only one \hi\ source with no known optical counterpart (in DSS image) amongst 199 detected sources \citep[][]{Popping11-2011A&A...528A..28P}. The reported source was detected at N(\hi) $\rm \sim10^{19}\,cm^{-2}$, with a narrow line-width of about 20 \kms. 
The dark \hi\ clouds that we identify in the filament are all detected at N(\hi)$\geq$ a few times $\rm 10^{20}\,cm^{-2}$, with \hi\ line widths $> 100$ \kms. 
Their   \hi\ masses are in the range $\rm 10^{8.5}-10^{9.5}\,M_{\odot}$, and their  $\rm M_{HI}/M_*$ lower limits   are within a range of $\sim 30-15500$, with a median of $\sim 900$. 
Nearby galaxies with $\rm M_* \gtrsim 10^8\,M_{\odot}$ typically have $\rm M_{HI}/M_* \lesssim 10$ \citep[see][for HIPASS and xGASS surveys]{Denes14-2014MNRAS.444..667D, Catenilla18-2018MNRAS.476..875C}. Galaxies with lower stellar masses  though are found to have higher \hi\ mass fractions. 
In particular, low-surface-brightness and irregular galaxies are shown to have $\rm M_{HI}/M_*$ as high as 100   \citep[see for e.g.,][]{Leisman17-2017ApJ...842..133L, Mahajan18-2018MNRAS.475..788M}.

The extremely  large \hi\ mass fractions that we obtain for the \hi\ clouds are comparable with, and in some cases exceed,  the highest  \hi\ mass fraction ever reported for dark systems, even those  in galaxy groups and clusters. 
Several  studies have reported $\rm M_{HI}/M_*<100$ for dark galaxy candidates \citep[e.g.,][]{Cannon15-2015AJ....149...72C, Janowiecki15-2015ApJ...801...96J, Bilek20-2020A&A...642L..10B, Roman21-2021A&A...649L..14R}. 
\citet[][]{Leisman21-2021AJ....162..274L} measured a $\rm M_{HI}/M_* \sim 100$ for an  extended \hi\  cloud with $\rm M_{HI}=10^{9.3}\,M_{\odot}$. 
Even  a threshold of $\rm M_{HI}/M_*>200$, defines eight  out of  the eleven  \hi\ clouds we have detected in the  filament  as  dark. This only decreases  to six systems with a threshold of $\rm M_{HI}/M_*>900$. Such large  \hi\ mass ratios for dark clouds have only been reported in  dense environments such as   galaxy groups and clusters.  
\citet[][]{Jozsa22-2022ApJ...926..167J} detected a massive chain of seven \hi\ clouds in a galaxy group   and found $\rm M_{HI}/M_* \sim 1200$, still smaller than the \hi\ mass fractions that we measure for half of the \hi\ clouds  in the narrow filament. 
\citet[][]{Lee-Waddell14-2014MNRAS.443.3601L} reported the presence  of a \hi\ cloud with $\rm M_{HI}=10^{9.3}\,M_{\odot}$ and  $\rm M_{HI}$ $\rm M_{HI}/M_* \sim 2000$ (though with a velocity width of only a few tens of \kms) in a galaxy group \citetext{\citealp[see also][]{Wong21-2021MNRAS.507.2905W} for   dark clouds in  galaxy groups, \citealp[and][]{Kent07-2007ApJ...665L..15K, Kent10-2010ApJ...725.2333K,  Sorgho17-2017MNRAS.464..530S} for dark clouds in the Virgo cluster}.   
Very recently \citet[][]{{Jones24-2024ApJ...966L..15J}} identified an extremely gas-rich,  low-mass, star-forming galaxy in the Virgo cluster, formed  via extreme ram pressure stripping events, with an extreme \hi\ mass fraction of $\sim 20000$. 
All these studies argue that  the detected  dark  \hi\ clouds within galaxy groups/clusters   are   the remnants of interacting systems or  tidal features that  survive in  high density environments.

Although our \hicm\ line observations are not sensitive enough to detect the diffuse \hi\ gas, the dark \hi\ clouds we have detected  must be  resembling  the peaks of the large-scale \hi\ distribution in the  filament. 
Such \hi\ peaks are in fact predicted by simulations to exist around the filament spines {at high redshifts} \citep[for e.g., see Figure 1. of][{for high resolution simulations of  filaments at $z\sim4$}]{Lu24-2024MNRAS.52711256L}. These massive gas clouds, with no stars, are expected to reside in the dark matter halos along the filaments.  
Most of the \hi\ clouds we detect in the filament    have large \hicm\ line-widths  for their masses. This together with their darkness could indicate that they belong to haloes that recently made it  to the cold-core  of the filament and   contain gas that might not yet have virialized - thus not forming stars, as predicted for high redshift filaments  \citep[e.g.,][]{Lu24-2024MNRAS.52711256L}.  
Note though that the \hi\ clouds we detect should also contain gas in  `cold neutral phase' with temperatures of $\lesssim 600$ K given their high \hi\ column densities \citep[N(\hi)$\geq$ a few times $\rm 10^{20}\,cm^{-2}$;][]{Kanekar11-2011ApJ...737L..33K}.  
We emphasise that the  properties of the filament we have identified in the local Universe can in principle be very different from those at high redshifts. Nevertheless, the  detection of these  dark/dim \hi\ clouds  along a filament  confirms  what has been predicted by simulations.

\subsection{Interactions  in the filament}
The \hi\ properties of filament galaxies have been characterised through a couple  of recent studies. 
\citet[][]{Kleiner17-2017MNRAS.466.4692K} found  that $\rm 10^{9} < M_* < 10^{11}\, M_{\odot}$ filament galaxies  have \hi\ mass fractions similar to those of control sample galaxies, galaxies  at  distances $\geq$ 5 Mpc from filament spines and selected in isolation. They also reported that  filament galaxies with $\rm  M_* > 10^{11}\, M_{\odot}$ have significantly higher gas fractions compared to control sample galaxies, suggesting that the most massive galaxies are accreting cold gas from the intrafilament medium.  
\citet{Castignani22-2022A&A...657A...9C}  found an increase  in \hi\ mass and a decrease  in \hi-deficiency with  increasing distance from the filament spine for filament galaxies with $\rm M_* \gtrsim 10^{9}\, M_{\odot}$. 
They suggested the observed \hi\ deficiency  to be due to tidal interactions or ram pressure stripping since  the  deficient systems were  preferentially found  in dense regions within filaments.

The  \hi\  mass measurements for  three of the galaxies in the narrow filament (presented in Table \ref{tab:HI}) shows that  two of them (G3 and G4) are  quite rich in their gas content   \citep[with   \hi\ masses above average for their stellar masses; see][for $\rm M_*-M_{HI}$ relation of nearby galaxies]{Catenilla18-2018MNRAS.476..875C}.
But we do   find signatures of interaction and mergers for these galaxies. A detailed study of the unusually disturbed  \hi\ in  G3  is presented in \citet[][]{Arabsalmani22-2022AJ....164...69A},  revealing  an uniquely unusual morphology and kinematics of \hi\  in this otherwise apparently normal galaxy.
This study explored several scenarios  to explain the unusual \hi\ structure and concluded that 
the most viable explanation was the penetrating passage of a satellite through the disk only a few Myr ago, redistributing the \hi\ in the galaxy without yet affecting its stellar distribution. The study mentioned how such disturbed gas  could play an important role in the formation of massive stars like the progenitor of GRB 171205A. 
We also find the distribution of \hi\ in G4 disturbed, and detect a tail for the \hi\ disk  of G5. 
A detailed study of the perturbed/unusual systems in the filament    will be presented  in an upcoming study (Arabsalmani et al., in prep.).

\citet[][]{Tempel15-2015MNRAS.450.2727T} studied the alignment between satellites and surrounding large-scale structure using the SDSS and found  a significant alignment between satellite galaxy position and filament axis. Similar results were found  in the  Millennium simulations \citep[see][and the references therein]{Tempel15-2015MNRAS.450.2727T}. 
The disturbed distribution of gas in the star-forming galaxies in the filament are therefore likely to be due to interaction with gas-rich systems in the filament. 
Such systems are believed to aid the formation of stars in galaxies by providing them the required gas reservoirs and by potential interactions and mergers with them. 
This scenario is supported by the presence of the  dark clouds that we have detected in the filament. 
Deeper \hicm\ emission line observations are planned  to explore the presence of low-mass, gas-rich systems in the vicinity of the star-forming galaxies in the filament.

\section{Summary}
\label{sec:sum}

We have identified an extremely  straight and narrow filament of galaxies in the nearby Universe. The filament has a length of 5 Mpc. The eight galaxies with spectroscopic measured redshift on the filament are within  very small distances from the filament axis, varying from 7 kpc to 104 kpc.  The redshifts of these galaxies are in a small range between $z = 0.03609$ and $z = 0.03702$. So their radial velocities with respect to their median redshift varies from $\sim -210$ \kms\ to $\sim 70$ \kms, similar to the proper motions of galaxies. Most of these galaxies are actively forming stars, unlike what is expected from the statistical studies of filament galaxies in the nearby Universe which find the majority of galaxies close to the filament spine to be passive galaxies. Their stellar masses are within a range of $\rm 10^{9.1}-10^{10.7}\,M_{\odot}$. We search in the filament catalogues built based on TNG simulations in order to find filaments with similar geometrical and galaxy properties. We find no direct match. This, however, might be due to the methods used in building up the simulated catalogues which likely introduce certain biases in the catalogues. 
We further use the  full-sky mock galaxy catalogue of the MillenniumTNG simulation project and search for galaxy distributions with  similar geometrical properties. We find that fairly straight filaments can be found and are expected in the $\Lambda$CDM galaxy distribution as well. Our analysis shows that the identified  filament is indeed unusual and rare, but  not exceptionally uncommon.

We study the distribution of cold gas in a 1.3 Mpc long section of the filament through \hicm\ emission line observations and find about a dozen of \hi\ sources. Their \hi\ masses varies between $\rm\sim 10^{8.5}-10^{9.5}\,M_{\odot}$ and they are all detected at \hi\ column densities N(\hi)$\rm \gtrsim 10^{20}\,cm^{-2}$. All of these detected \hi\ sources have  no confirmed optical counterparts in our deep CFHT observations. Their extreme  $\rm M_{HI}/M_*$ values imply that they are very dark. This, together with the large velocity widths of the \hicm\ emission lines of the clouds, indicates that they might not yet be virialized and hence are not yet forming stars.  
The presence of such clouds in filaments, though predicted by simulations, had never been confirmed with observations. This is the first time that the existence of such clouds in a filament is observationally confirmed. 
We suspect that these clouds mark the peaks of the distribution of \hi\ over large scales within the filament as predicted by simulations. Further observations are required to confirm this.


\section*{Acknowledgments}
M.A. thanks   Lister Staveley-Smith, Nirupam Roy, and Nicola Malasavi for helpful discussions.  
The presented study is funded by the Deutsche Forschungsgemeinschaft (DFG, German Research Foundation) under Germany´s Excellence Strategy – EXC-2094 – 390783311. 
SR acknowledges support by DFG through project number 50082519, and Australian Research Council Centre of Excellence for All Sky Astrophysics in 3 Dimensions (ASTRO 3D) through project number CE170100013. 
The National Radio Astronomy Observatory is a facility of the National Science Foundation operated under cooperative agreement by Associated Universities, Inc.
This work is partly based on tools and data products produced by GAZPAR operated by CeSAM-LAM and IAP. 
We are particularly grateful to Olivier Ilbert and Christophe Adami for their extensive support with GAZPAR.  
This research uses services or data provided by the Astro Data Lab, which is part of the Community Science and Data Center (CSDC) Program of NSF NOIRLab. NOIRLab is operated by the Association of Universities for Research in Astronomy (AURA), Inc. under a cooperative agreement with the U.S. National Science Foundation. 
F.R. acknowledges support provided by the University of Strasbourg Institute for Advanced Study (USIAS), within the French national programme Investment for the Future (Excellence Initiative) IdEx-Unistra.




\appendix
\label{sec:appendix}

\section{\hi\ data reduction and source identification}
\label{app:data}

We used the L-band receivers of the Karl J. Jansky Very Large Array (JVLA) in B and C configurations to map the \hicm\ emission in the field. The observations in B-configuration  were carried out on 09-March-2019, 05-May-2019, and  16-October-2022 for a total time of $\sim$ 10.5 hours (proposal ID: VLA/2018-07-058; PI: Arabsalmani).  The C-configuration observations were carried out on   23-October-2022, and  28-October-2022 for a total time of $\sim$ 17.5 hours (proposal ID: VLA/2022-00-109; PI: Arabsalmani). 
``Classic'' {\sc aips} was used for the analysis of the data \citep[][]{Greisen03-2003ASSL..285..109G}. For each day's data, after initial data editing, flux and bandpass calibration, a ``channel-0'' visibility data set was created by averaging together line-free channels. A standard self-calibration, imaging  and data-editing loop was applied on the channel-0 data set, until no further improvement was seen in the continuum image on further self-calibration. 
At the end of the loop, the final antenna-based gains were applied to all the visibilities of the original multi-channel data set.The radio continuum image made using the line-free channels at the end of the self-calibration cycle, was used to subtract the continuum from the calibrated visibilities, using the tasks {\sc uvsub}, following which the task  {\sc uvlin} was also run. At this point, cumulative visibility datasets were created separately for B and C configurations by combining the residual visibilities from the daily observations (two and three days respectively). We also combined the residual visibilities from the full 5 days in order to create a combined B+C configuration dataset.

The three sets of combined visibilities (B-configuration, C-configuration, and B+C-configuration) were Fourier transformed to produce spectral cubes using the task {\sc imagr}. 
For each configuration, we created two data cubes, with optimized velocity resolutions of $\sim 34$ \kms\ and  $\sim 50$ \kms\ to improve the statistical significance of the detected H{\sc i} 21\,cm emission in independent velocity channels while still having sufficient velocity resolution to accurately trace the velocity field of any detected source.. 
In order to optimize between the signal-to-noise ratio of the detections and the spatial resolution,  robust factors of  of 0.5, $-$0.5, and 0.5  were  used for the B-configuration, C-configuration, and B+C-configuration visibilities respectively,  to create the cubes. 
The properties of the cubes produced are provided in Table \ref{tab:cubes}.  
We also applied the primary beam correction using the task {\sc pbcor}. The field of view of the JVLA has a full width at half power of $\sim$ 30\arcm\ at the frequency of our observations in the L band. We restrict ourselves to within the  HPBW where it has a relatively flat response, in order to only detect \hi\ emission with high sensitivity.

\begin{table*}
\begin{center}
\caption{Parameters of the JVLA \hi\ data cubes.}
\begin{tabular}{lccc}
\hline
Array & Synthesized beam & Channel width & Noise in line-free channel \\
& (\arcs $\times$ \arcs) & (\kms) & mJy Bm$^{-1}$\\
\hline
B	&	7.5$\times$6.4		&	34.2	& 0.20 \\
B	&	7.5$\times$6.4		&	51.3	& 0.17 \\
B+C	& 	10.5$\times$9.1		&	34.2	& 0.14 \\
B+C	&	10.5$\times$9.1		&	51.3	& 0.11 \\
C	&	20.3$\times$17.0	&	34.2	& 0.18 \\
C	&	20.3$\times$17.0	&	51.3	& 0.15 \\
\hline
\end{tabular}   
\end{center}
\flushleft{
}
\label{tab:cubes}
\end{table*}

We identified the  \hi\ sources in the field observed following a two-step process described below: 
\\
\textbf {(i)} we identified the possible \hi\ source candidates in the field  using the Source Finding Application (SoFiA) software, designed for finding \hi\ sources from 3-D cubes \citep[][]{Serra15-2015MNRAS.448.1922S, Westmeier21-2021MNRAS.tmp.1648W}. 
SoFiA uses a Smooth+Clip (S+C) finder where it creates multiple \hi\ cubes out of the original cube by smoothing it, and clips pixels below a certain absolute flux level in the (original and) smoothed cubes which are then merged over a pre-set merging lengths along the spatial and spectral dimensions to create a mask, which is then applied to the original cube to get the source fluxes out.
Other than the native spatial and spectral resolution cube, we instructed SoFiA to also spatially smooth the data with Gaussian filters 1 and 2 synthesized beams in size, and spectrally smooth the data with boxcar filters 3 and 5 channels in size.
The absolute clip flux density was set to four times the (smoothed) rms noise level, with local noise normalisation enabled given the varying noise across the field due to primary beam correction applied on the input cube.
For the pixels thus selected, we conservatively set the minimum merging lengths to be one pixel only along both spatial and spectral dimensions. 
Finally the reliability parameter which filters out potential real detections based on the density of positive and negative sources in a certain parameter space (please see references above for details), was also conservatively set to a high value of 0.8, {and the threshold of constant integrated signal-to-noise ratio in the search space was set to 5.}
\\
\textbf {(ii)} Next for each of the sources identified by SoFiA, we created two spectra from the 35 and 50 \kms, cubes using the task {\sc viewer} of CASA, the Common Astronomy Software Applications package. The spectra were extracted from the cubes in the regions defined based on the moment 0 maps created by SoFiA. 
{Only for one source  (G5) which was not identified by SoFIA  we use the moment maps created in {\sc aips}. This is one of the star-forming galaxies in the field located very close to the edge of the HPBW of JVLA. We apply the task {\sc momnt} to the spectral cube in order to obtain maps of the H{\sc i} total intensity and the intensity-weighted velocity field for this source. 
{\sc momnt} works by masking out pixels in the spectral data cube which lie below a threshold flux in a secondary data cube created by smoothing the original  cube both spatially and along the velocity axis -- the smoothing ensures that any localized noise peaks are ignored and only emission correlated spatially and in velocity is chosen.
We created the secondary data cube by applying Hanning smoothing across  blocks of three consecutive velocity channels, whereas spatially a Gaussian kernel of FWHM equal to twelve pixels (about twice the size of the synthesised beam) was applied.
The threshold flux used to select pixels was approximately 1.5 times the noise in a line-free channel of the original cube.}

\begin{figure*}
\centering
\begin{tabular}{cccc} 
\includegraphics[width=0.30 \textwidth]{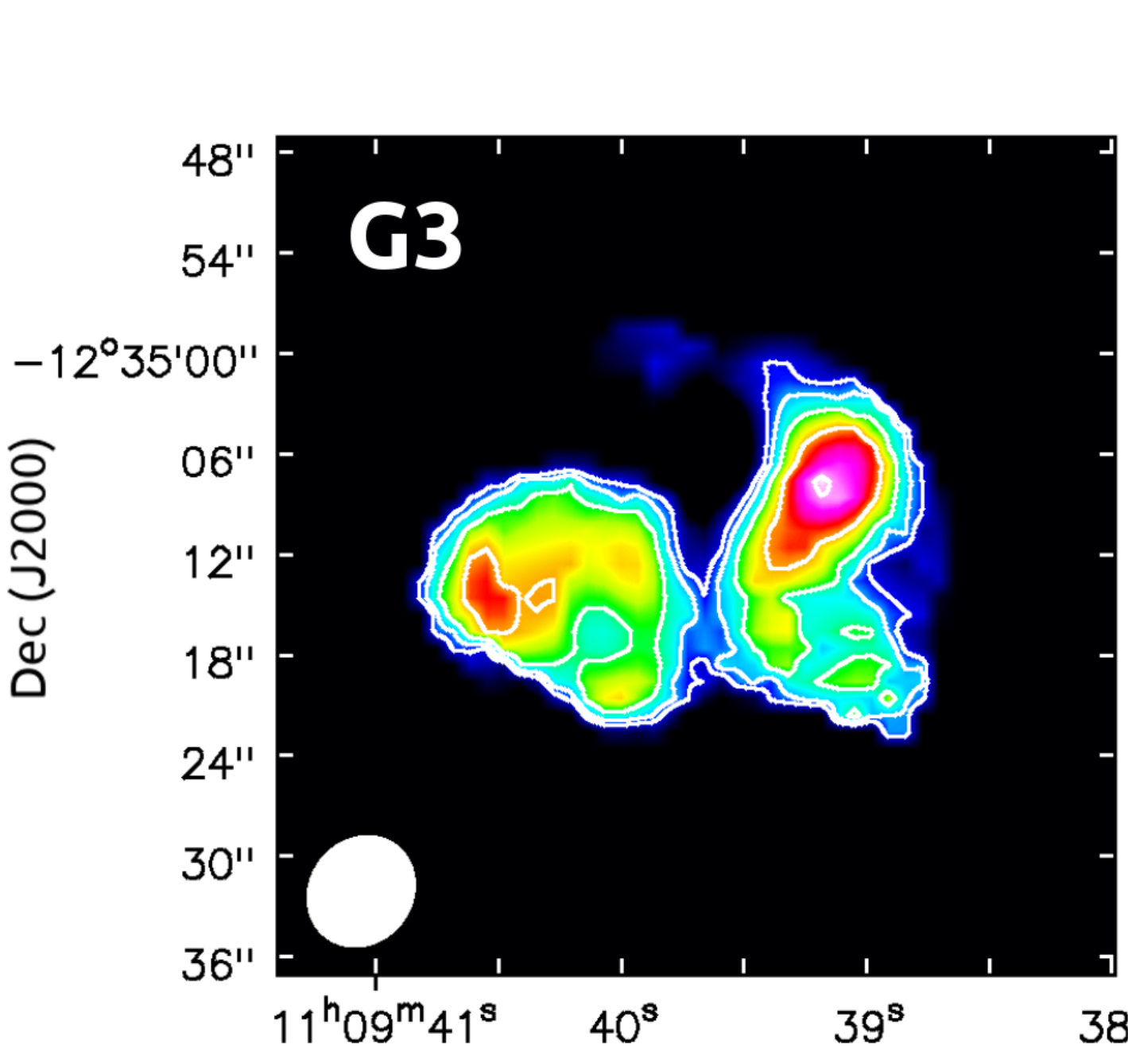}&
\includegraphics[width=0.30 \textwidth]{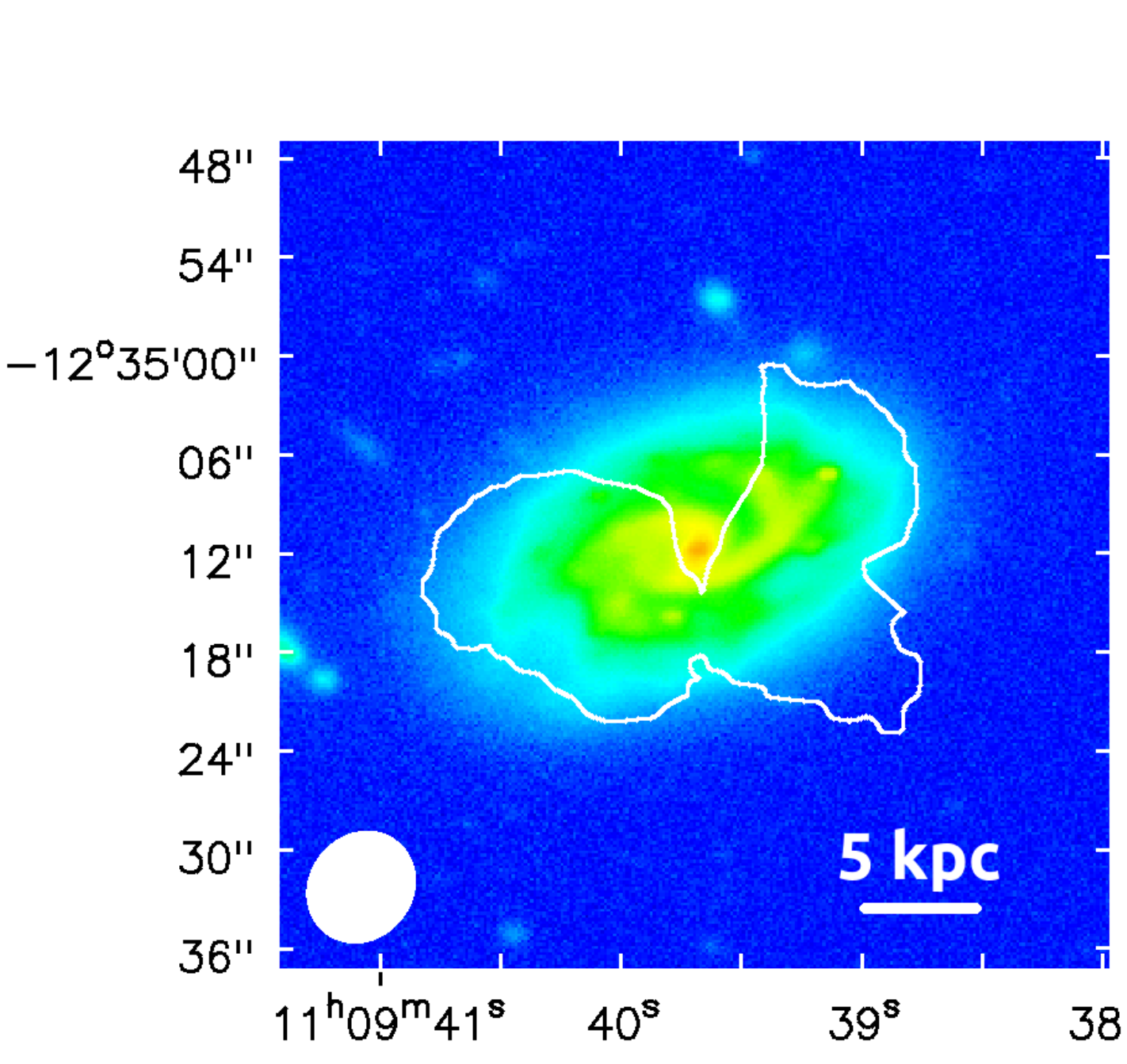}&
\includegraphics[width=0.30 \textwidth]{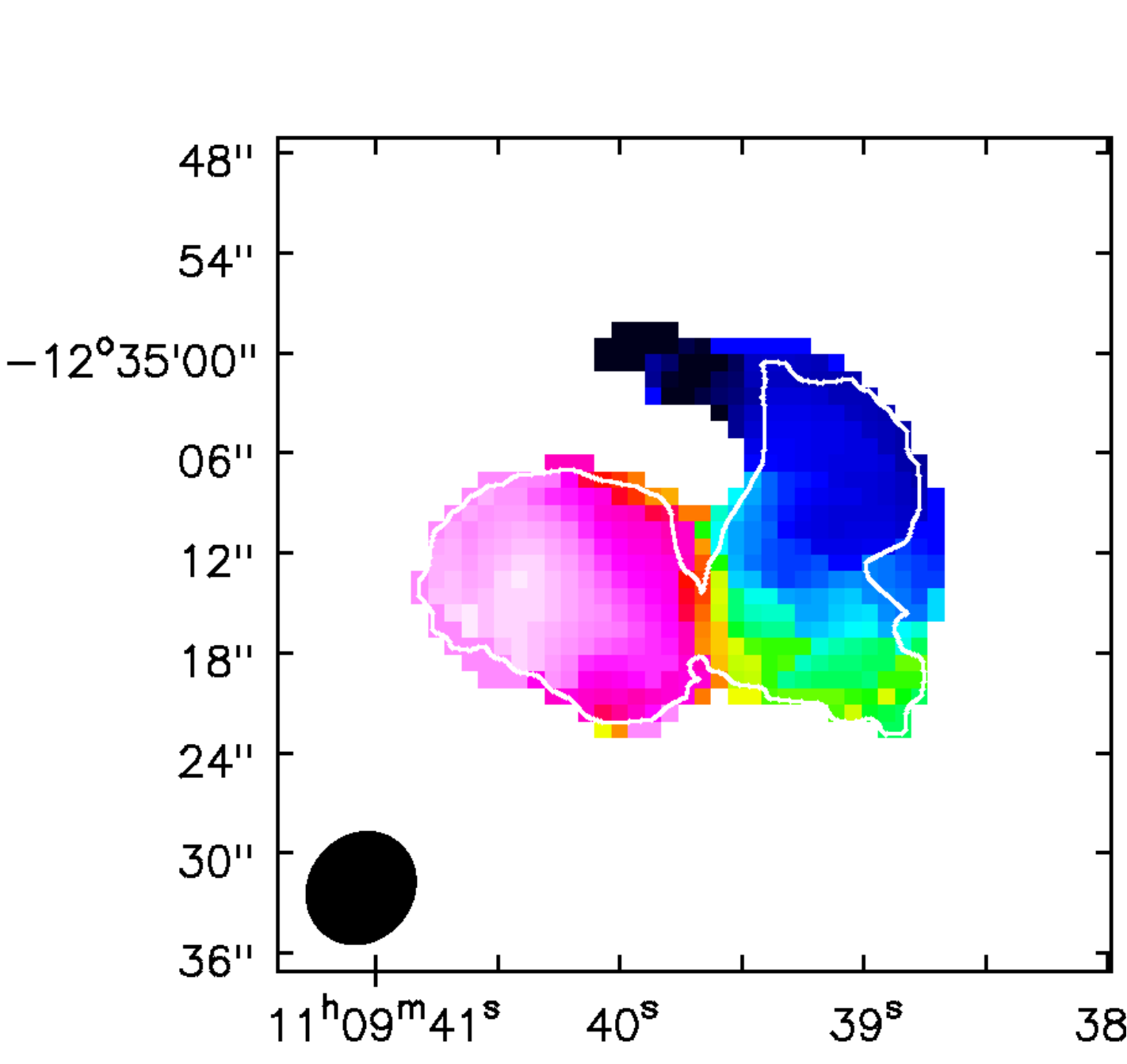}&
\includegraphics[width=0.08 \textwidth]{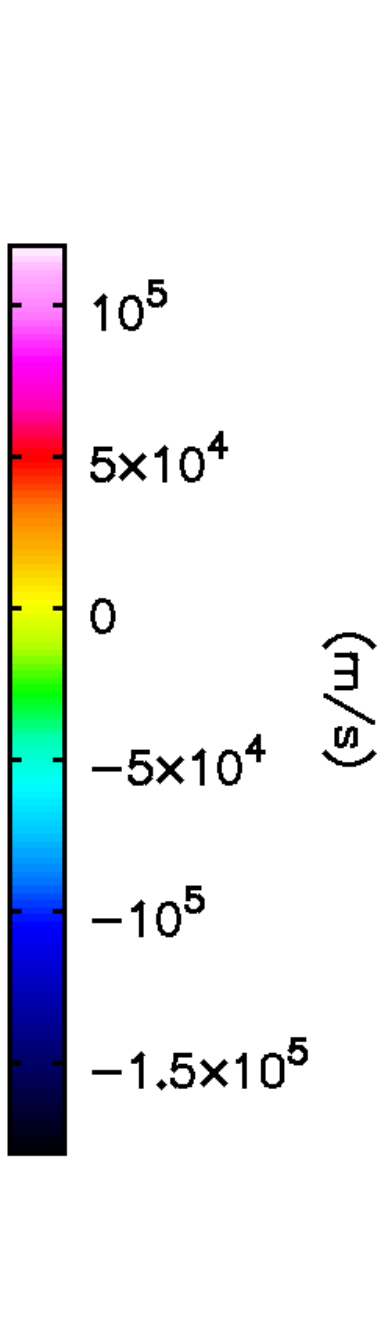}
\\
\includegraphics[width=0.30 \textwidth]{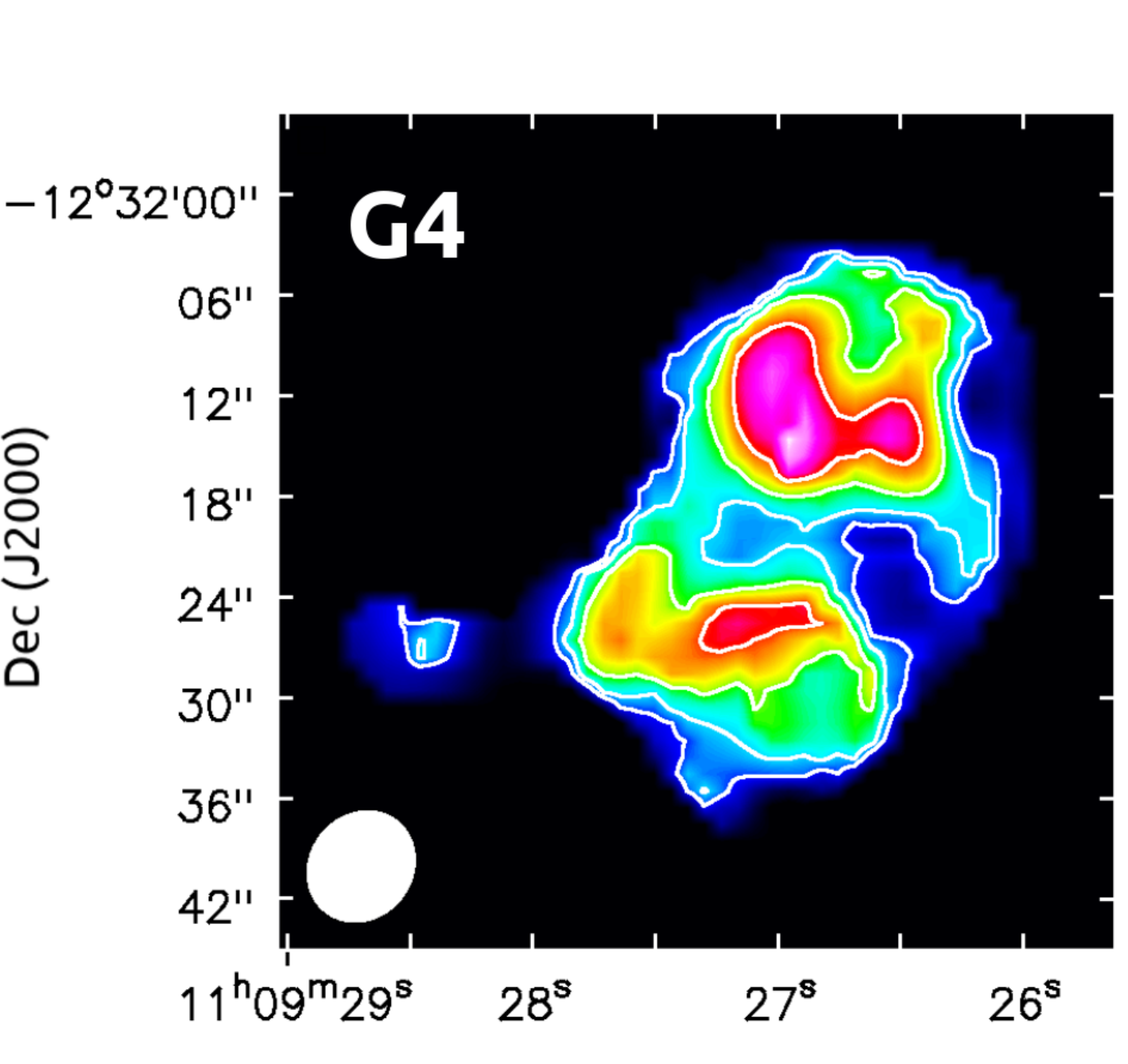}&
\includegraphics[width=0.30 \textwidth]{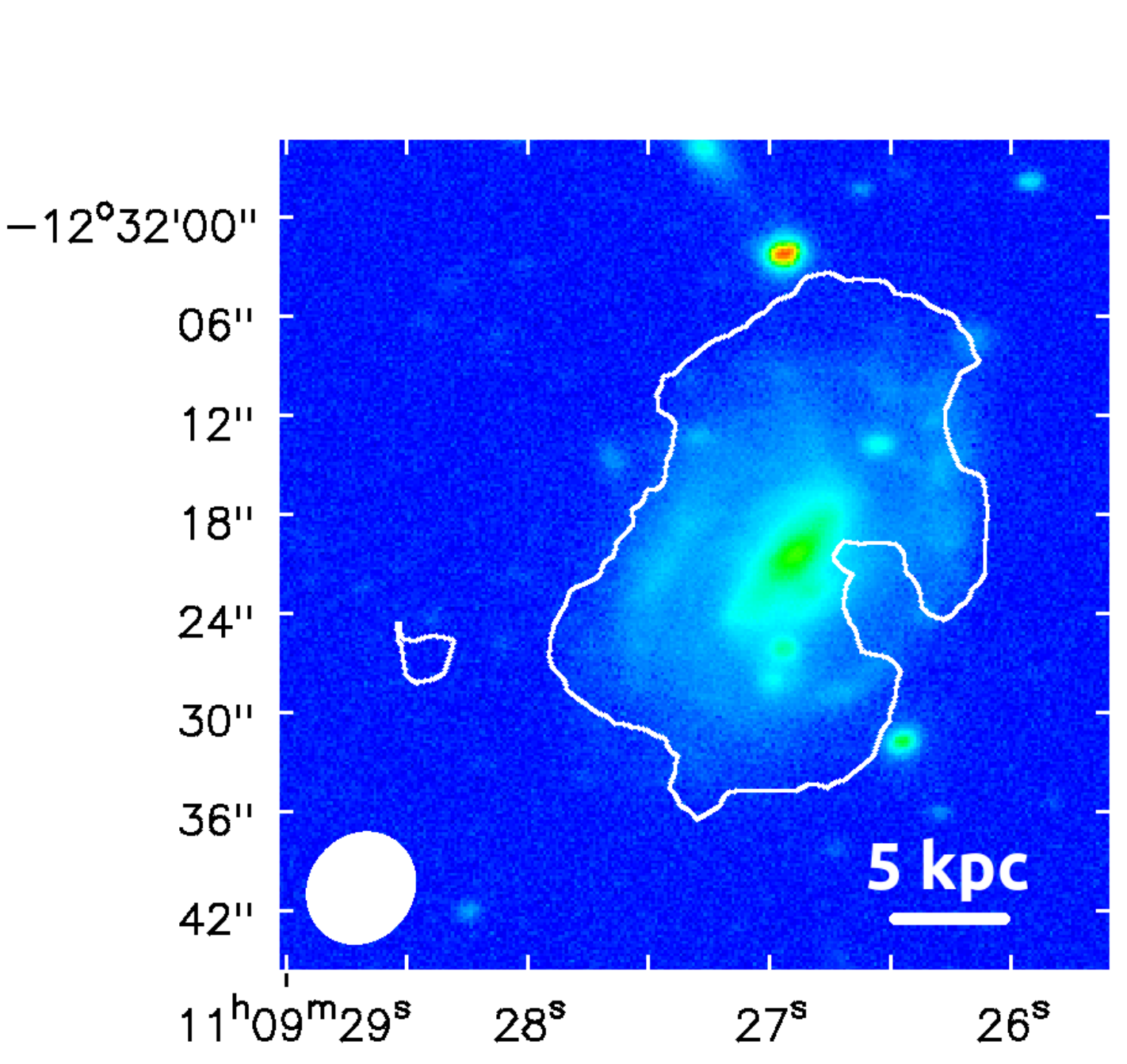}&
\includegraphics[width=0.30 \textwidth]{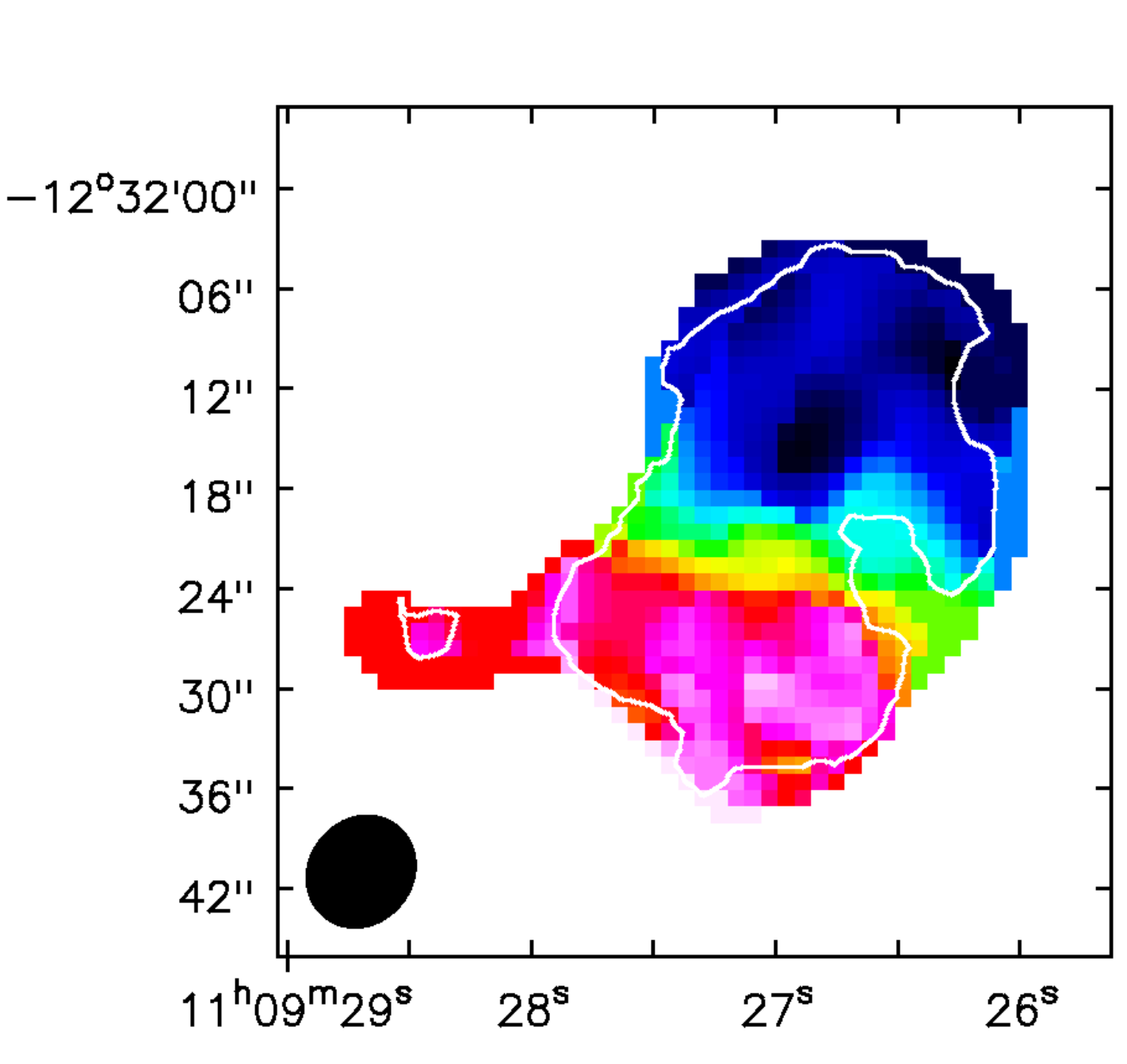}&
\includegraphics[width=0.08 \textwidth]{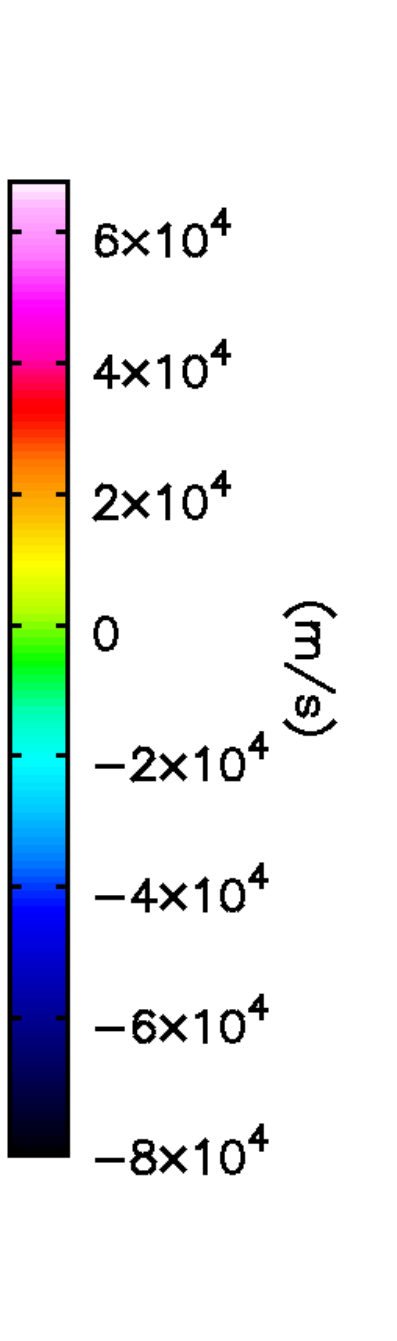}
\\
\includegraphics[width=0.30 \textwidth]{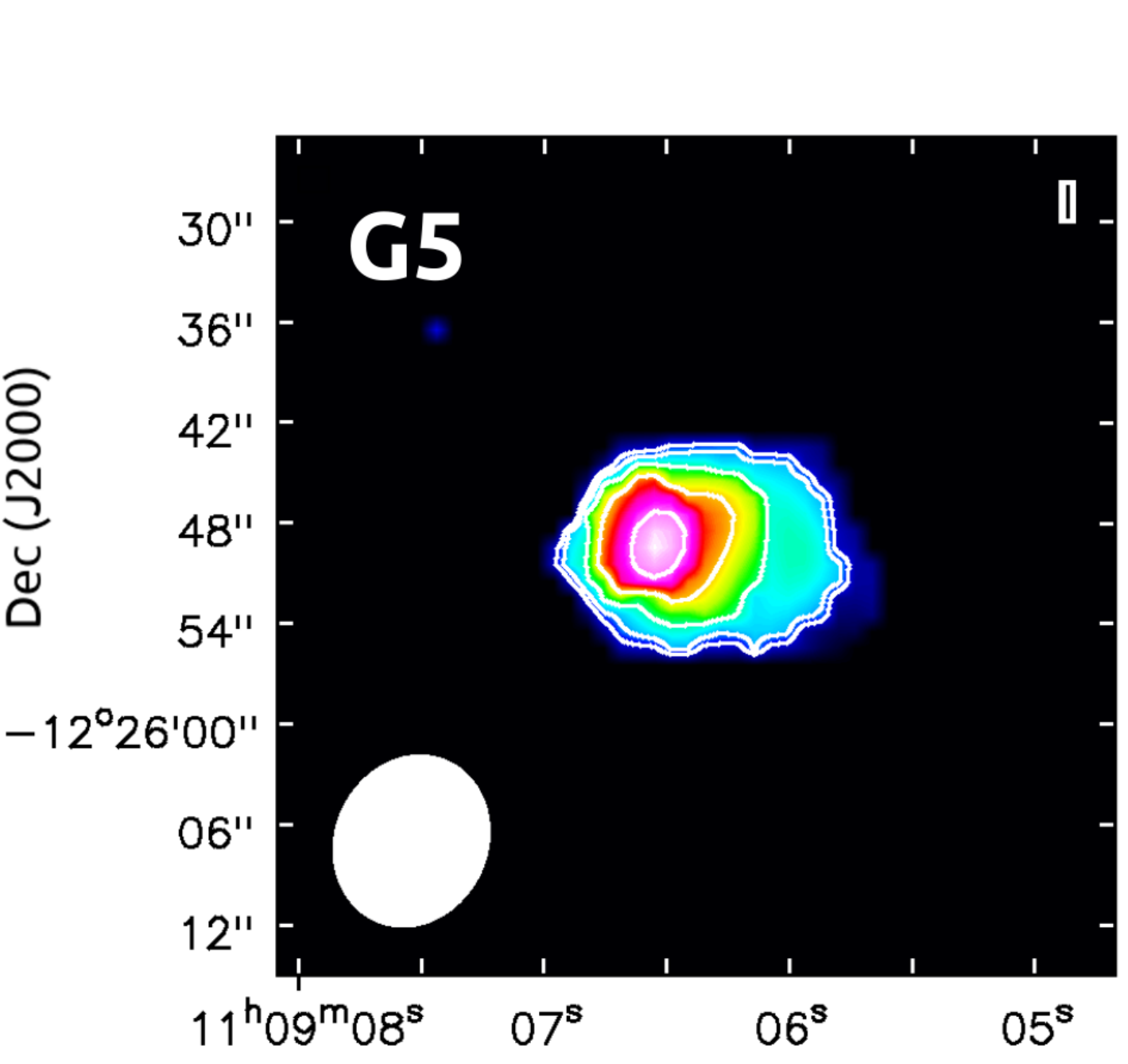}&
\includegraphics[width=0.30 \textwidth]{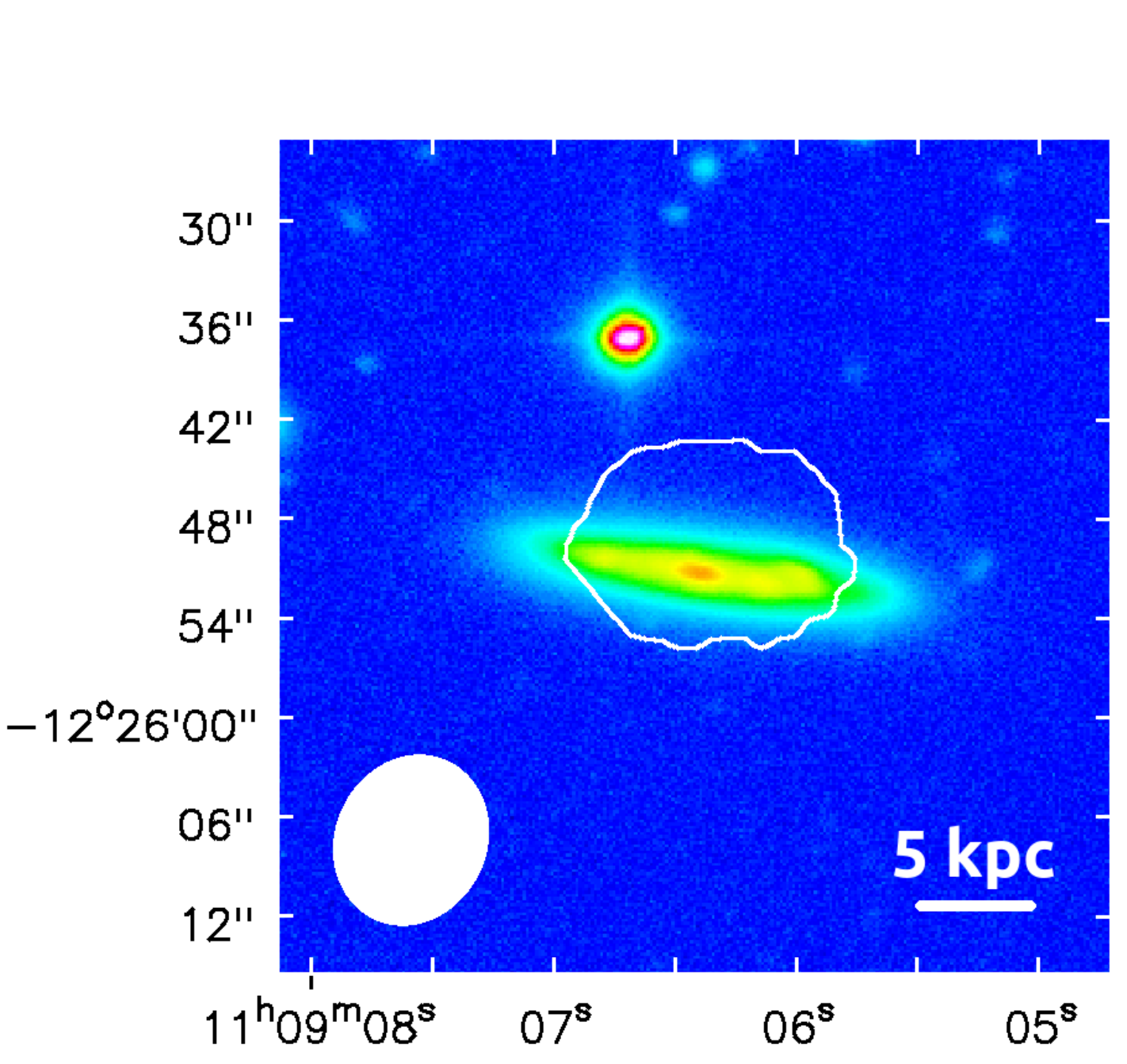}&
\includegraphics[width=0.30 \textwidth]{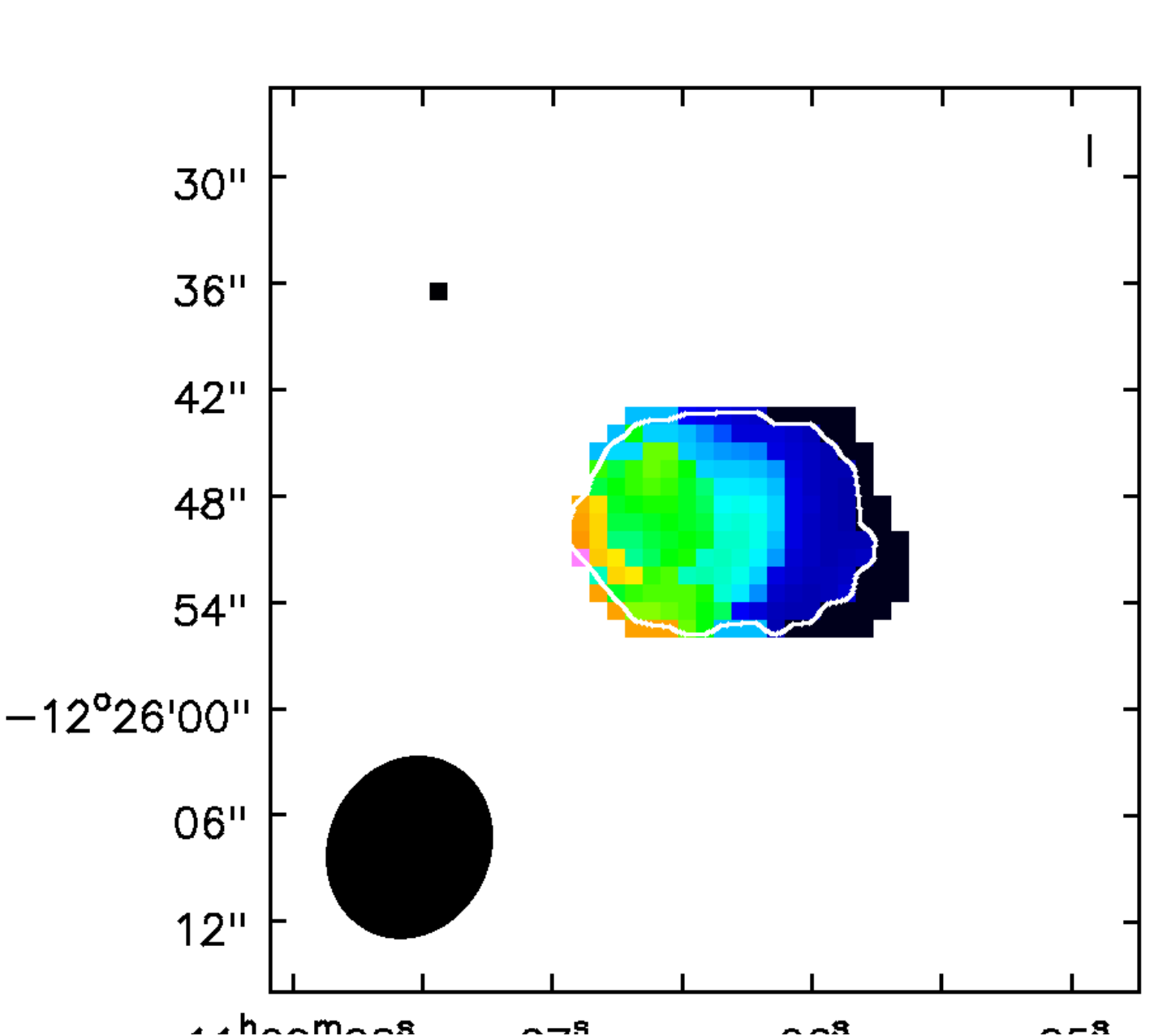}&
\includegraphics[width=0.08 \textwidth]{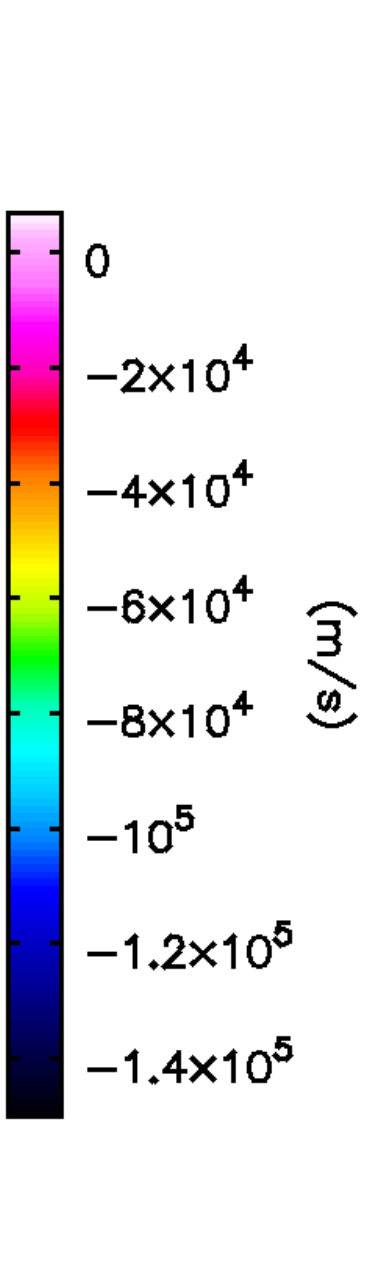}
\\
\includegraphics[width=0.30 \textwidth]{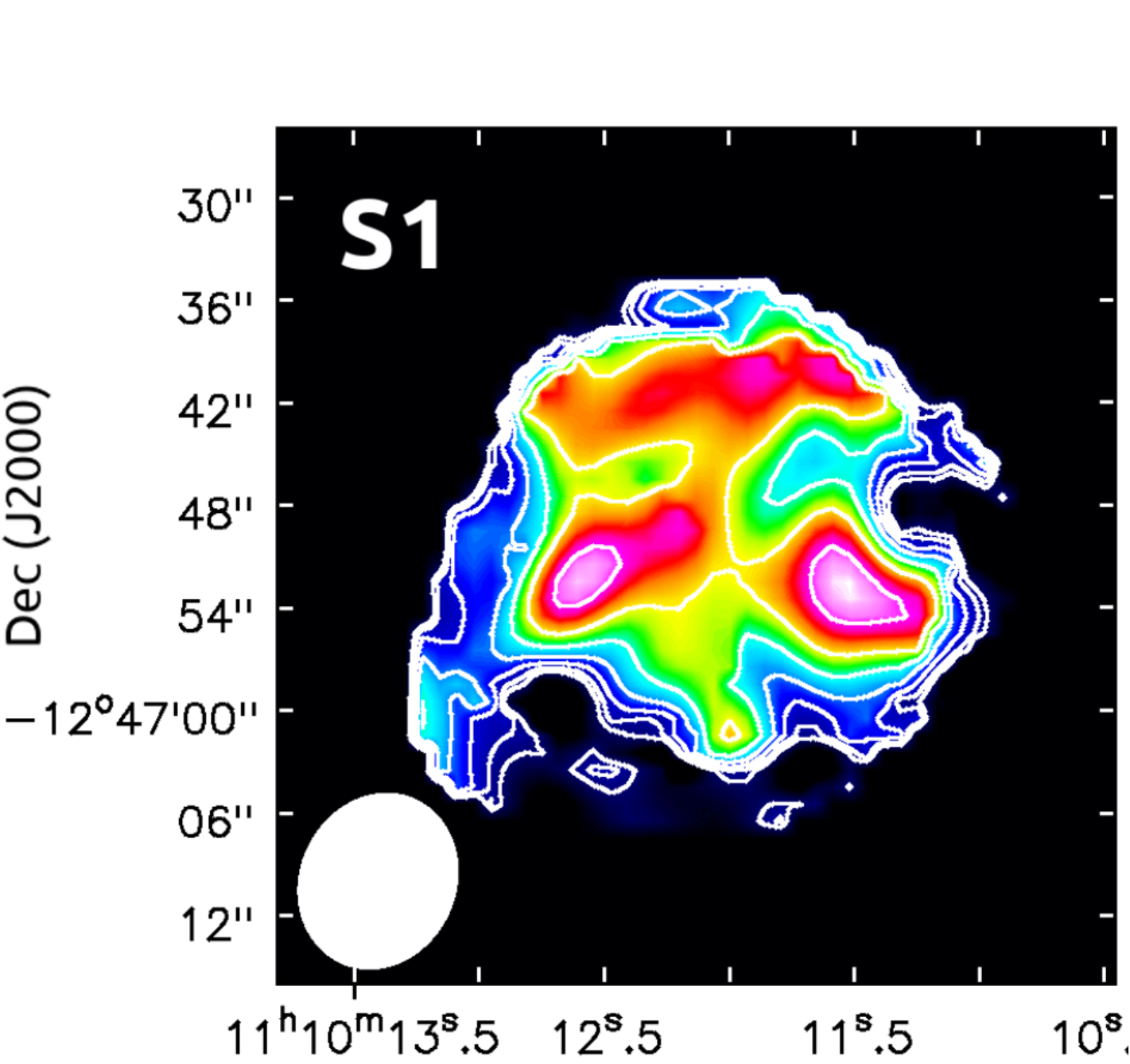}&
\includegraphics[width=0.30 \textwidth]{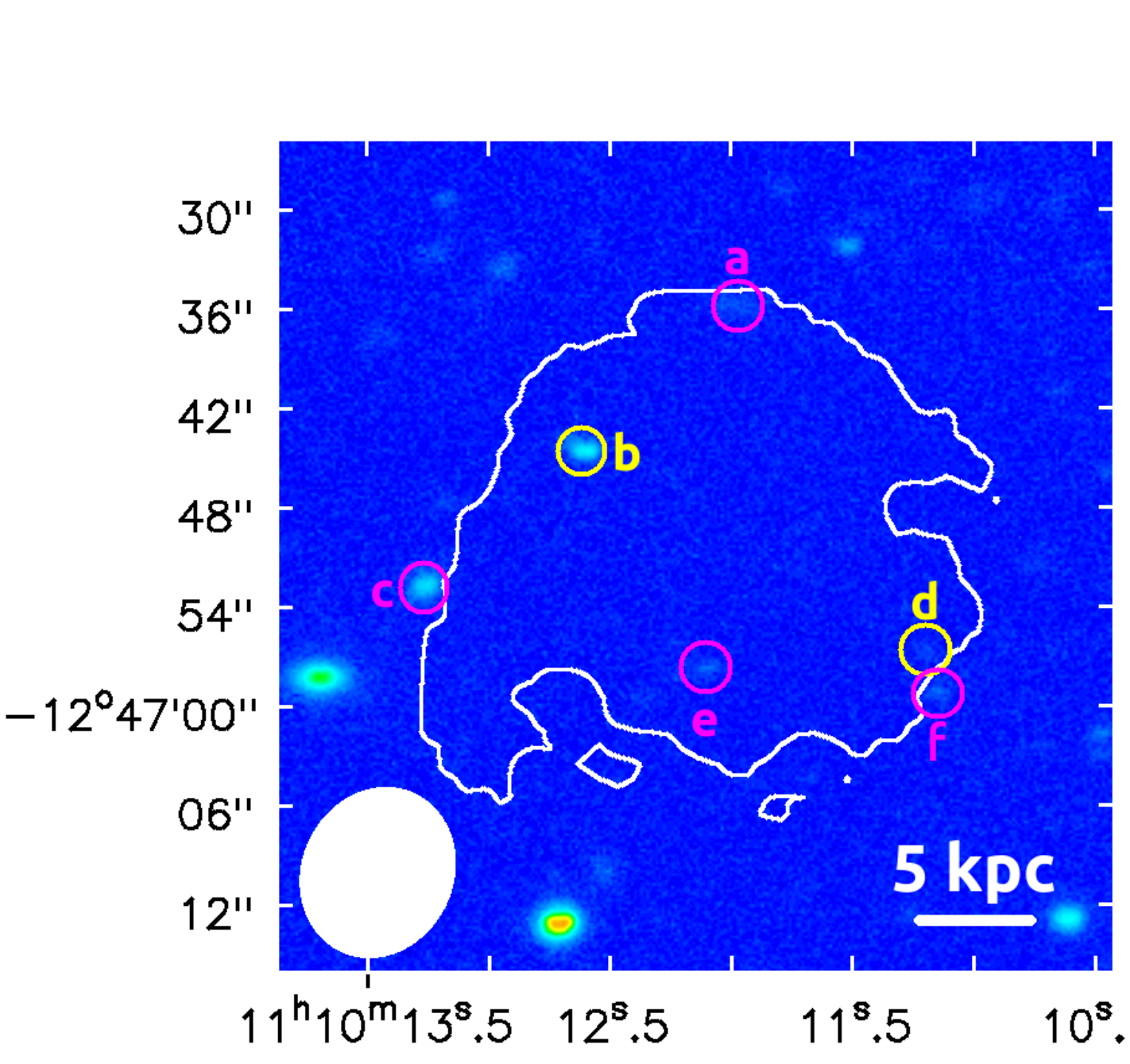}&
\includegraphics[width=0.30 \textwidth]{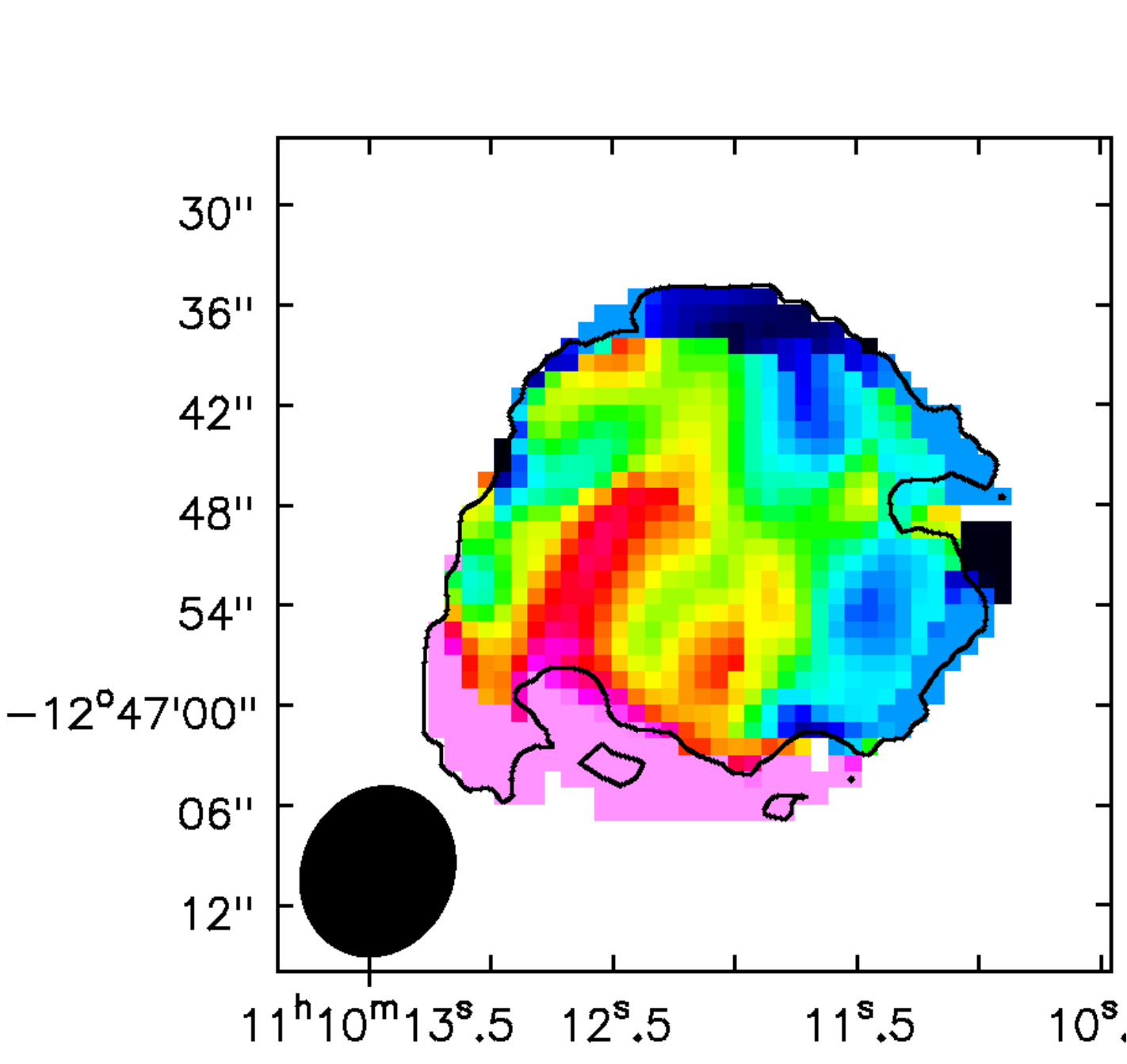}&
\includegraphics[width=0.08 \textwidth]{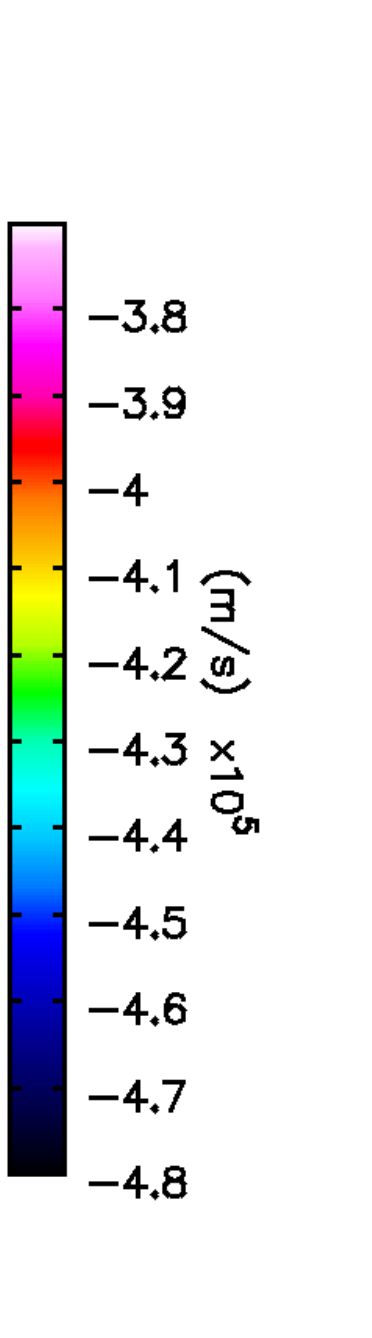}
\\
\hskip 1 cm \includegraphics[width=0.07 \textwidth]{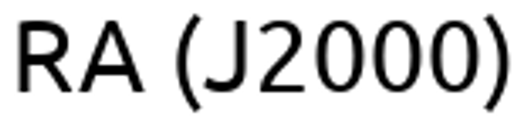}&
\hskip 1 cm \includegraphics[width=0.07 \textwidth]{figs/RA.pdf}&
\hskip 1 cm \includegraphics[width=0.07 \textwidth]{figs/RA.pdf}&
\end{tabular}
\caption{ 
Each row presents a source listed in Table \ref{tab:HI}.
\textbf {Left:} Total intensity map of \hicm\ emission line in color, overlaid with contours of \hicm\ emission line. The outermost contour marks the $3\sigma$ level of emission in a single channel with a velocity width of 34 \kms\  (equivalent to the \hi\ column densities provided in  Table  \ref{tab:HI}). Each subsequent contour is in multiples of $\sqrt 2$. 
\textbf {Middle:} The r-band CFHT image, overlaid with the $3\sigma$ contour of \hicm\ emission line. The detected optical sources are marked with their identification numbers and circles. For sources marked with red colour, association with the filament is ruled out based on photometric redshift measurements. For the  sources marked with yellow colour (listed in Table \ref{tab:optical}) $z=0.037$ is not ruled out.  
\textbf {Right:} The intensity-weighted velocity map of \hicm\ emission line in color, overlaid with the $3\sigma$ contour of \hicm\ emission line. 
The size of the images in all panels are 50\arcs $\times$ 50\arcs except for S12 where the image sizes are 100\arcs $\times$ 100\arcs. The synthesized beam of JVLA observations is shown in the bottom-left corners of the  maps.
\label{fig:maps}}
\end{figure*}

\addtocounter{figure}{-1}

\begin{figure*}
\centering
\begin{tabular}{cccc} 
\includegraphics[width=0.30 \textwidth]{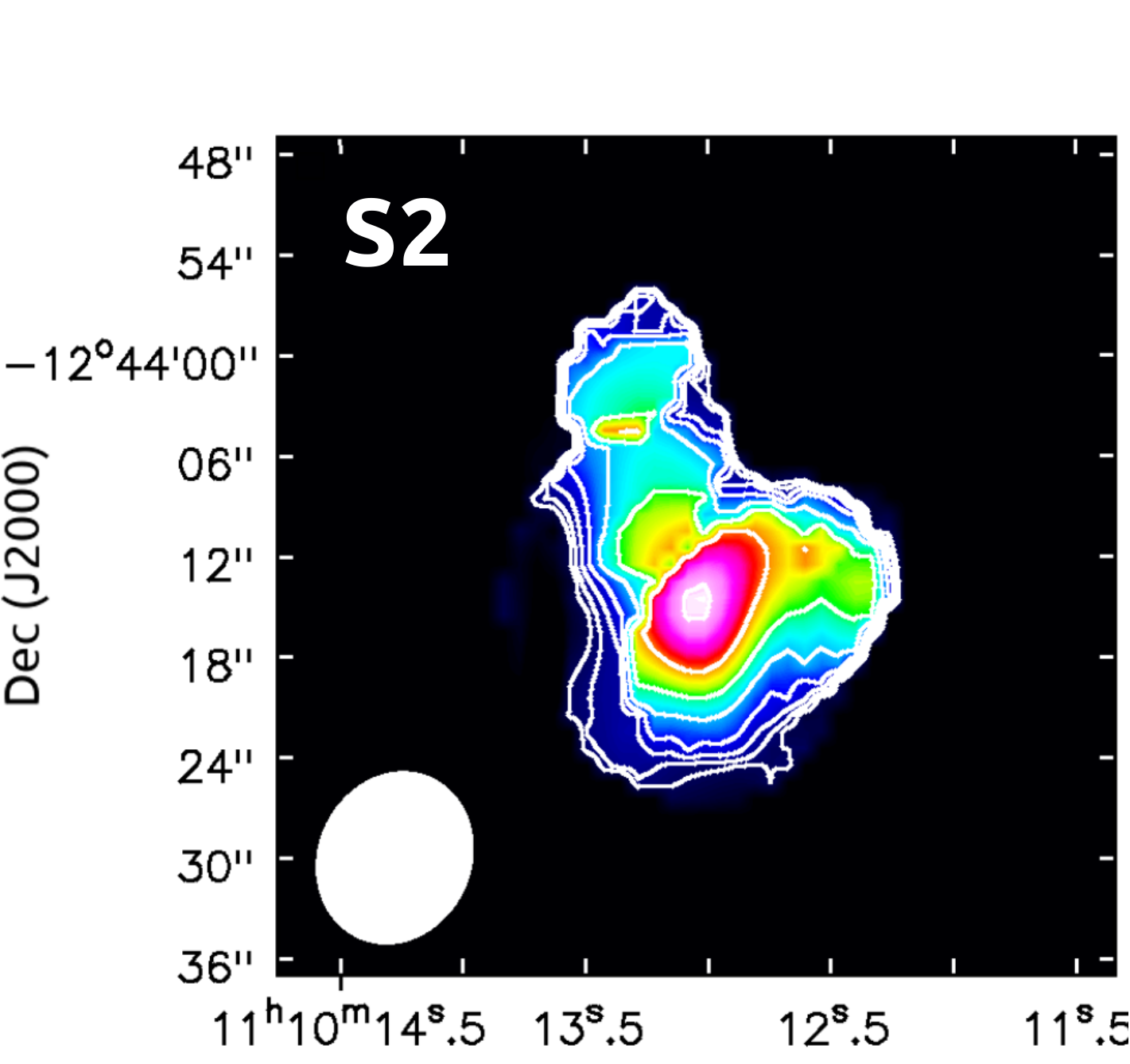}&
\includegraphics[width=0.30 \textwidth]{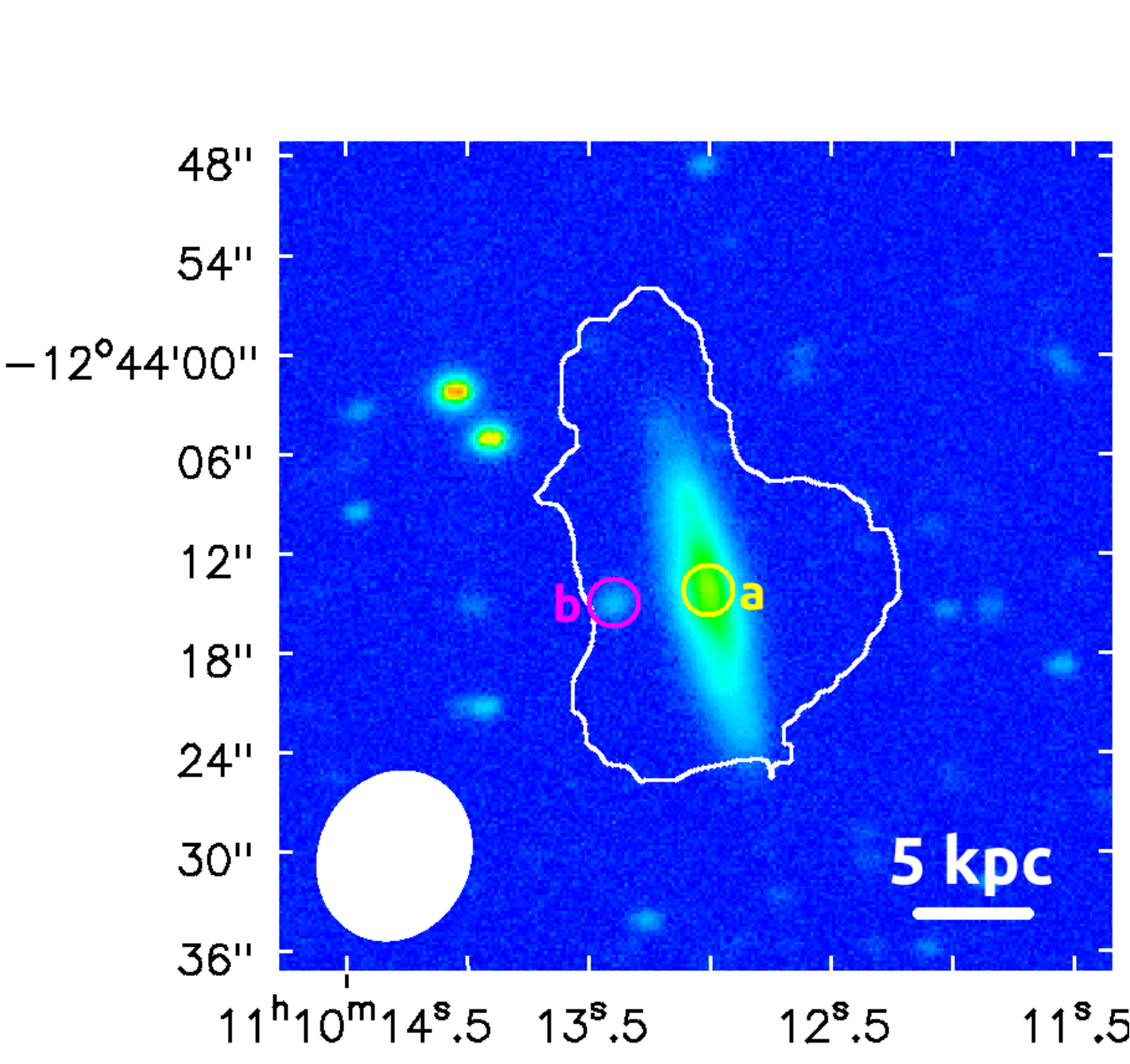}&
\includegraphics[width=0.30 \textwidth]{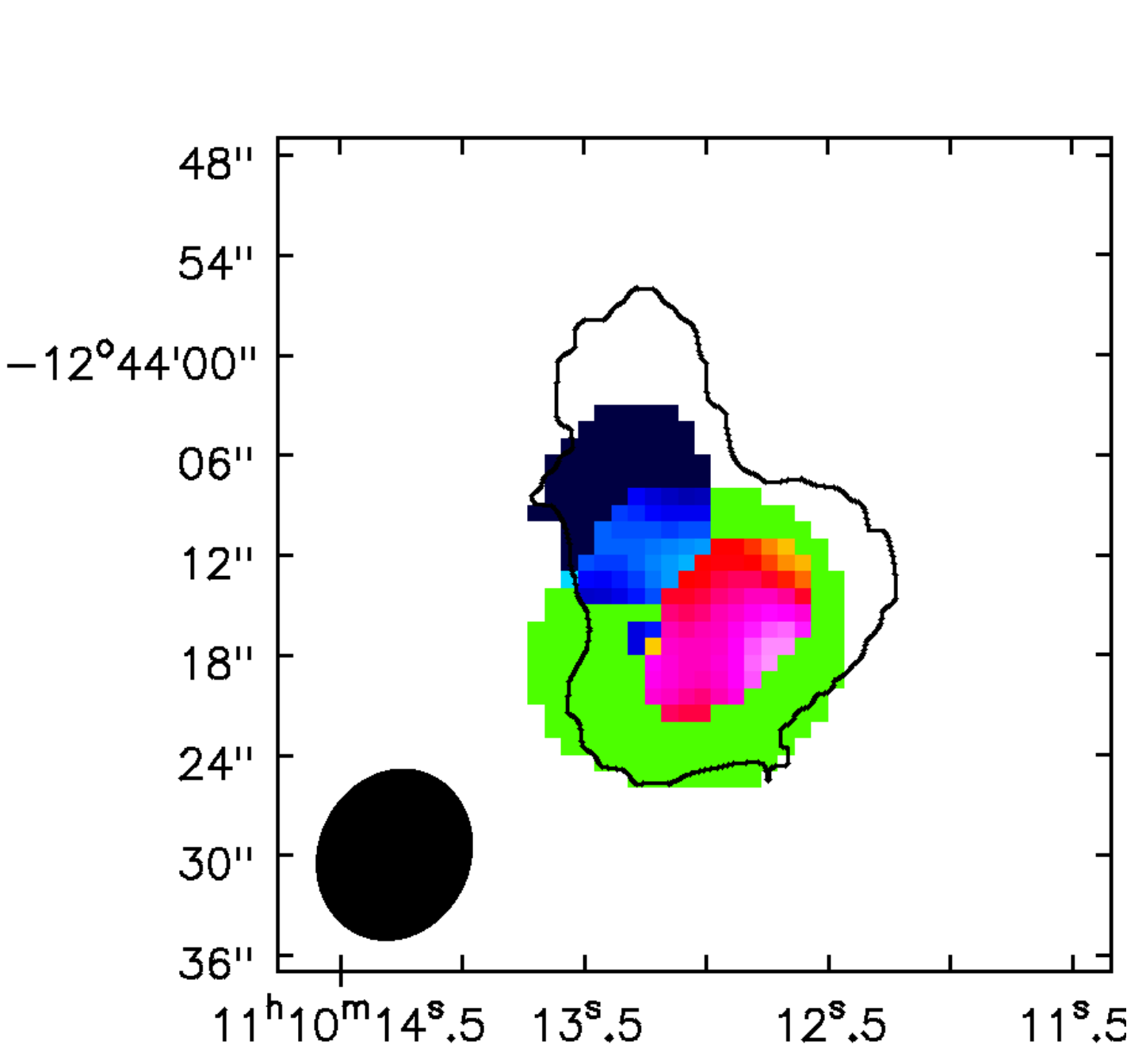}&
\includegraphics[width=0.08 \textwidth]{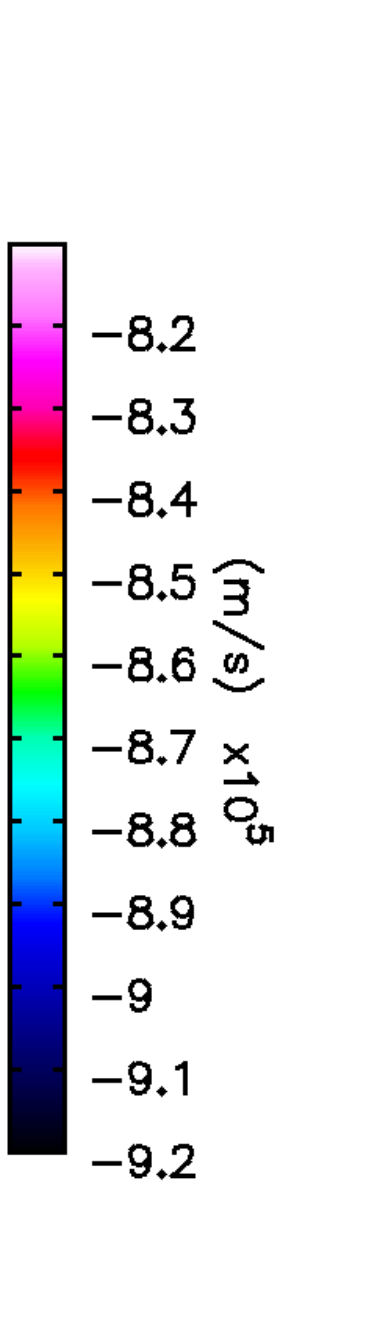}
\\
\includegraphics[width=0.30 \textwidth]{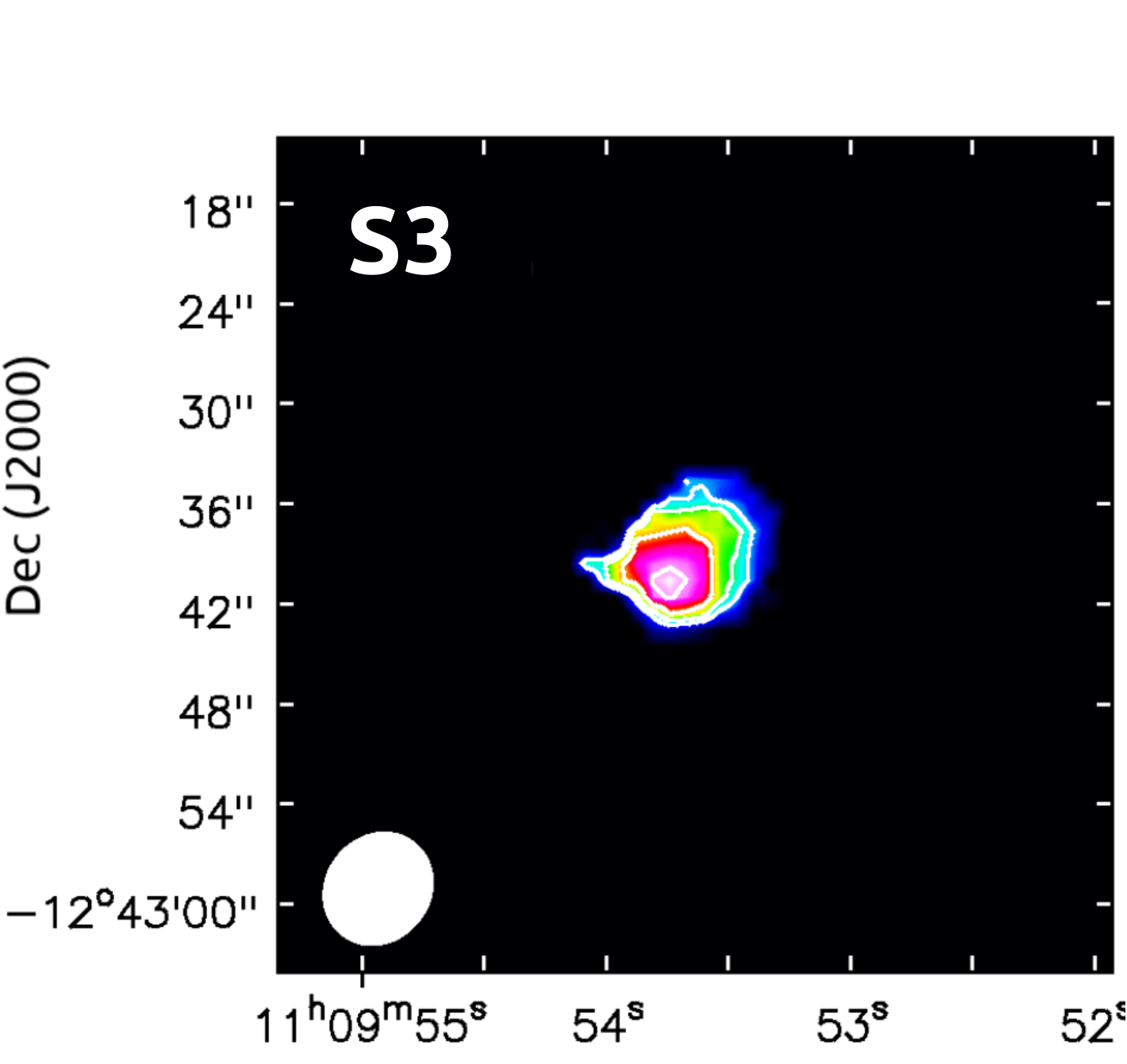}&
\includegraphics[width=0.30 \textwidth]{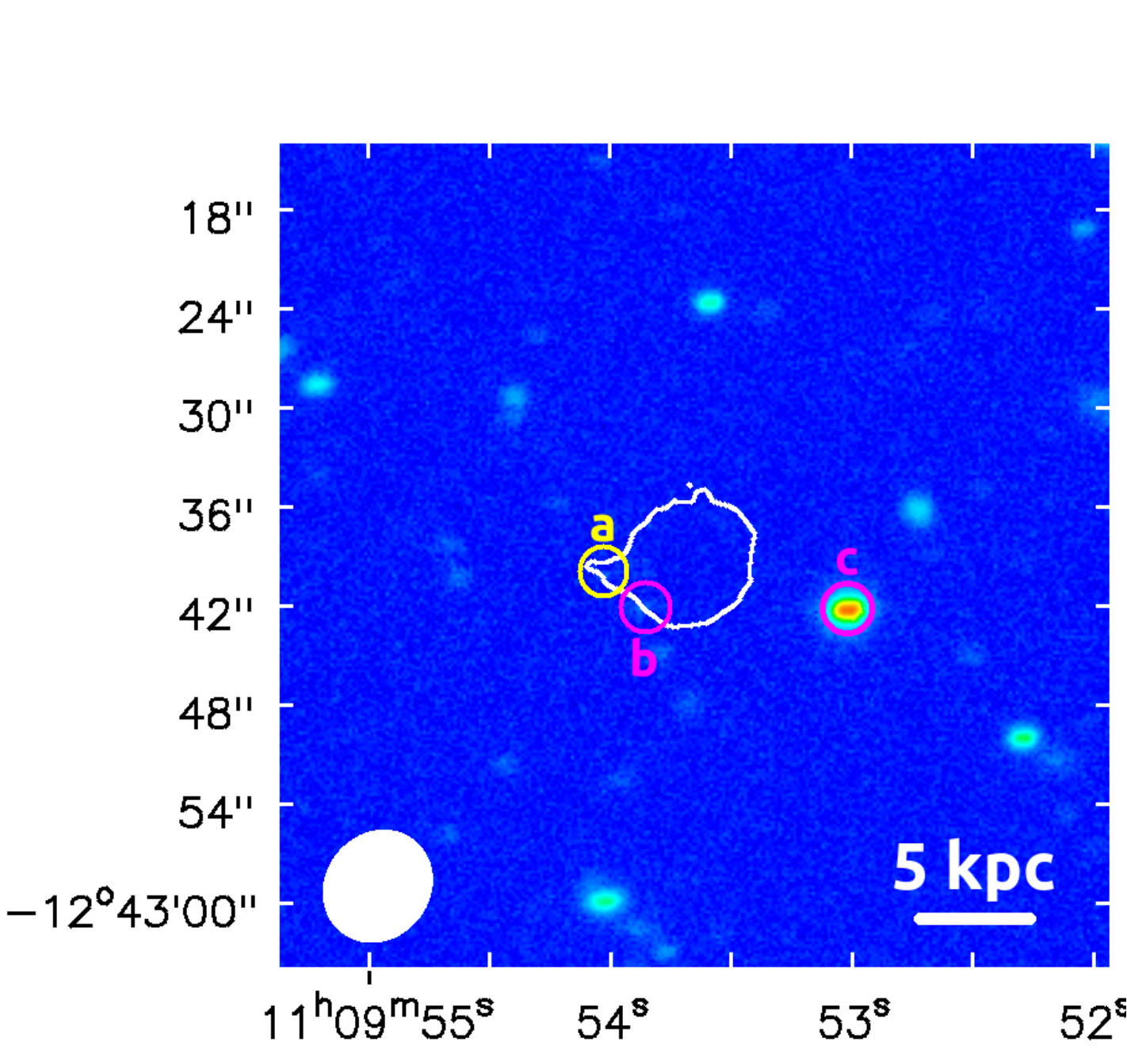}&
\includegraphics[width=0.30 \textwidth]{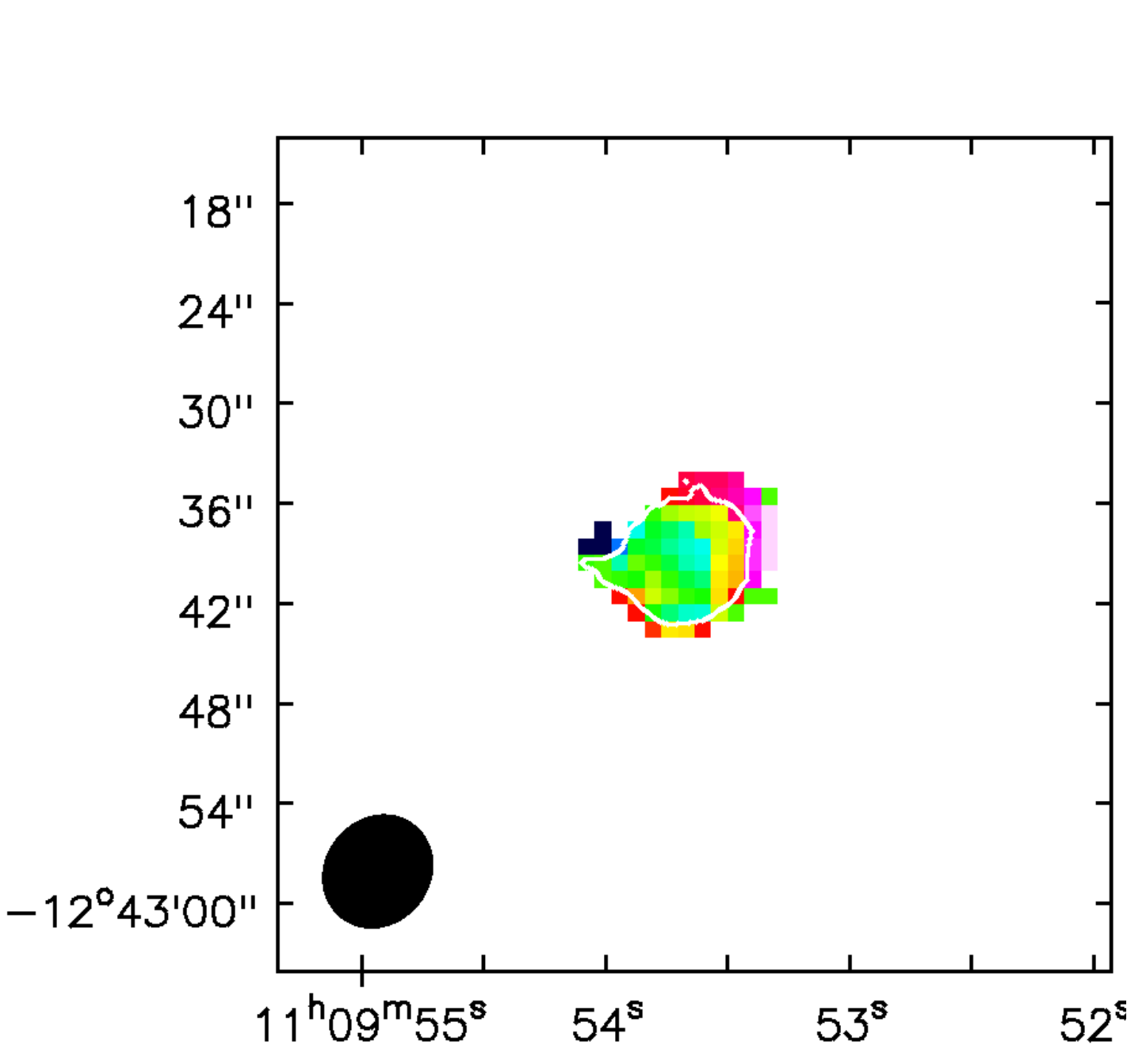}&
\includegraphics[width=0.08 \textwidth]{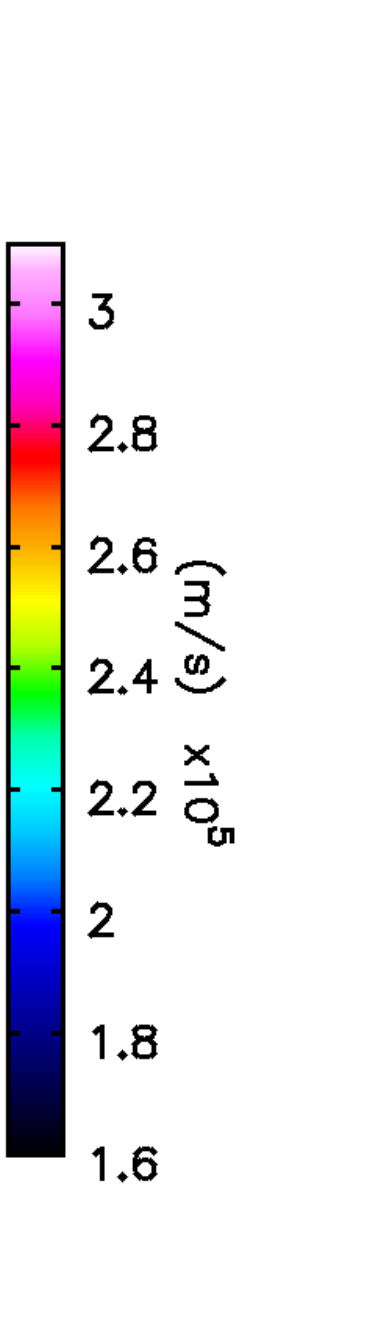}
\\
\includegraphics[width=0.30 \textwidth]{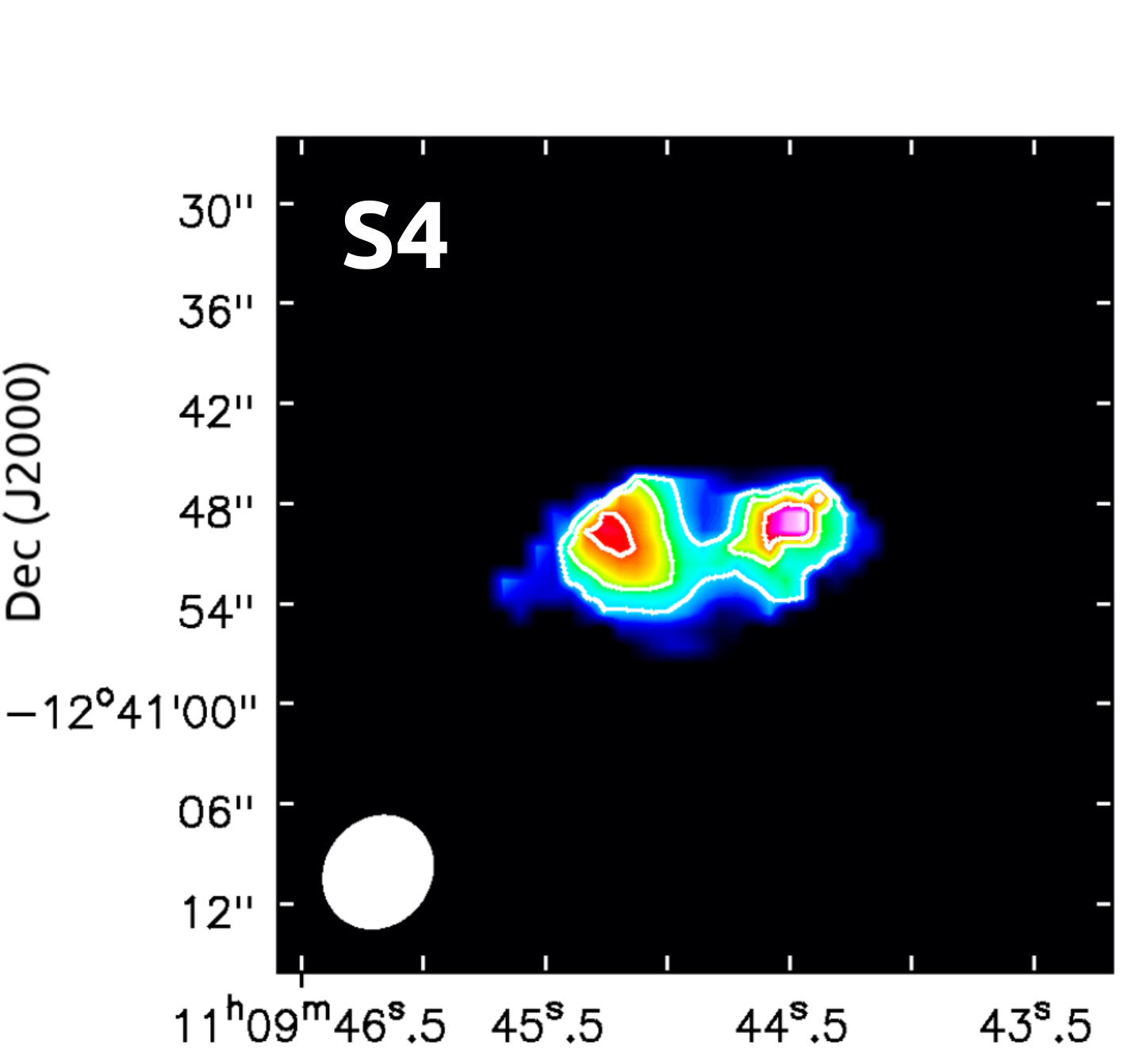}&
\includegraphics[width=0.30 \textwidth]{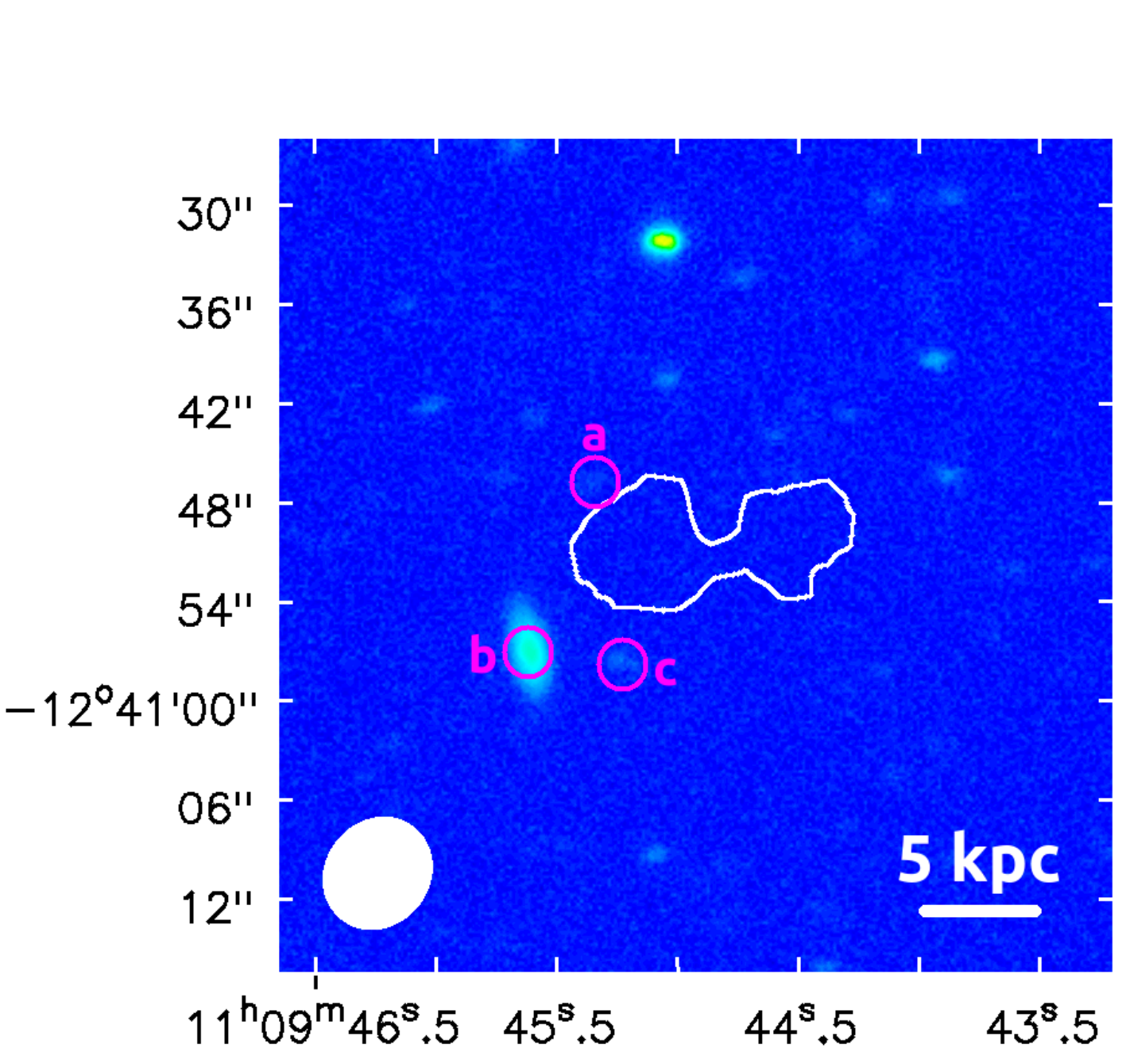}&
\includegraphics[width=0.30 \textwidth]{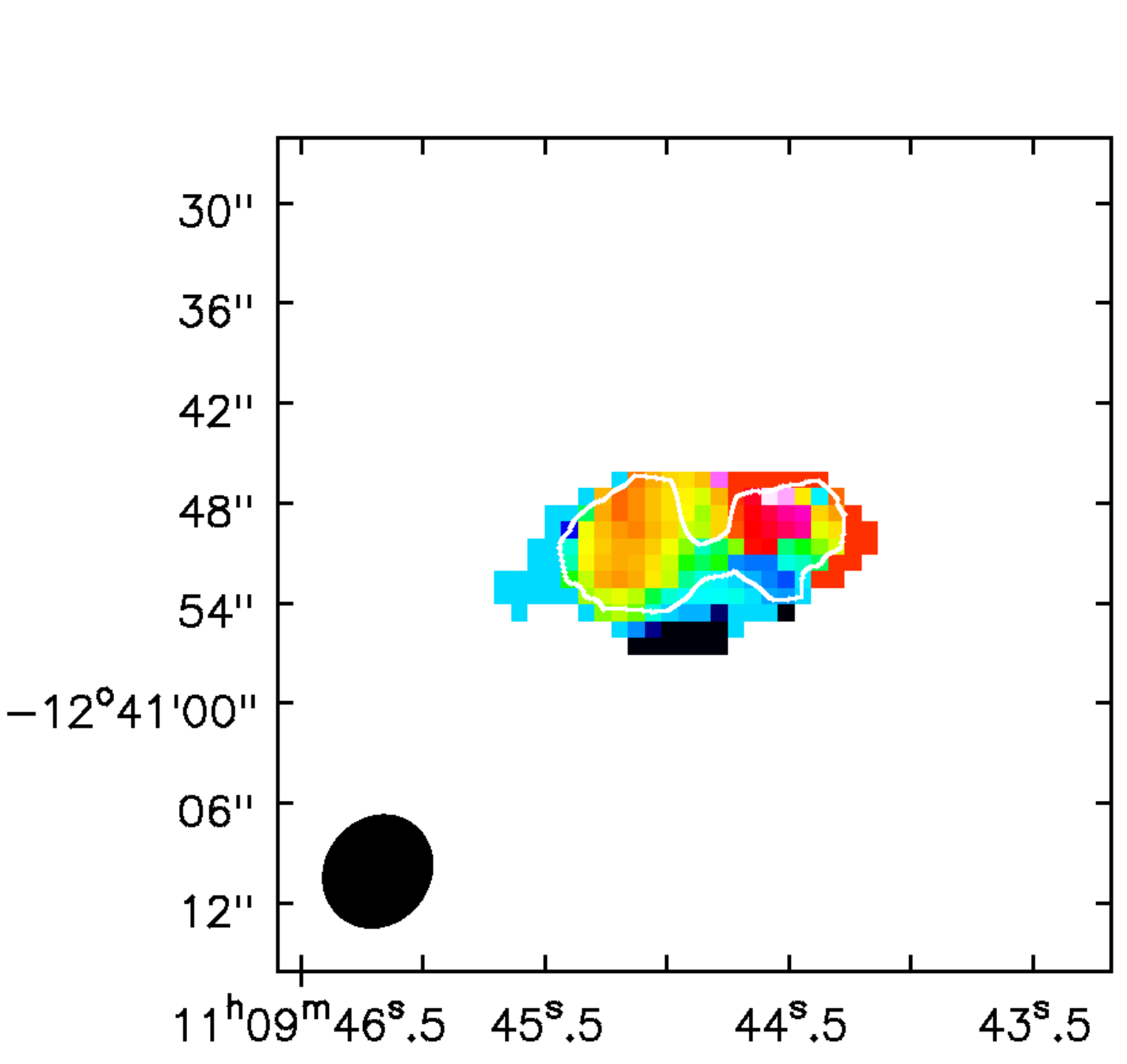}&
\includegraphics[width=0.08 \textwidth]{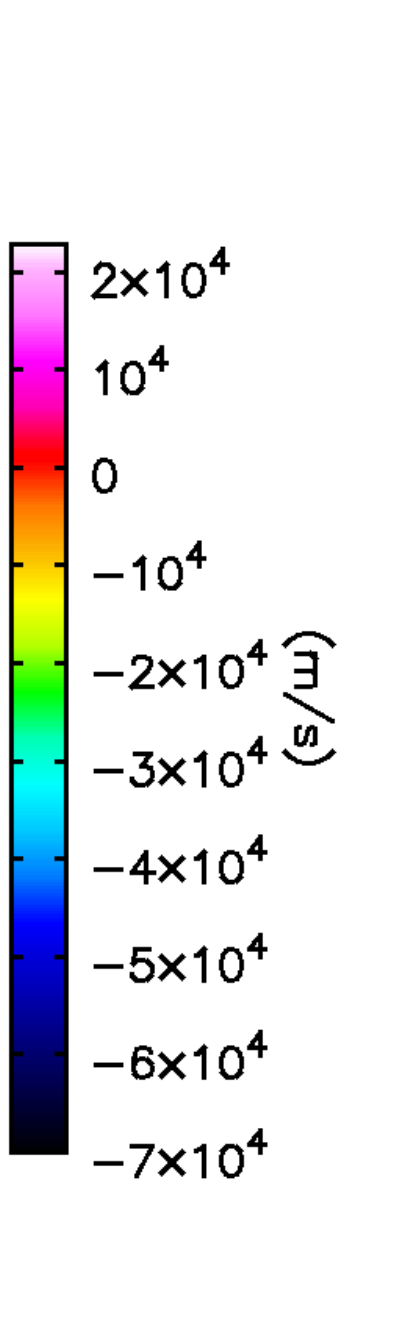}
\\
\includegraphics[width=0.30 \textwidth]{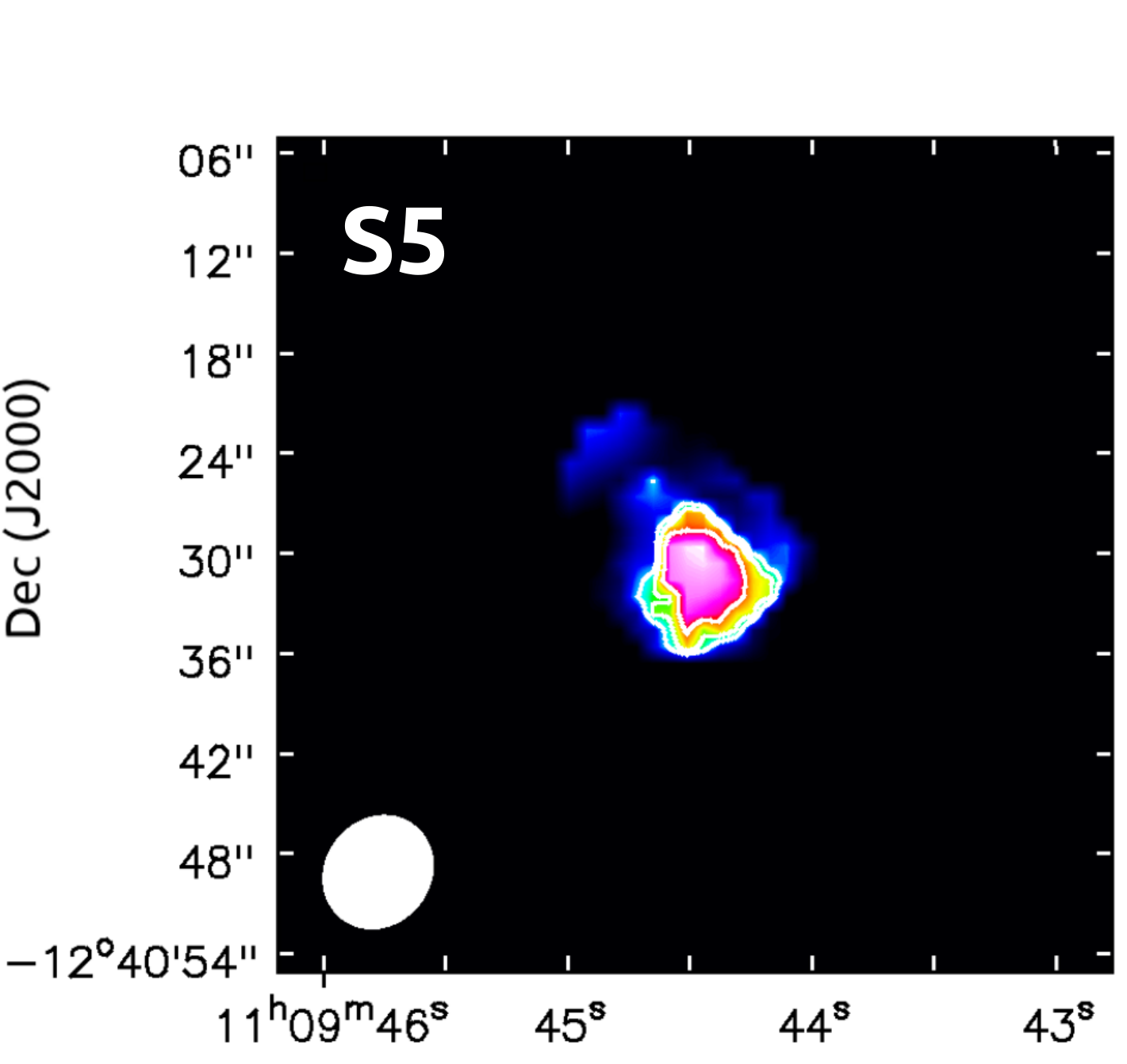}&
\includegraphics[width=0.30 \textwidth]{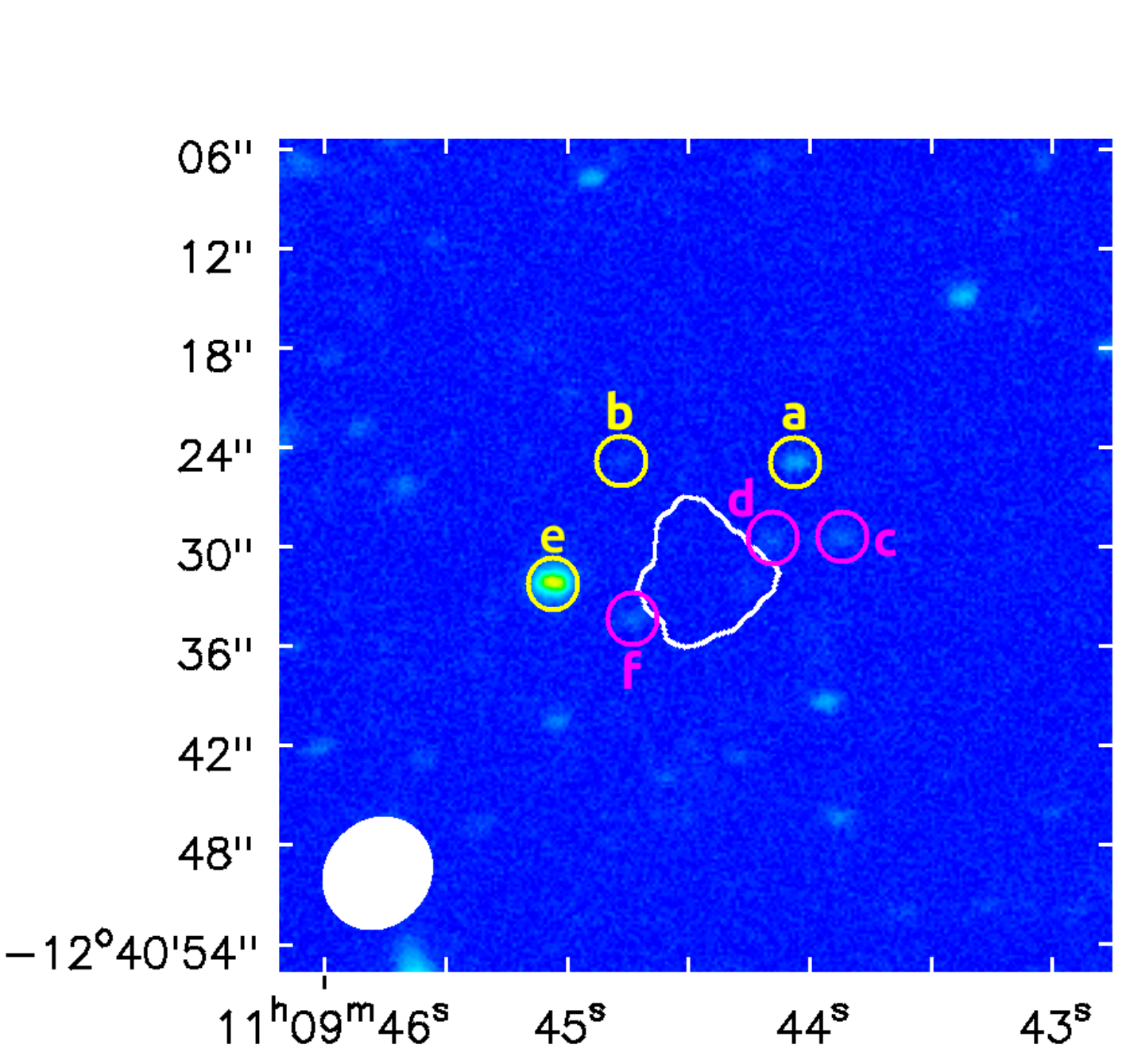}&
\includegraphics[width=0.30 \textwidth]{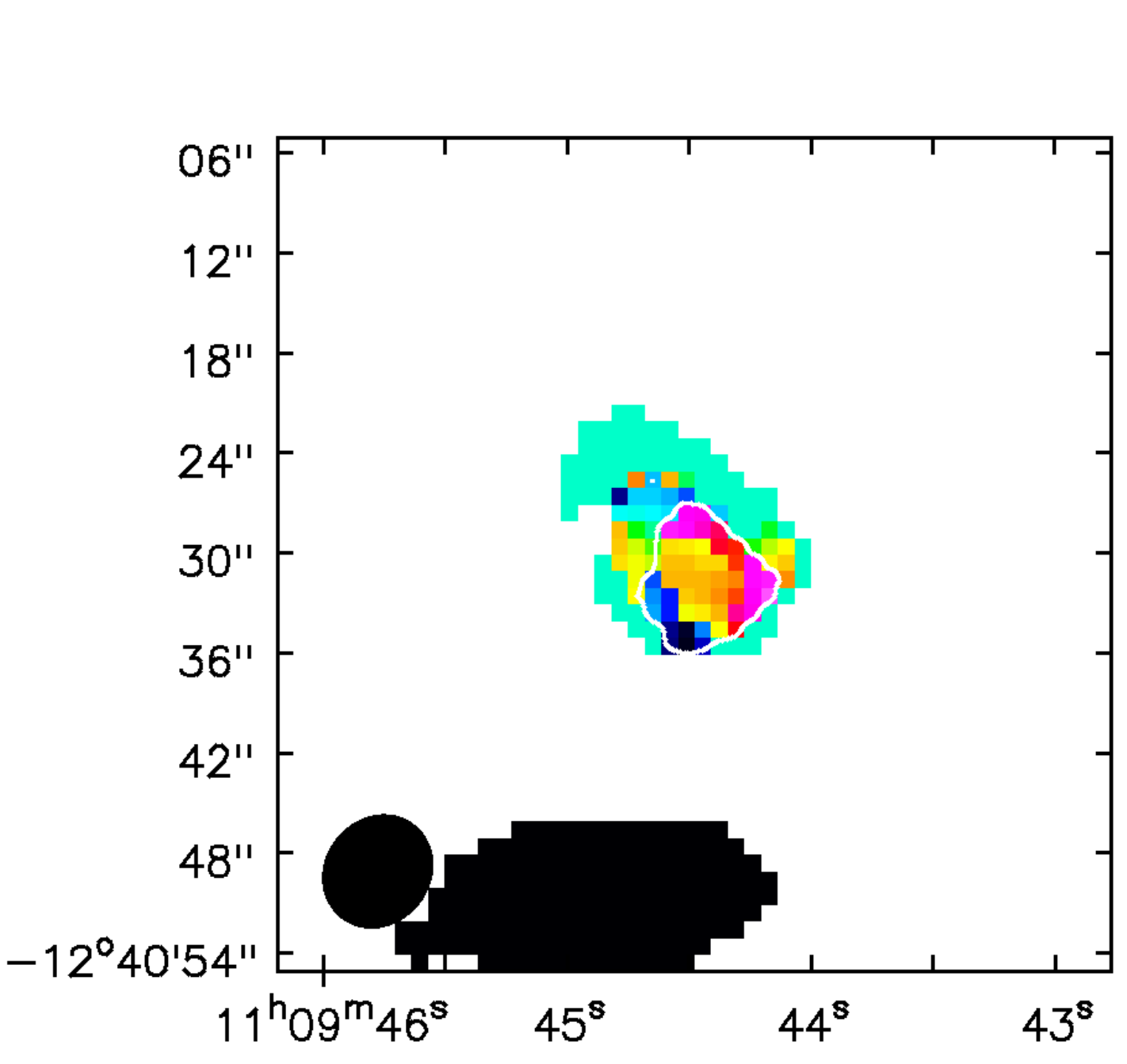}&
\includegraphics[width=0.08 \textwidth]{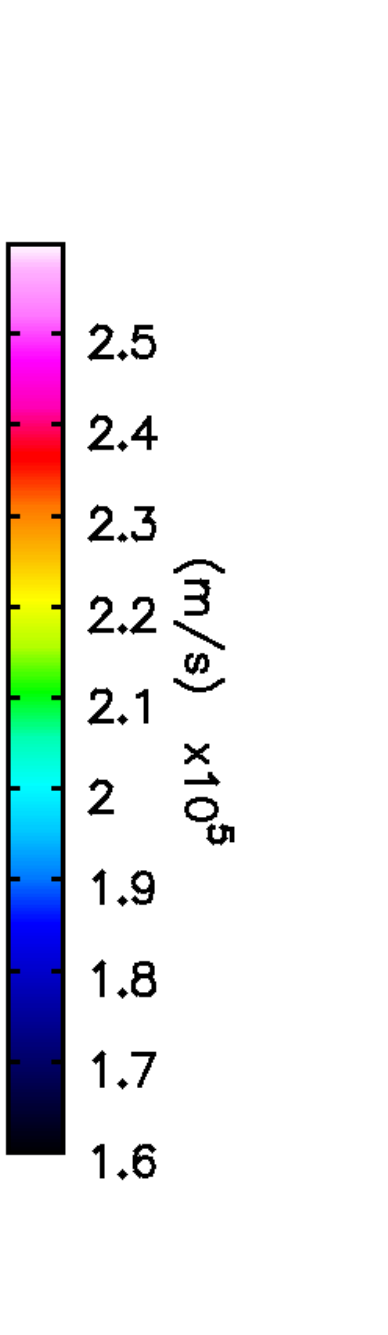}
\\
\hskip 1 cm \includegraphics[width=0.07 \textwidth]{figs/RA.pdf}&
\hskip 1 cm \includegraphics[width=0.07 \textwidth]{figs/RA.pdf}&
\hskip 1 cm \includegraphics[width=0.07 \textwidth]{figs/RA.pdf}&
\end{tabular}
\caption{ 
\label{fig:maps} Continued}
\end{figure*}

\addtocounter{figure}{-1}

\begin{figure*}
\centering
\begin{tabular}{cccc} 
\includegraphics[width=0.30 \textwidth]{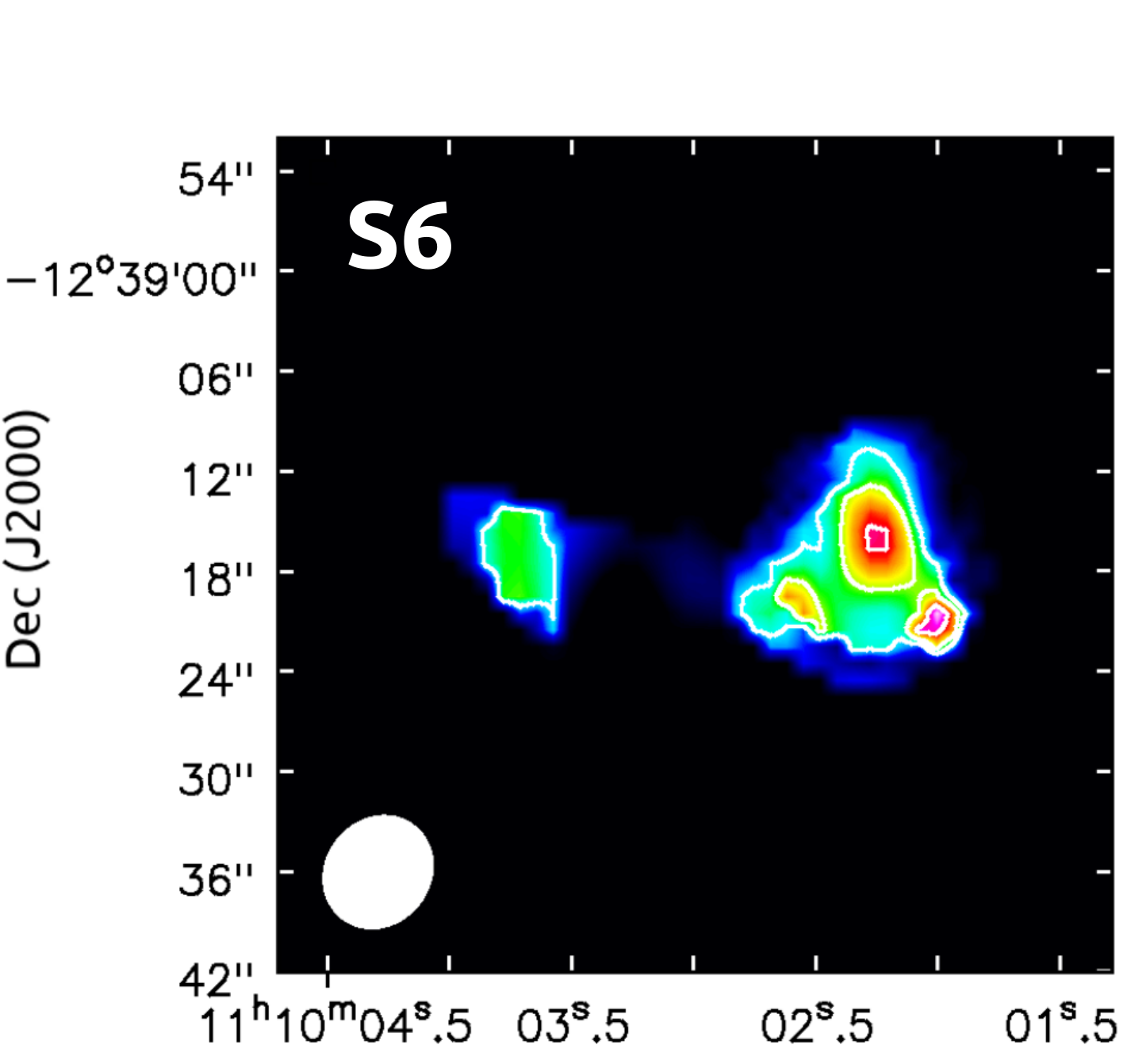}&
\includegraphics[width=0.30 \textwidth]{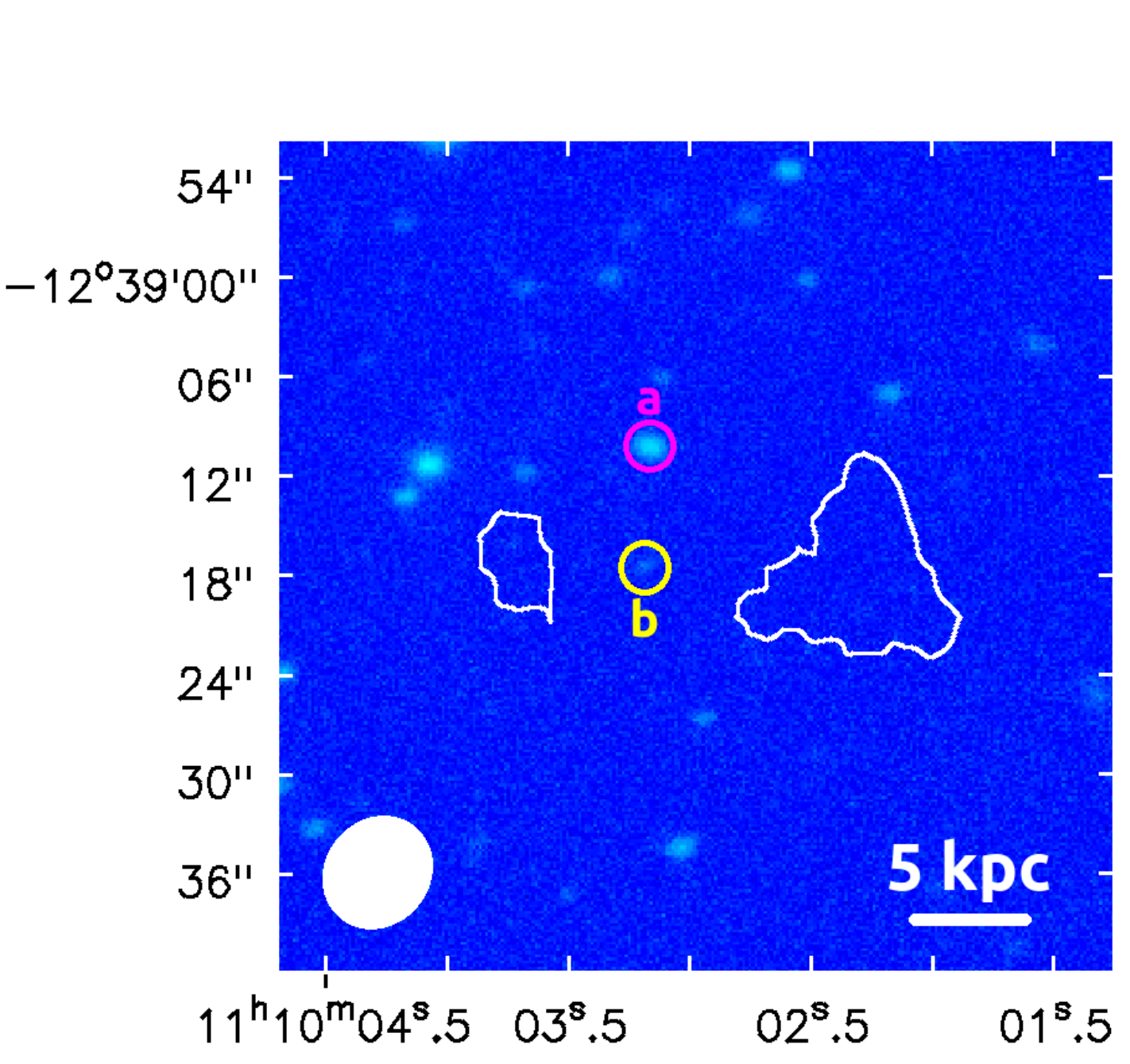}&
\includegraphics[width=0.30 \textwidth]{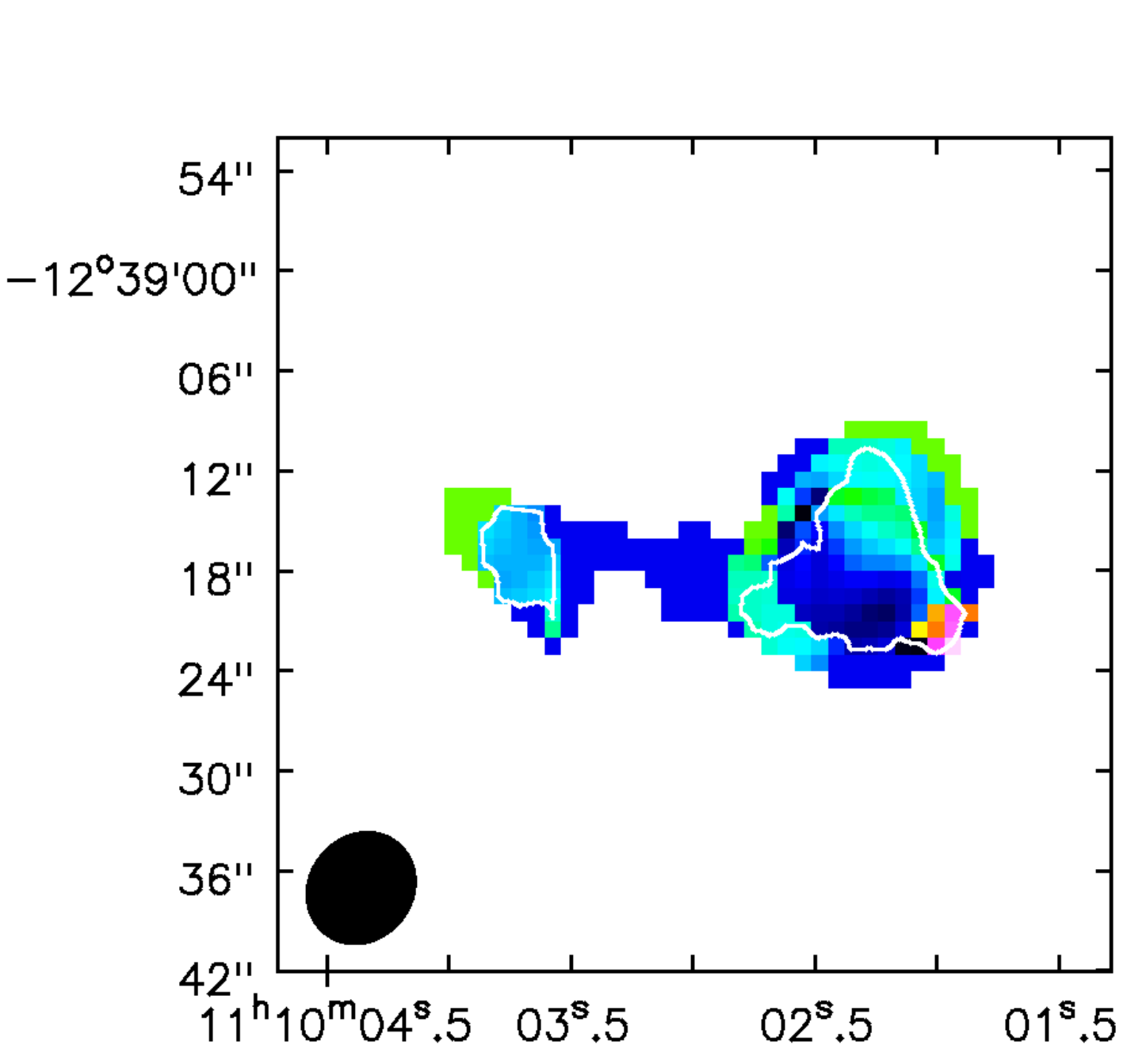}&
\includegraphics[width=0.08 \textwidth]{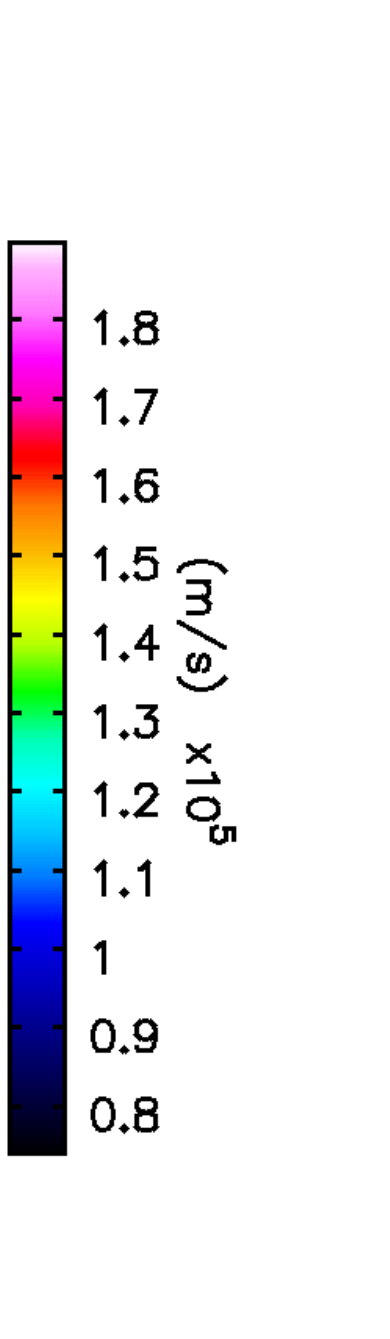}
\\
\includegraphics[width=0.30 \textwidth]{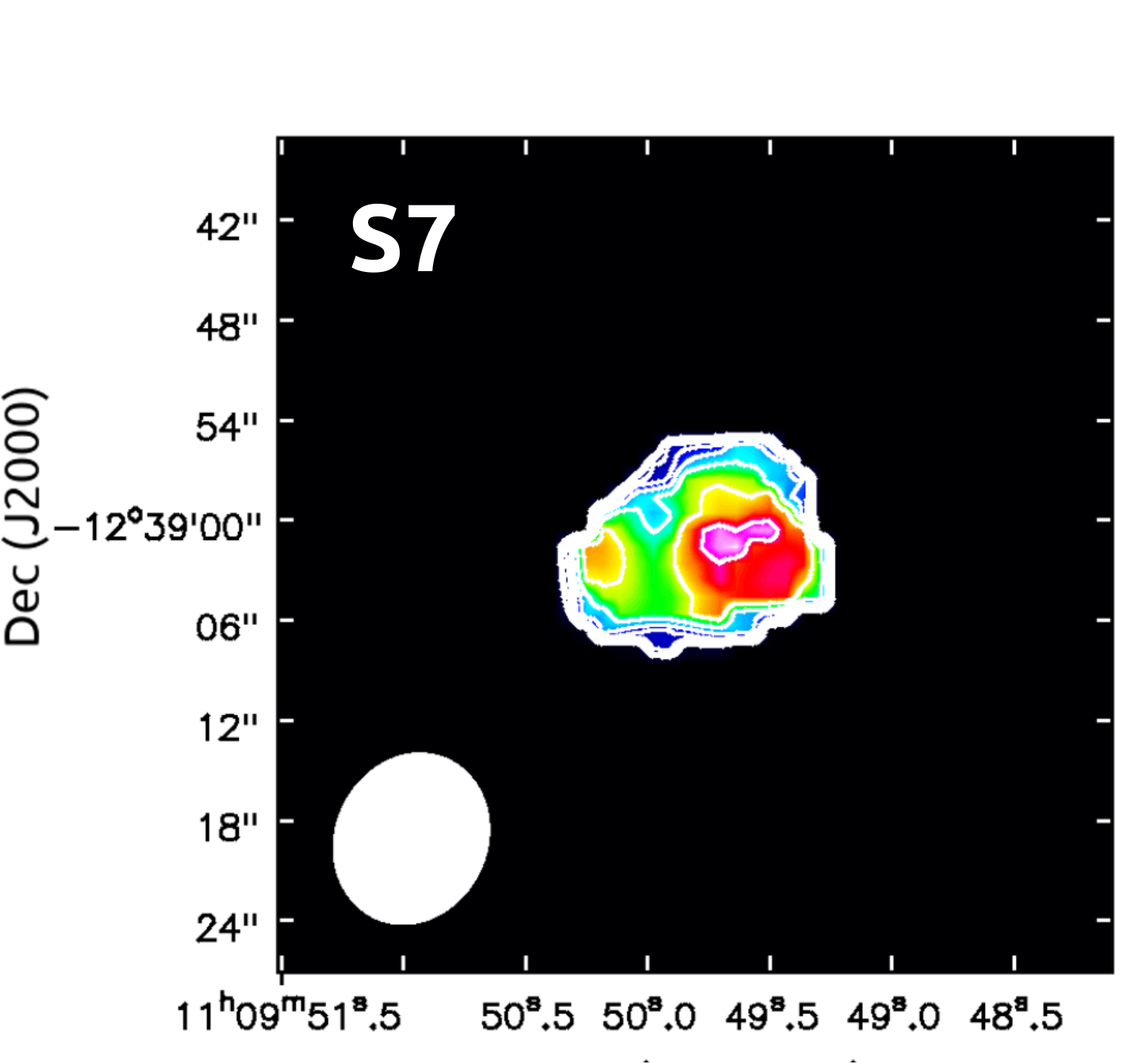}&
\includegraphics[width=0.30 \textwidth]{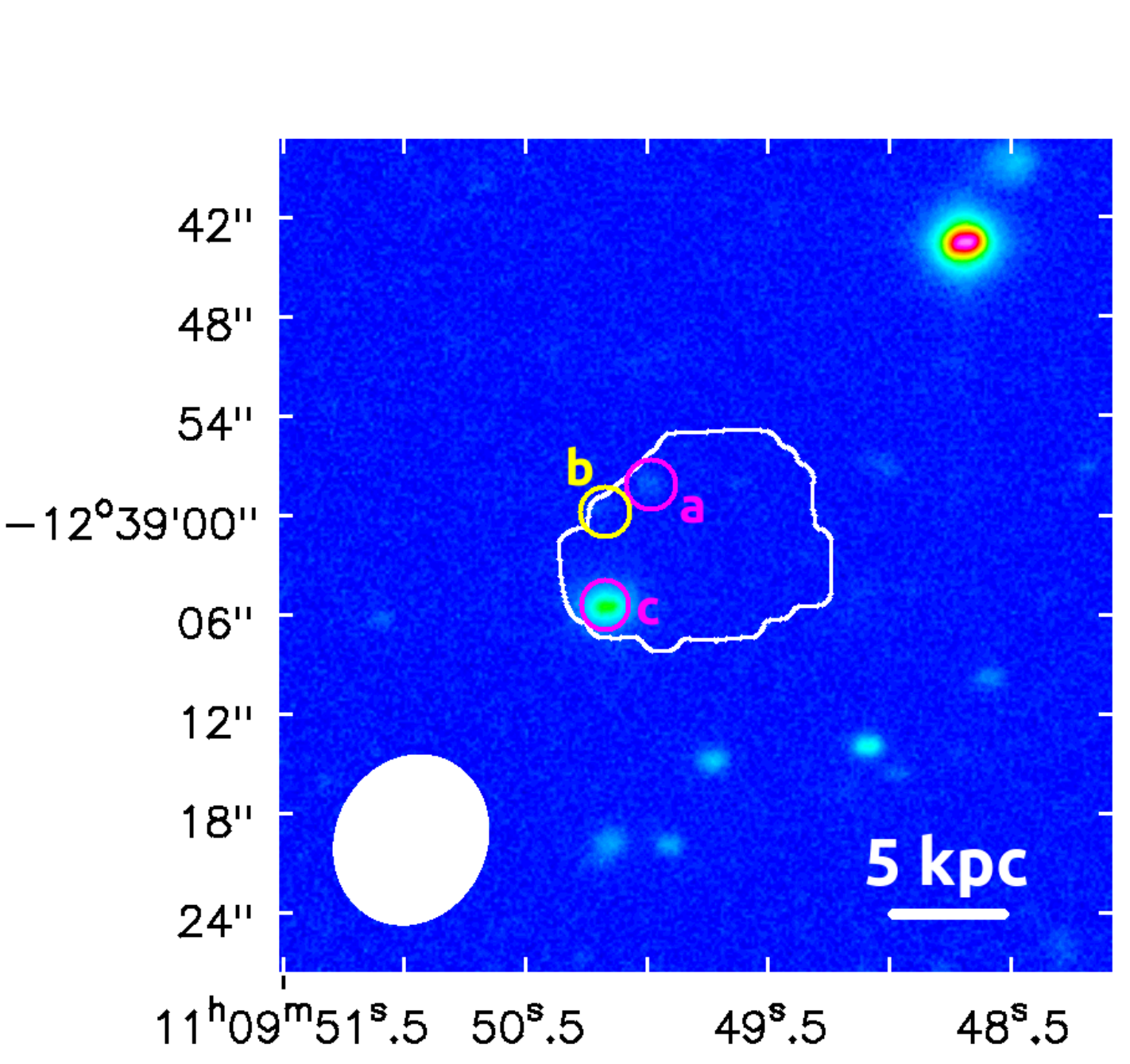}&
\includegraphics[width=0.30 \textwidth]{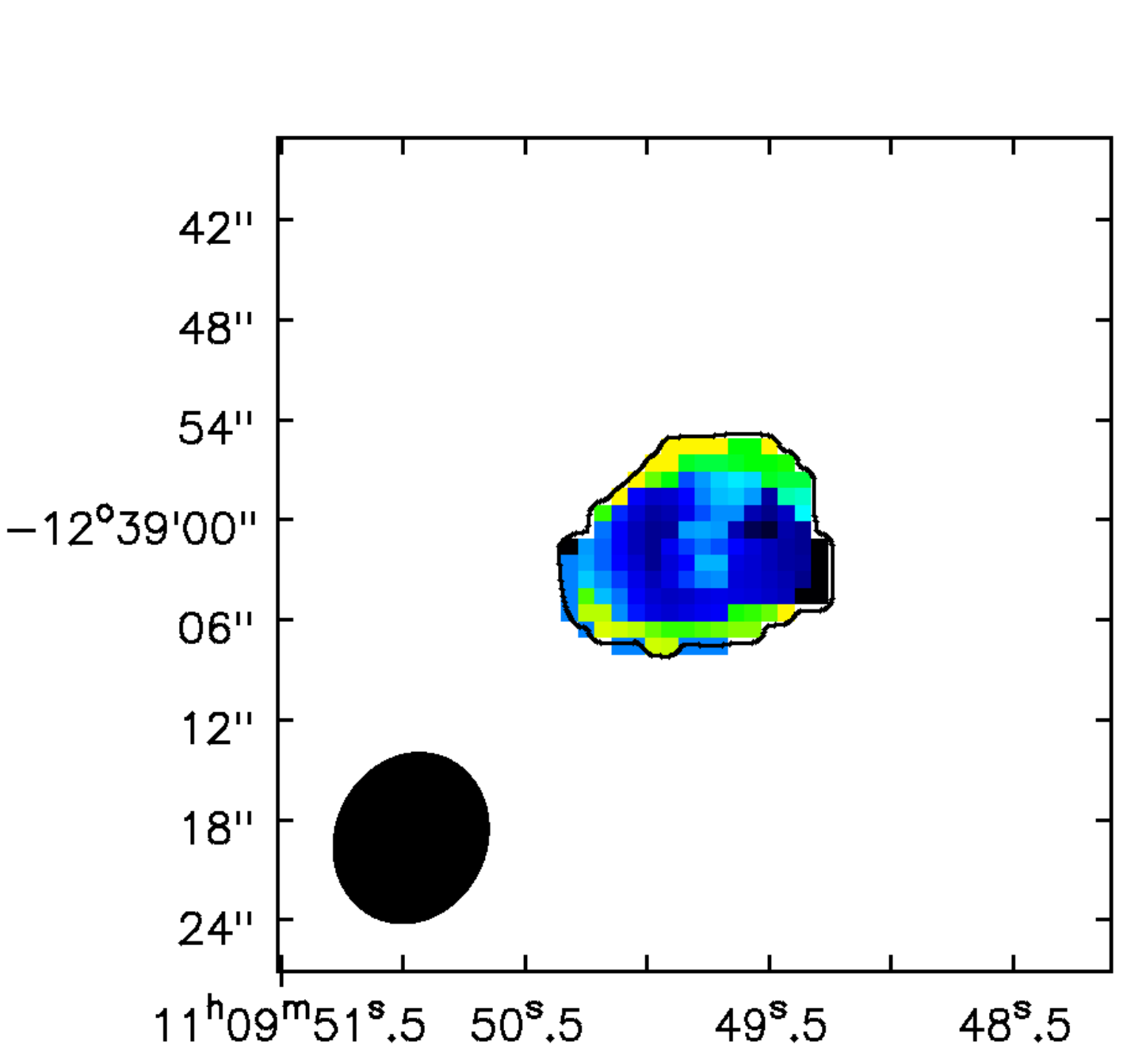}&
\includegraphics[width=0.08 \textwidth]{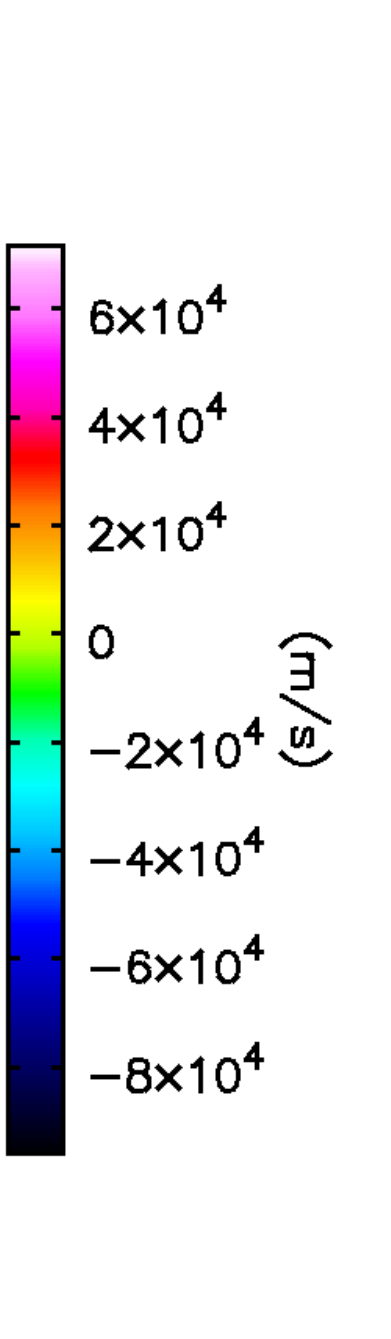}
\\
\includegraphics[width=0.30 \textwidth]{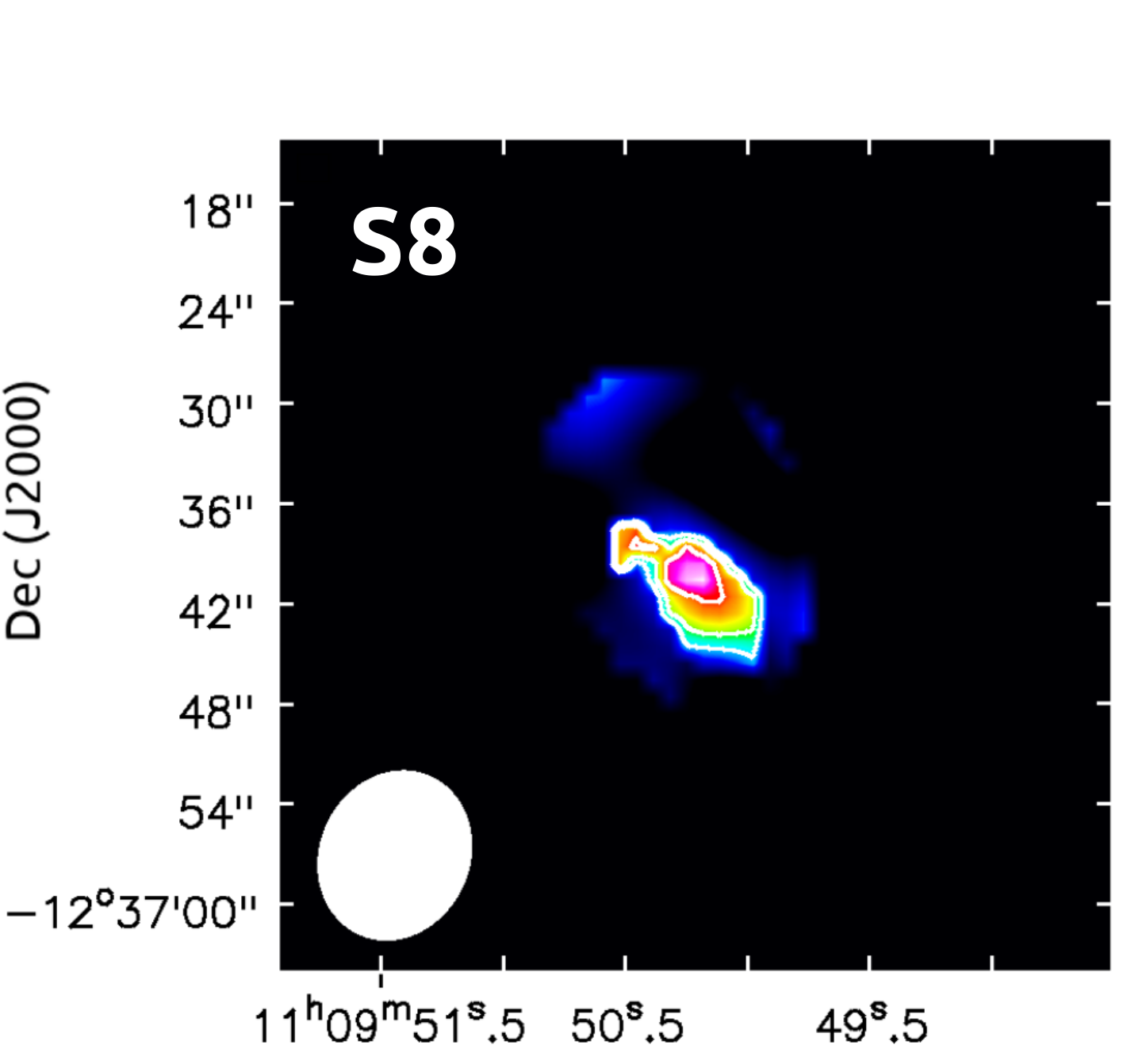}&
\includegraphics[width=0.30 \textwidth]{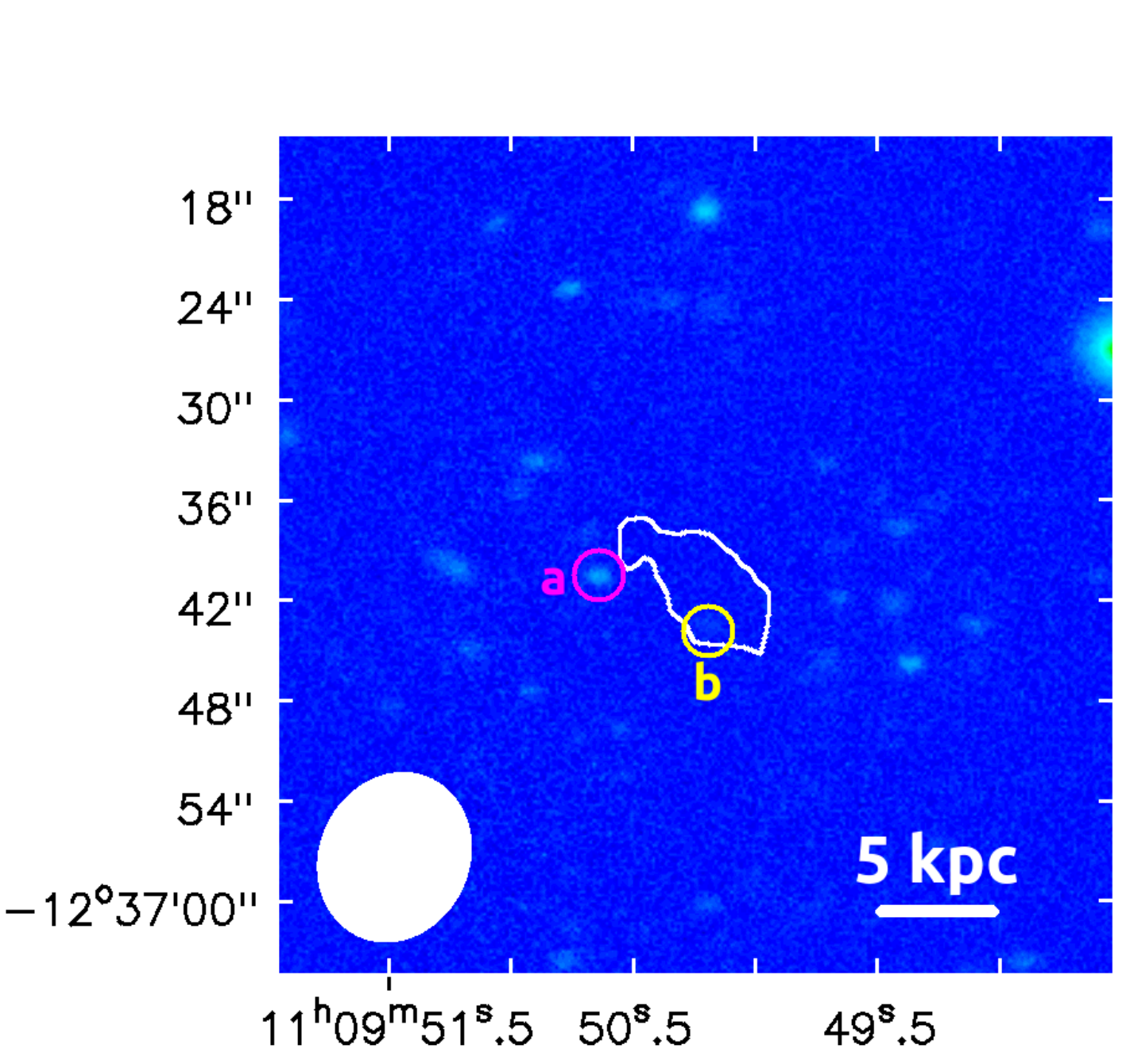}&
\includegraphics[width=0.30 \textwidth]{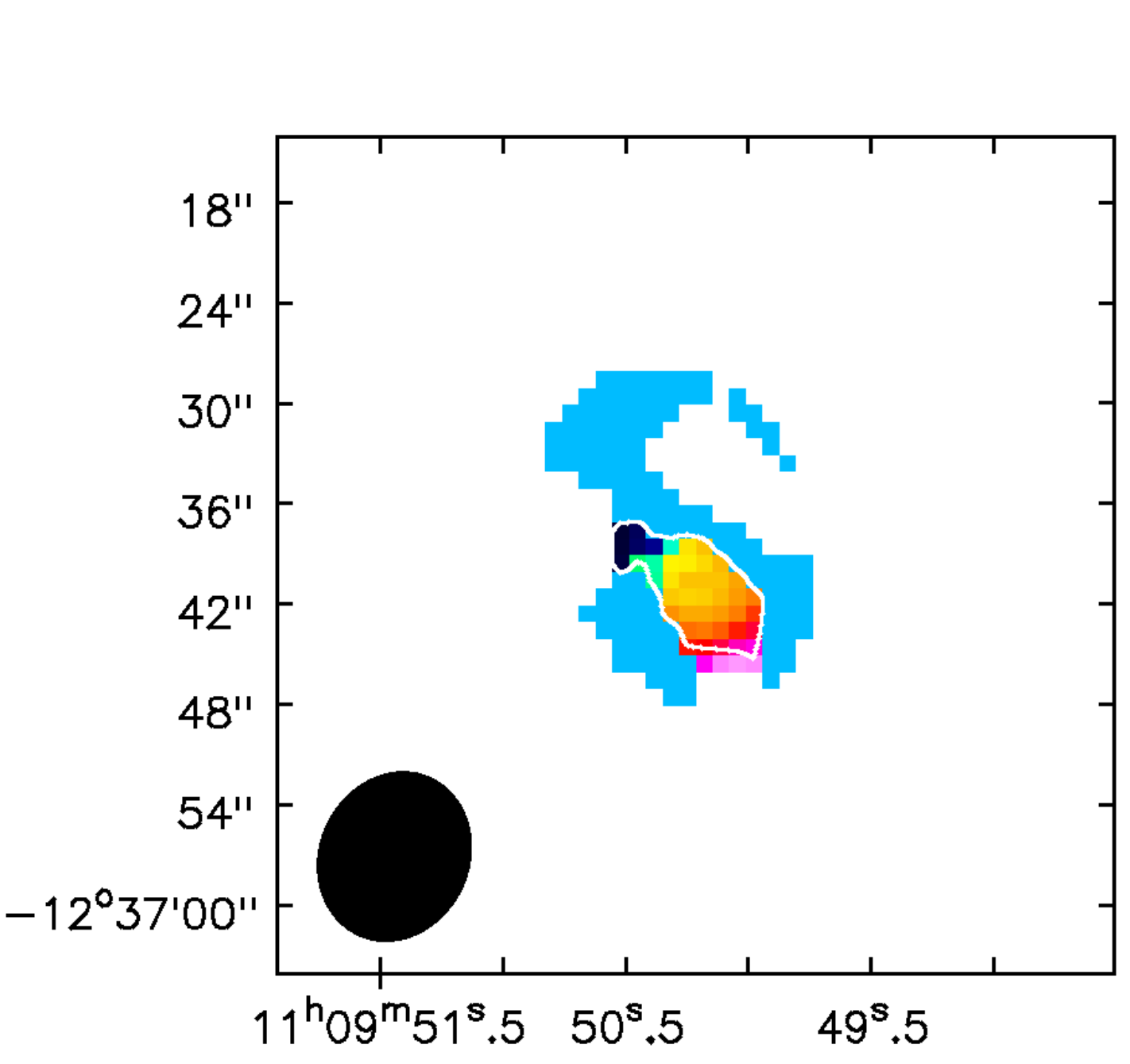}&
\includegraphics[width=0.08 \textwidth]{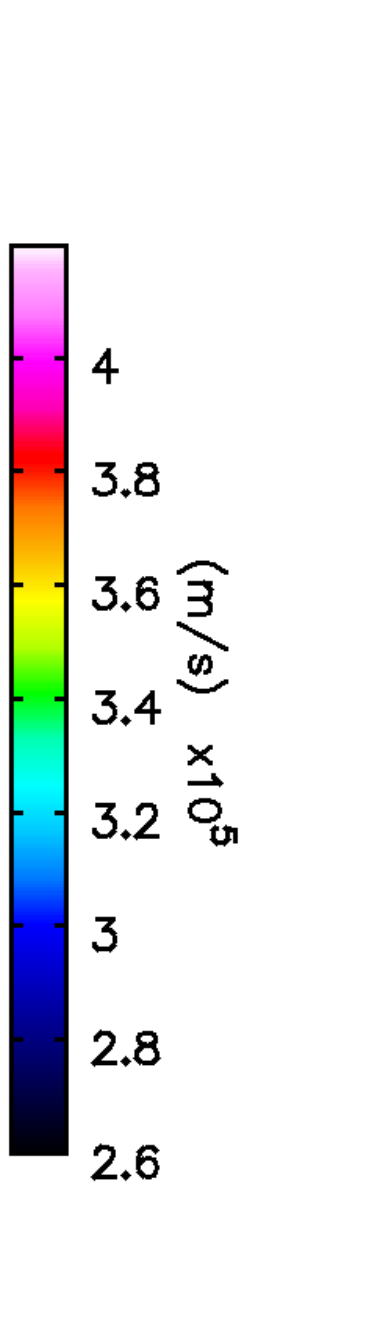}
\\
\includegraphics[width=0.30 \textwidth]{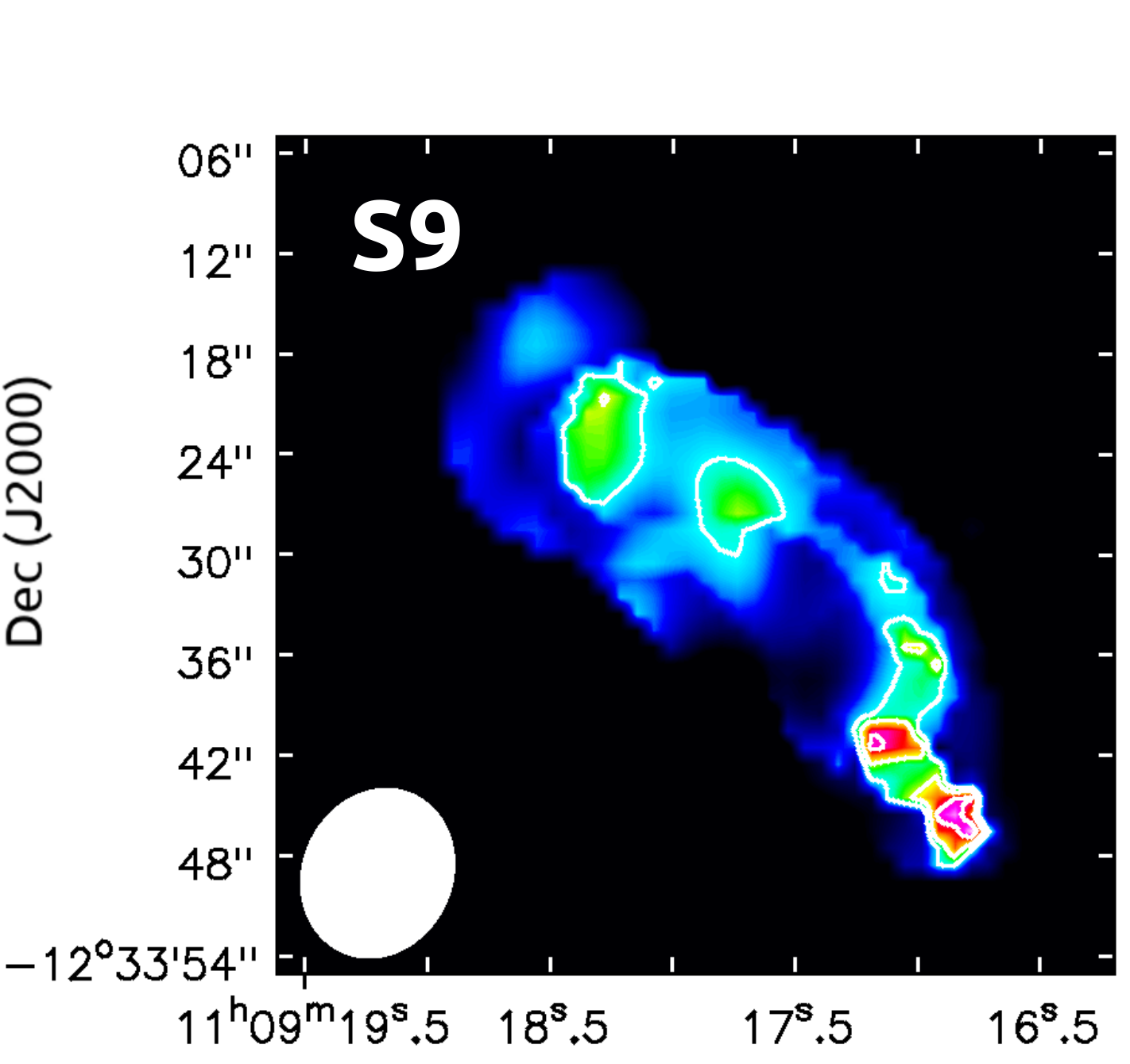}&
\includegraphics[width=0.30 \textwidth]{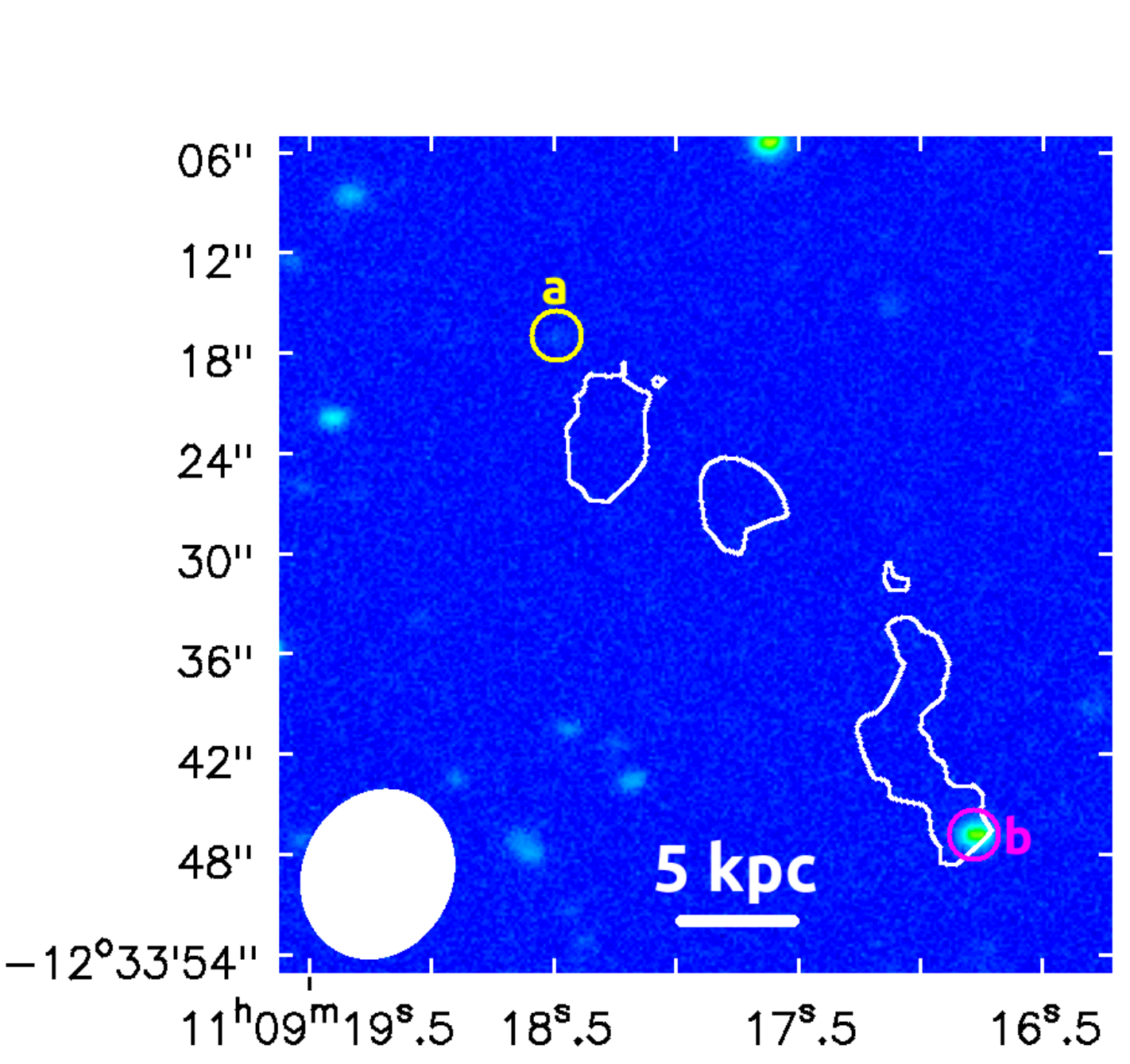}&
\includegraphics[width=0.30 \textwidth]{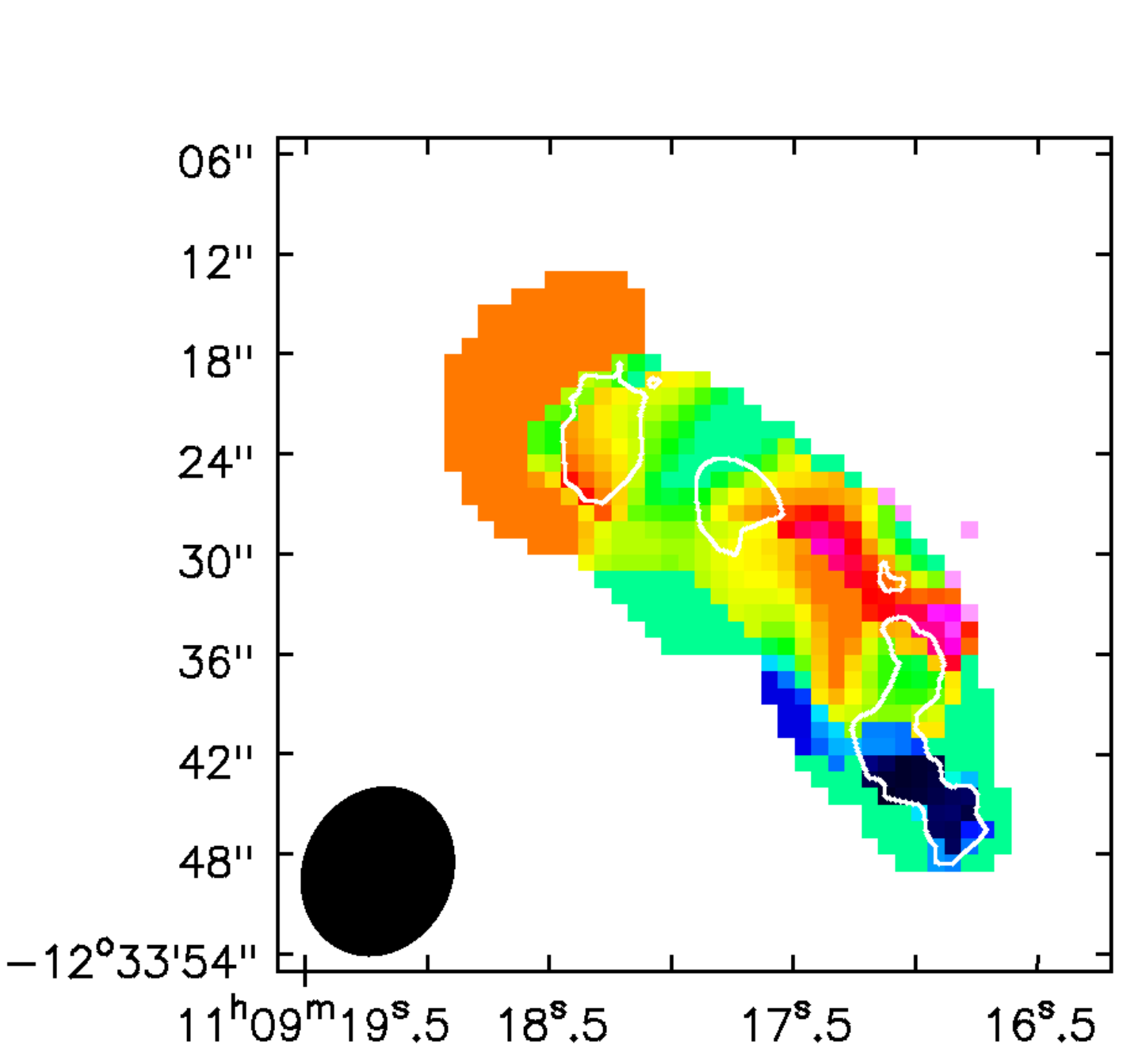}&
\includegraphics[width=0.08 \textwidth]{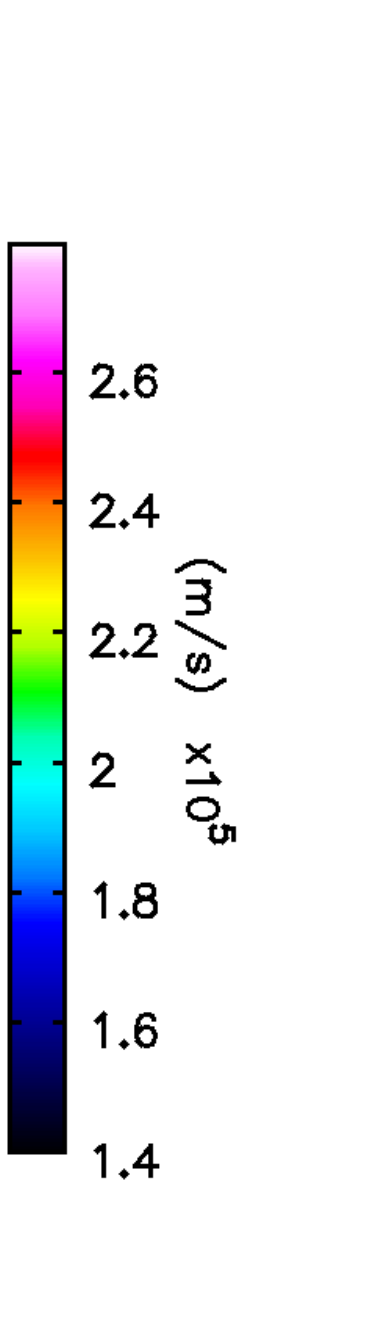}
\\
\hskip 1 cm \includegraphics[width=0.07 \textwidth]{figs/RA.pdf}&
\hskip 1 cm \includegraphics[width=0.07 \textwidth]{figs/RA.pdf}&
\hskip 1 cm \includegraphics[width=0.07 \textwidth]{figs/RA.pdf}&
\end{tabular}
\caption{ 
\label{fig:maps} Continued}
\end{figure*}

\addtocounter{figure}{-1}

\begin{figure*}
\centering
\begin{tabular}{cccc} 
\includegraphics[width=0.30 \textwidth]{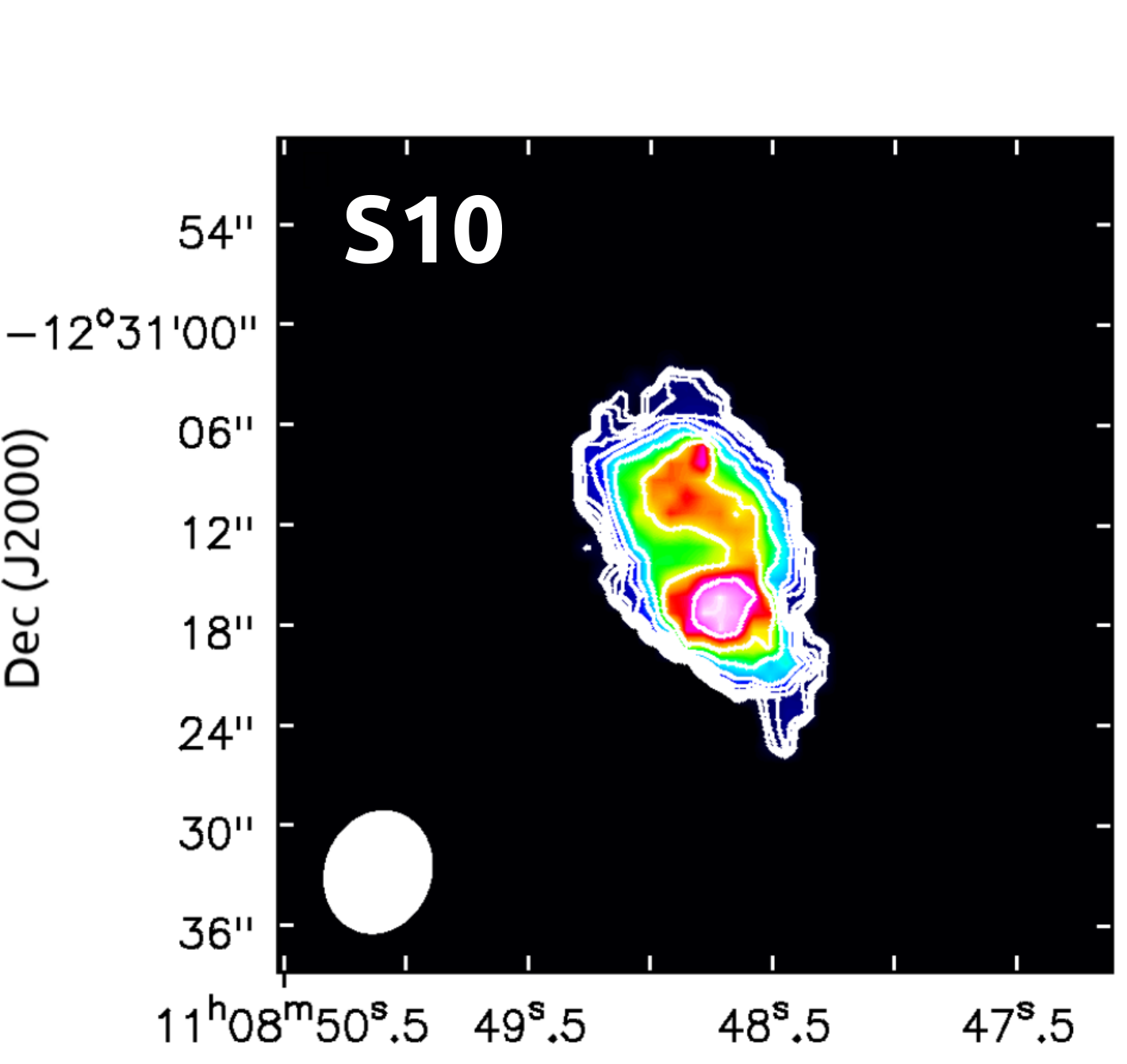}&
\includegraphics[width=0.30 \textwidth]{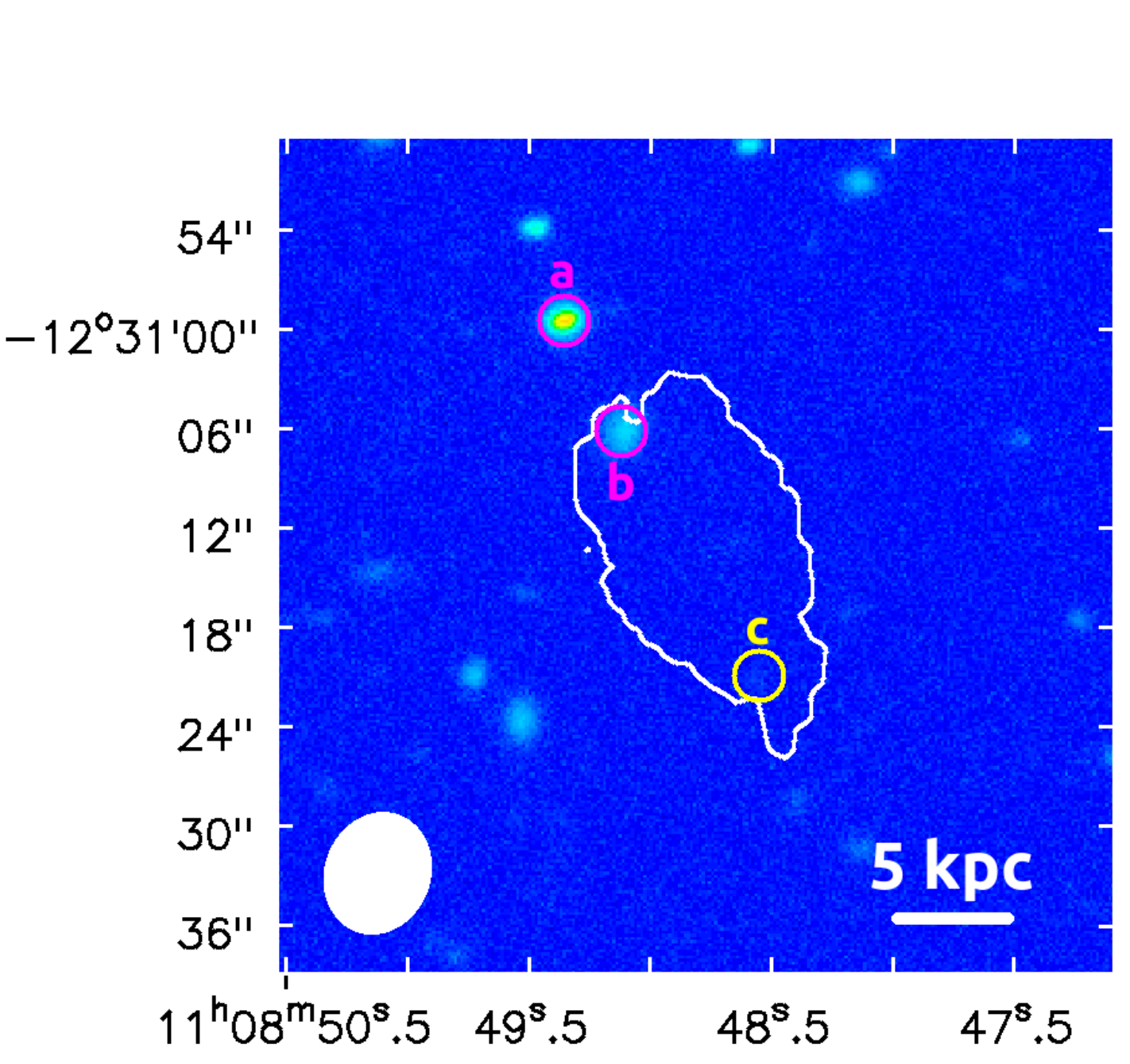}&
\includegraphics[width=0.30 \textwidth]{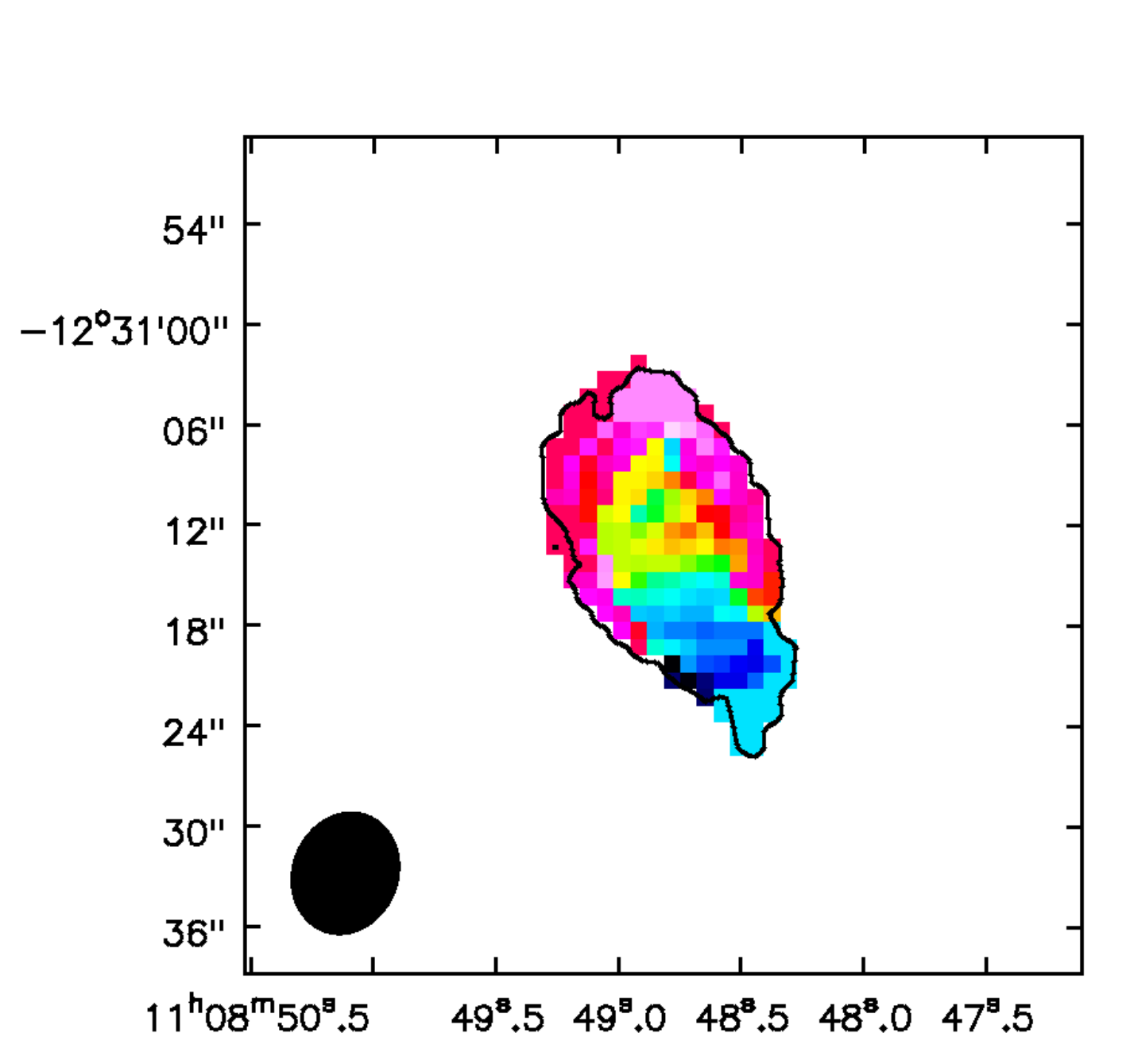}&
\includegraphics[width=0.08 \textwidth]{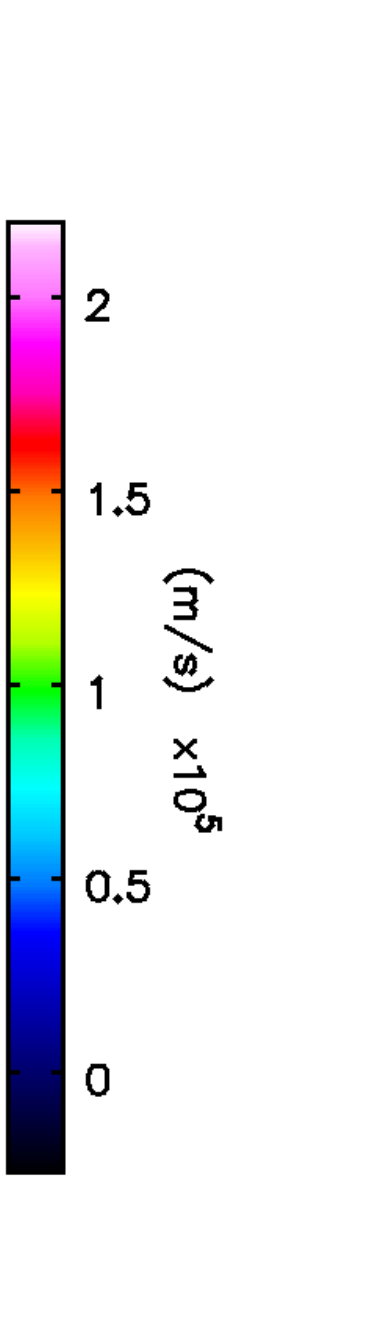}
\\
\includegraphics[width=0.30 \textwidth]{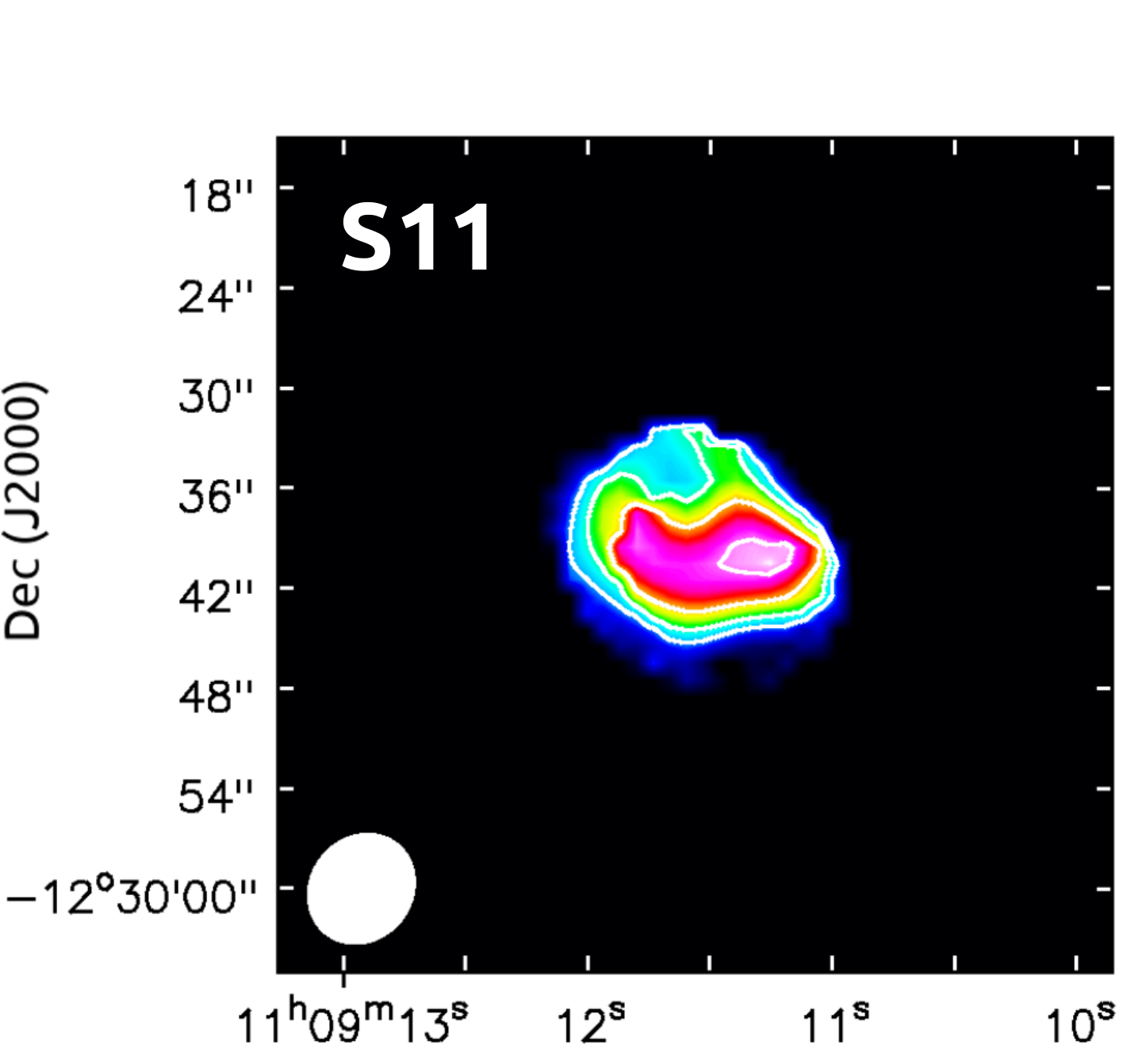}&
\includegraphics[width=0.30 \textwidth]{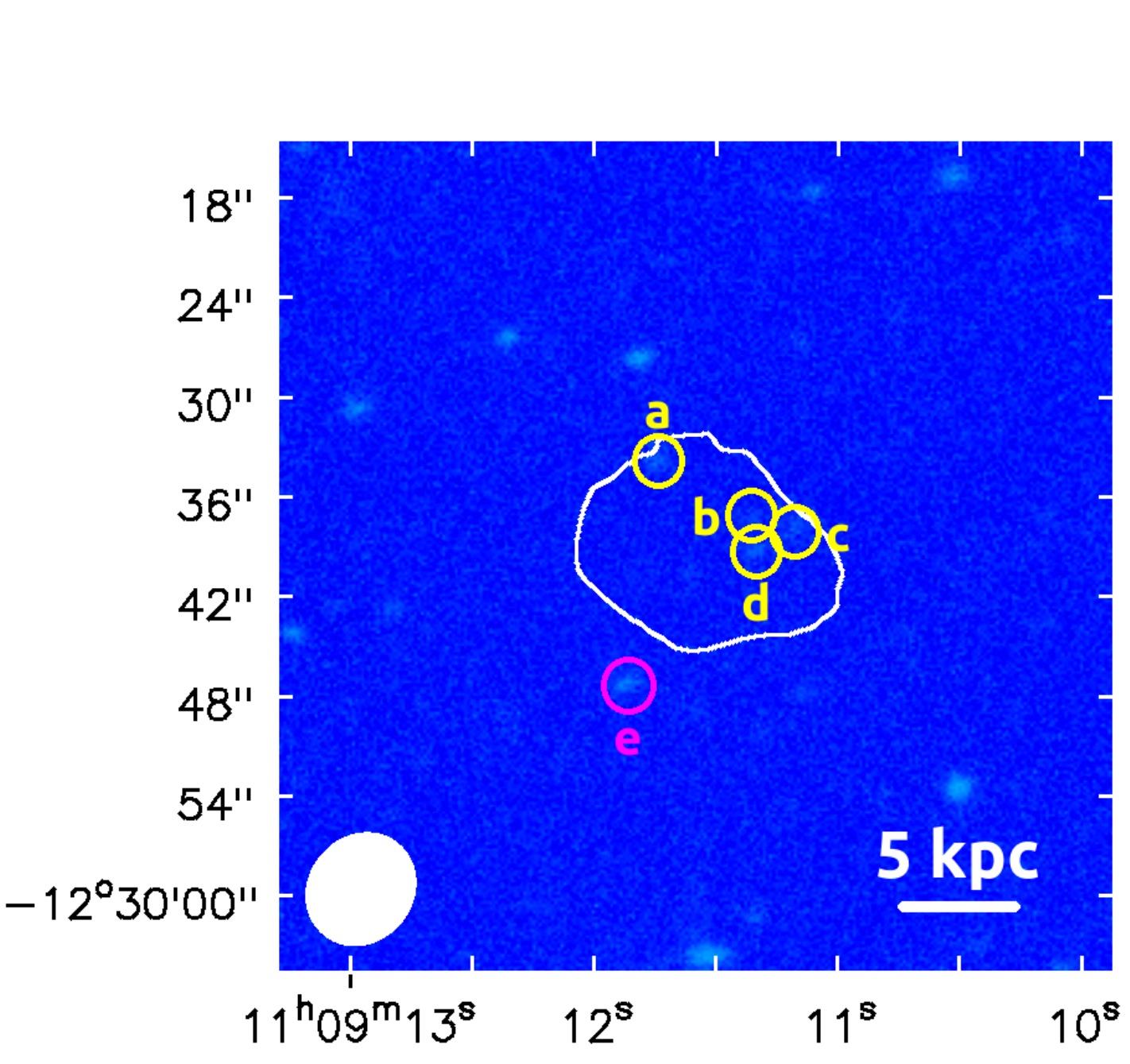}&
\includegraphics[width=0.30 \textwidth]{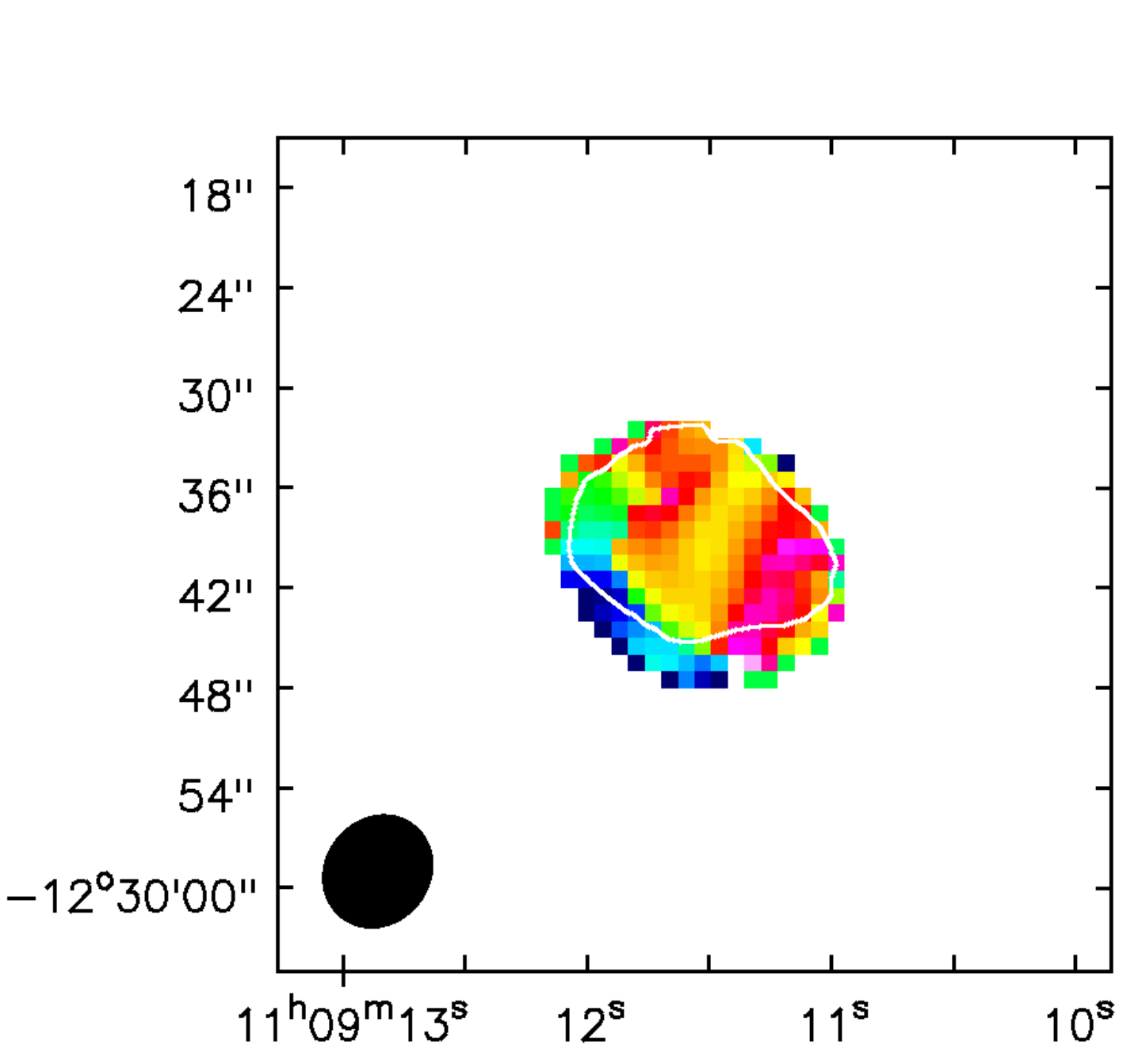}&
\includegraphics[width=0.08 \textwidth]{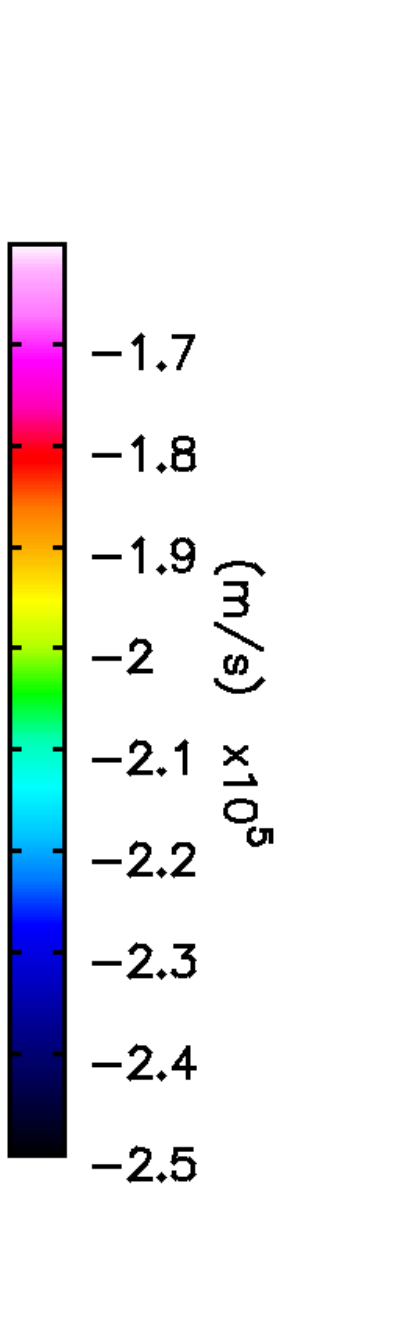}
\\
\includegraphics[width=0.30 \textwidth]{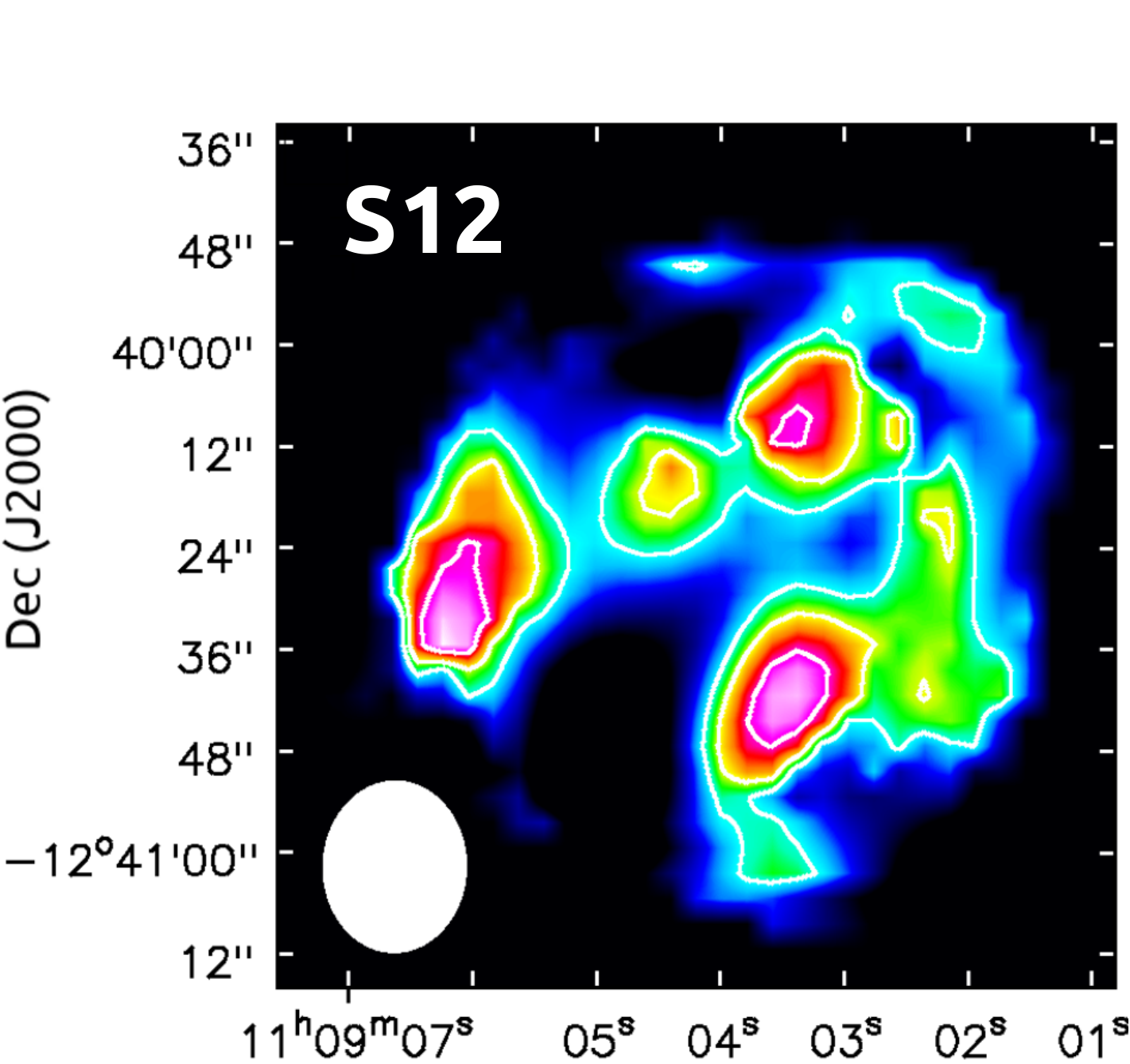}&
\includegraphics[width=0.30 \textwidth]{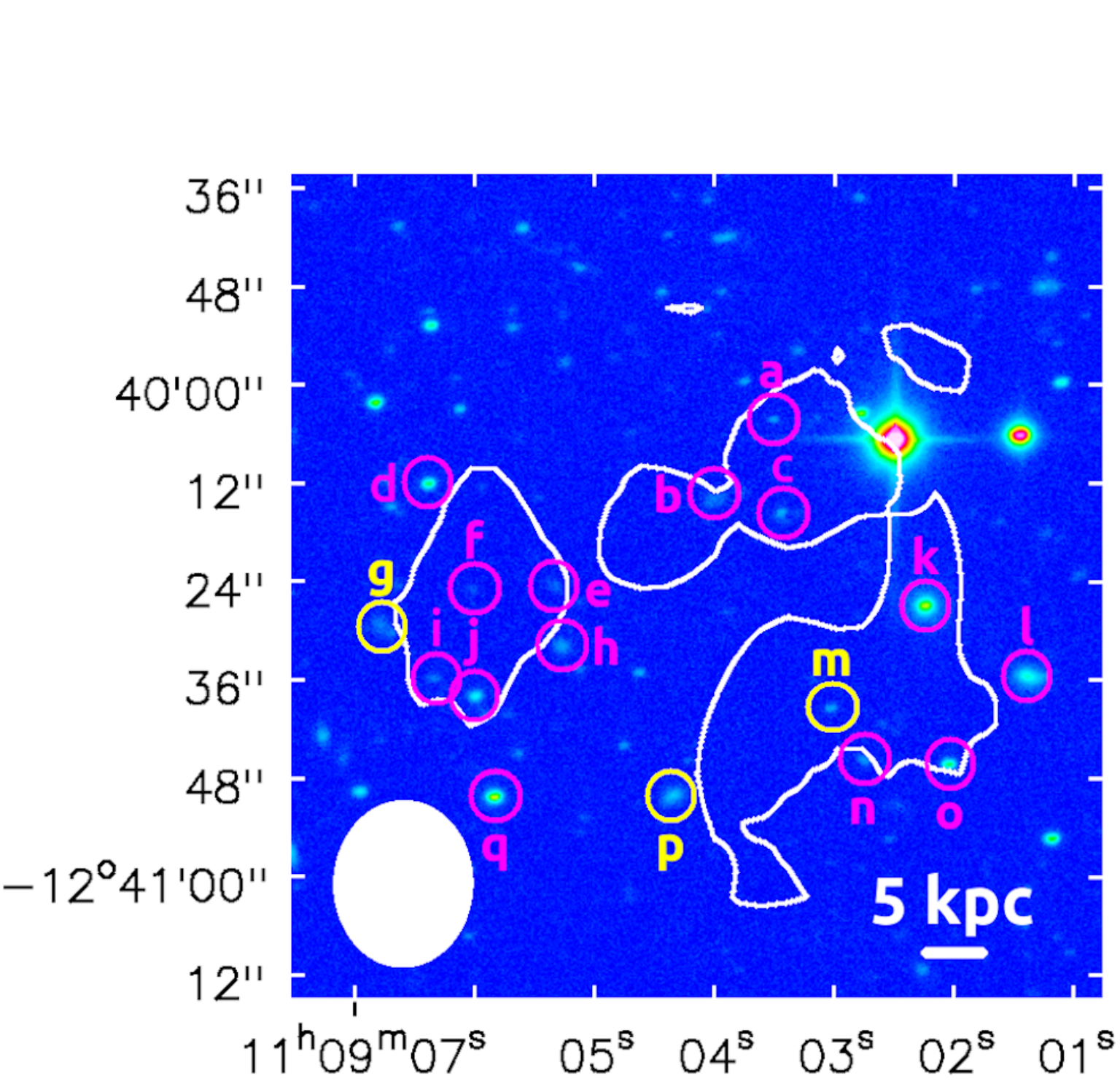}&
\includegraphics[width=0.30 \textwidth]{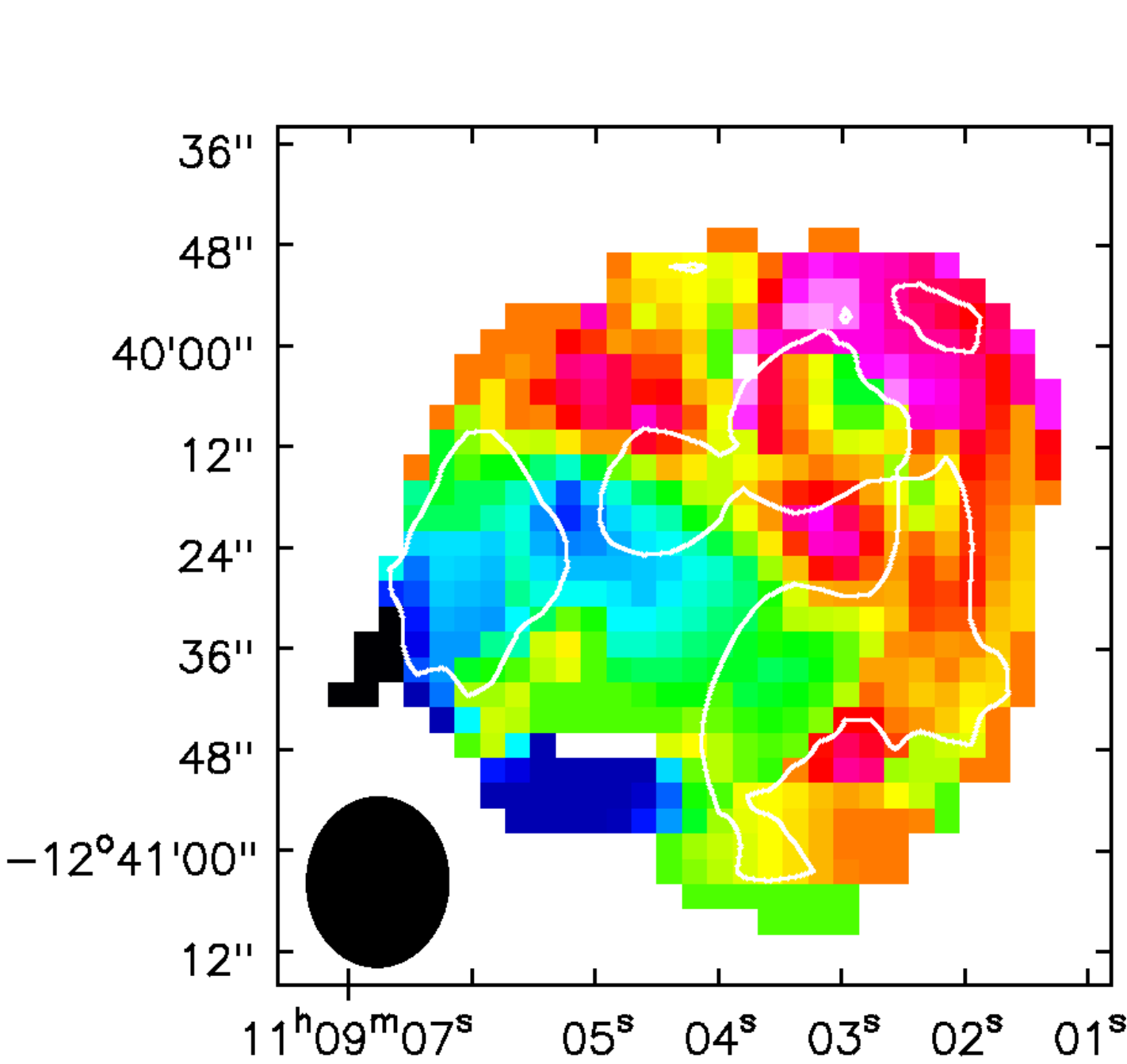}&
\includegraphics[width=0.08 \textwidth]{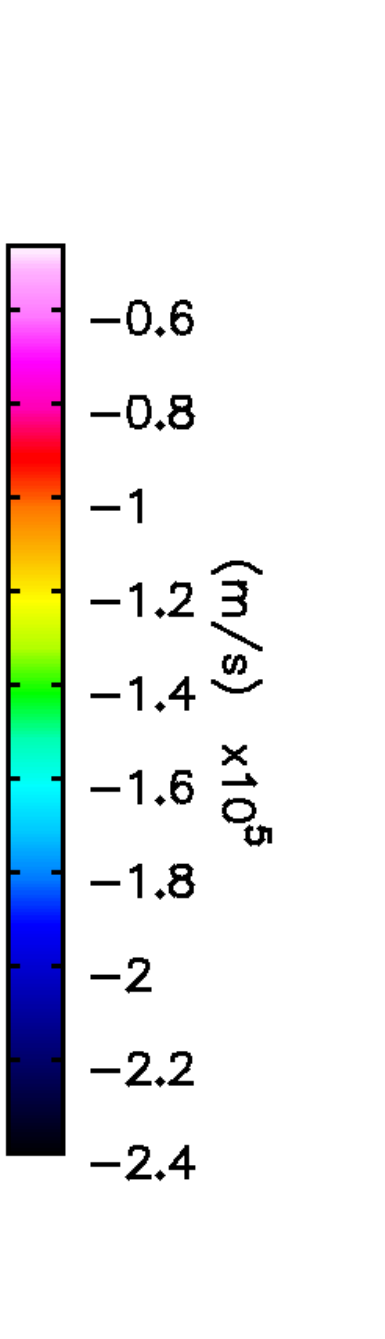}
\\
\hskip 1 cm \includegraphics[width=0.07 \textwidth]{figs/RA.pdf}&
\hskip 1 cm \includegraphics[width=0.07 \textwidth]{figs/RA.pdf}&
\hskip 1 cm \includegraphics[width=0.07 \textwidth]{figs/RA.pdf}&
\end{tabular}
\caption{ 
\label{fig:maps} Continued}
\end{figure*}

\begin{figure*}
\centering
\begin{tabular}{ccc} 
\includegraphics[width=0.33 \textwidth]{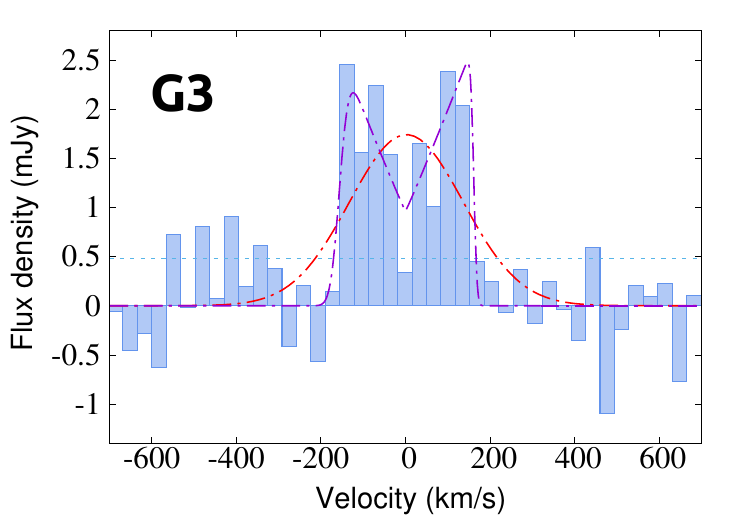}&
\includegraphics[width=0.33 \textwidth]{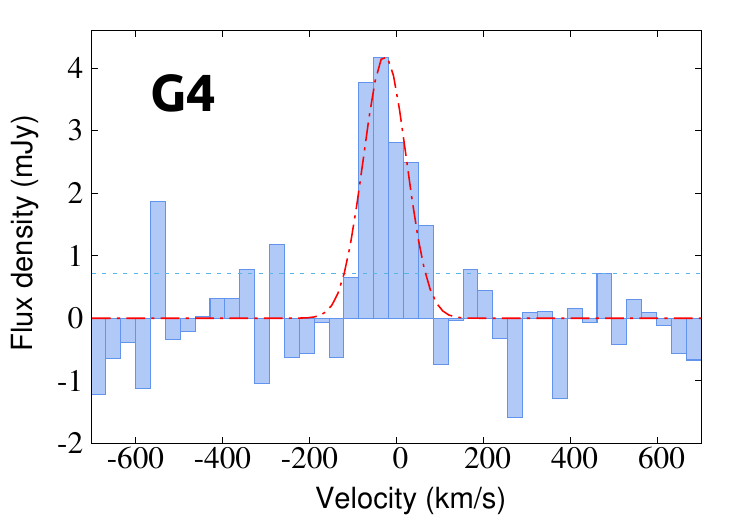}&
\includegraphics[width=0.33 \textwidth]{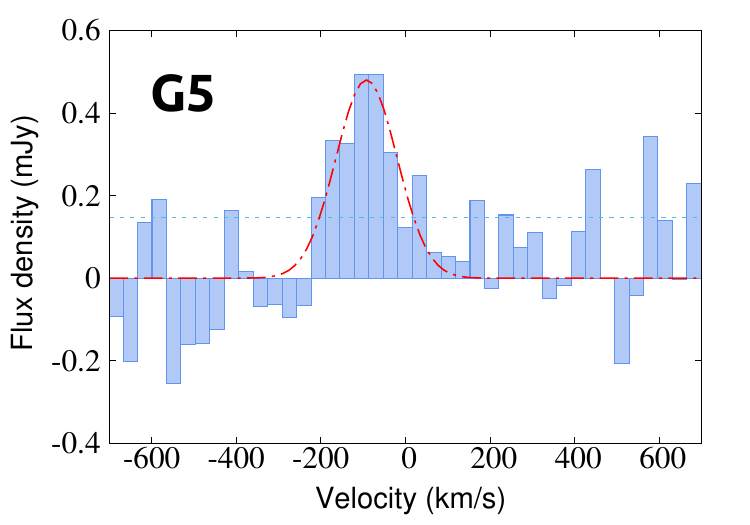}
\\
\includegraphics[width=0.33 \textwidth]{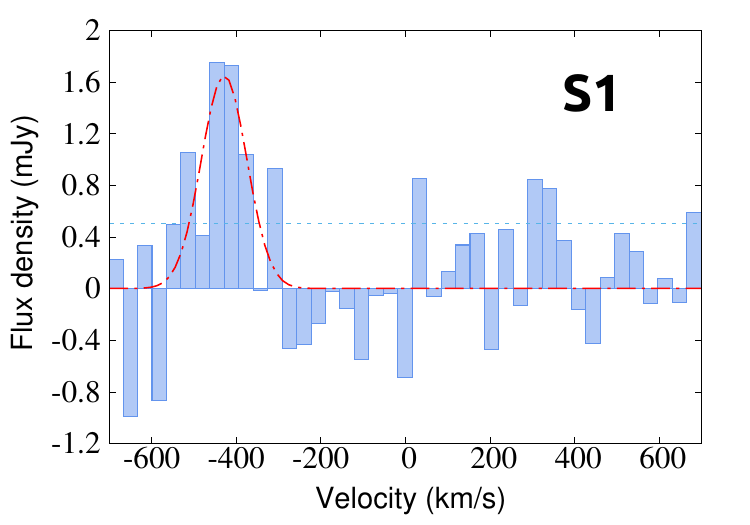}&
\includegraphics[width=0.33 \textwidth]{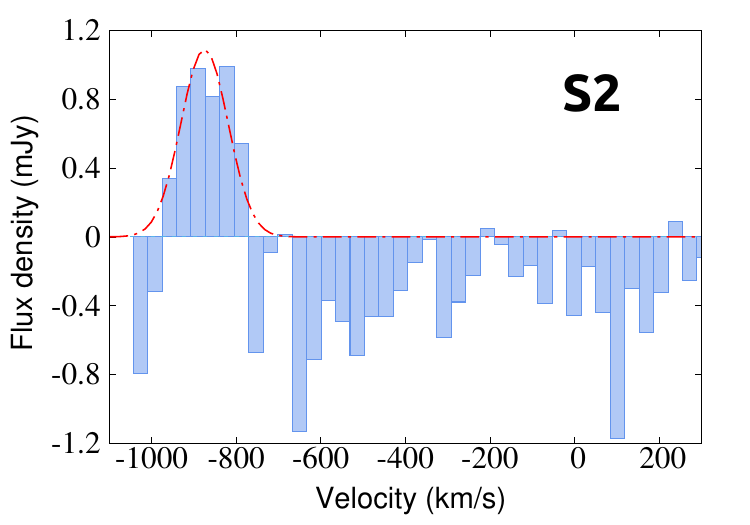}&
\includegraphics[width=0.33 \textwidth]{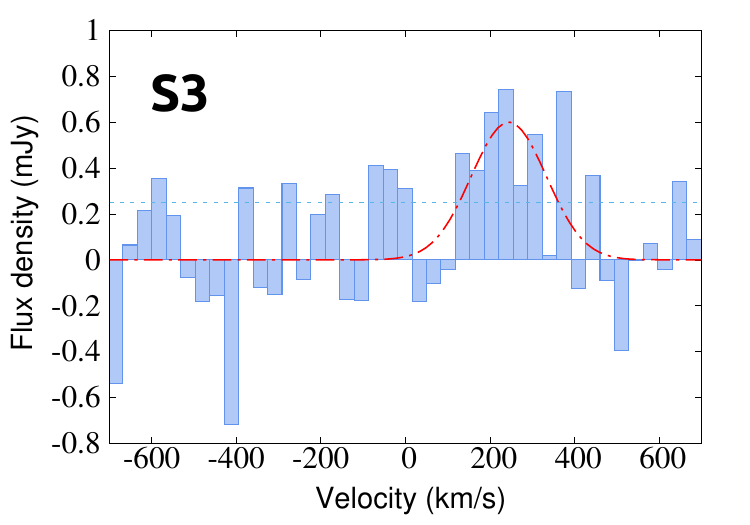}
\\
\includegraphics[width=0.33 \textwidth]{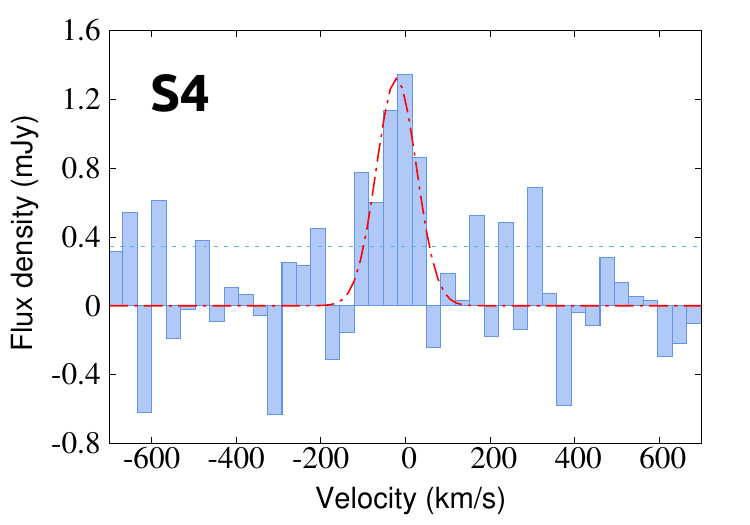}&
\includegraphics[width=0.33 \textwidth]{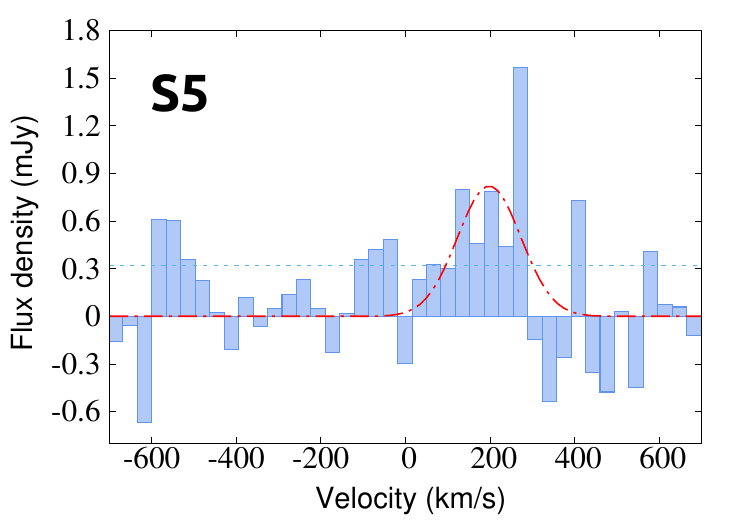}&
\includegraphics[width=0.33 \textwidth]{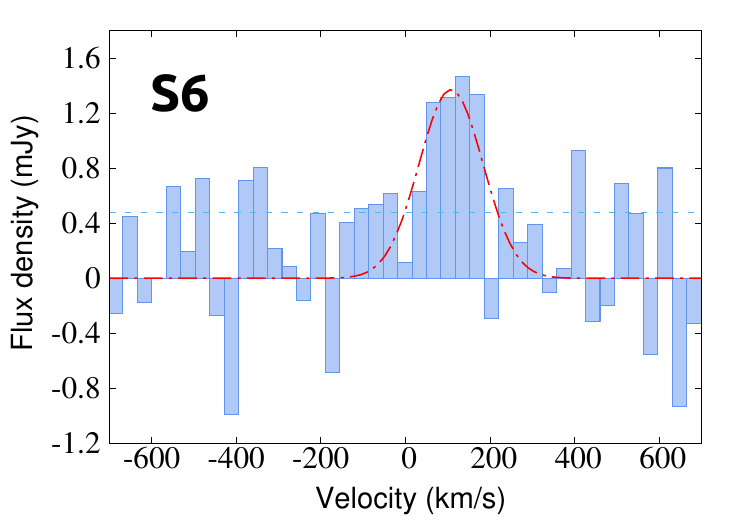}
\\
\includegraphics[width=0.33 \textwidth]{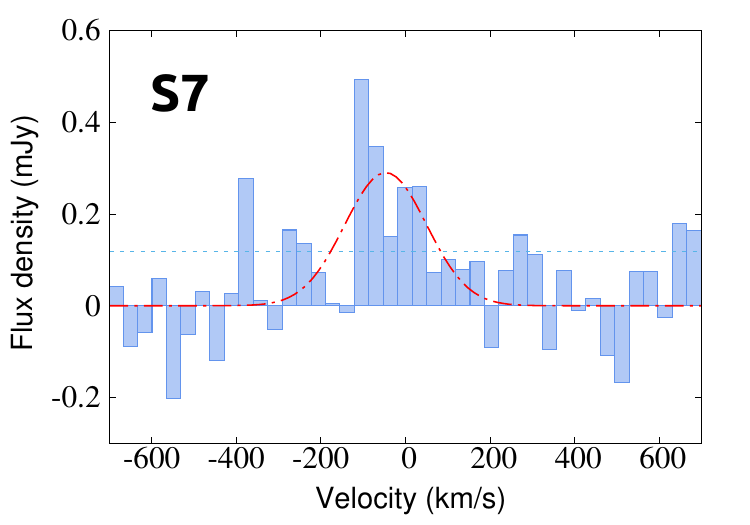}&
\includegraphics[width=0.33 \textwidth]{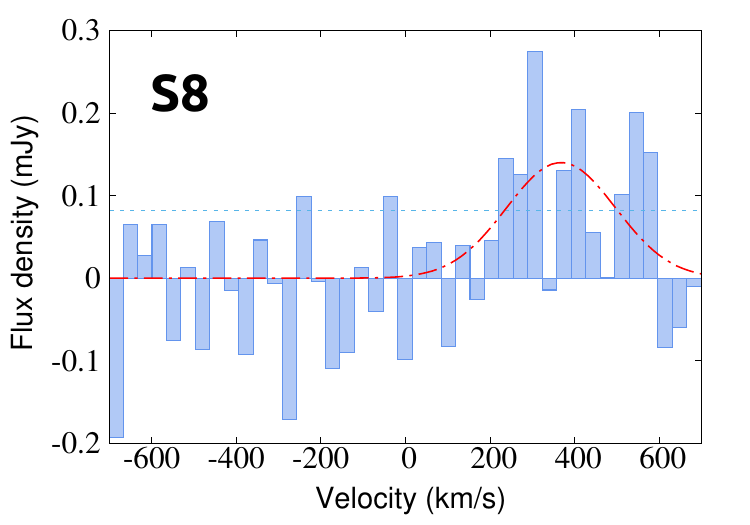}&
\includegraphics[width=0.33 \textwidth]{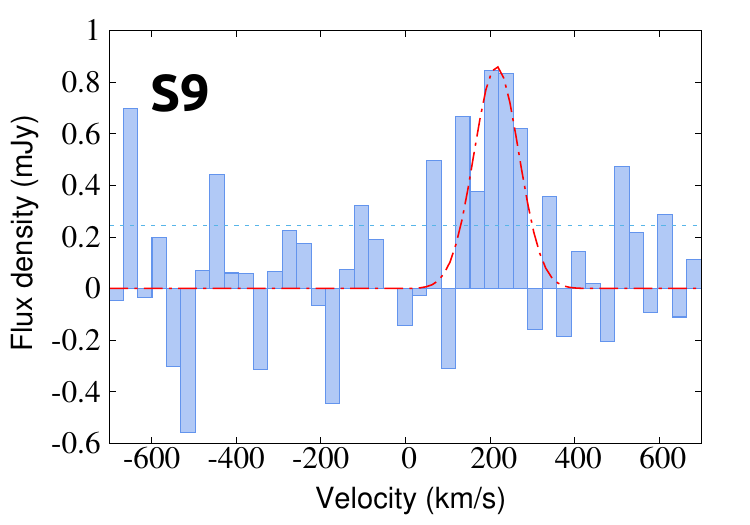}
\\
\includegraphics[width=0.33 \textwidth]{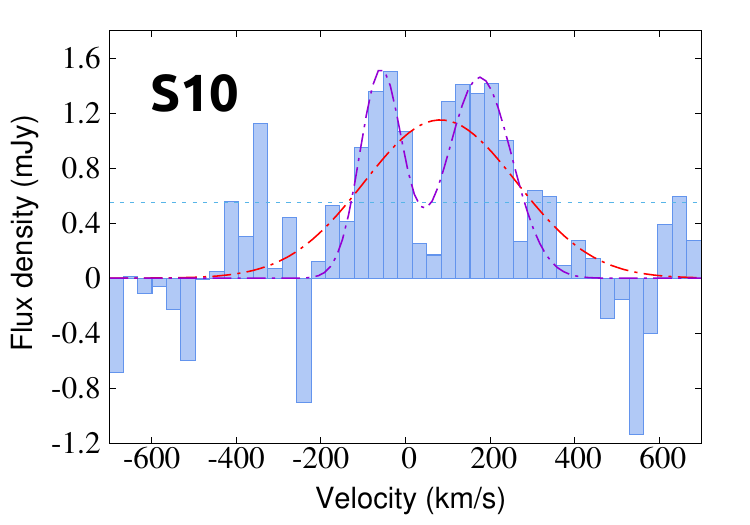}&
\includegraphics[width=0.33 \textwidth]{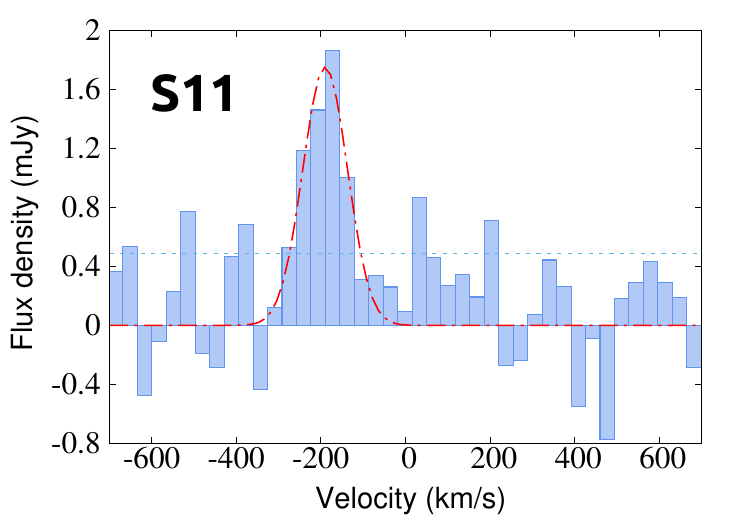}&
\includegraphics[width=0.33 \textwidth]{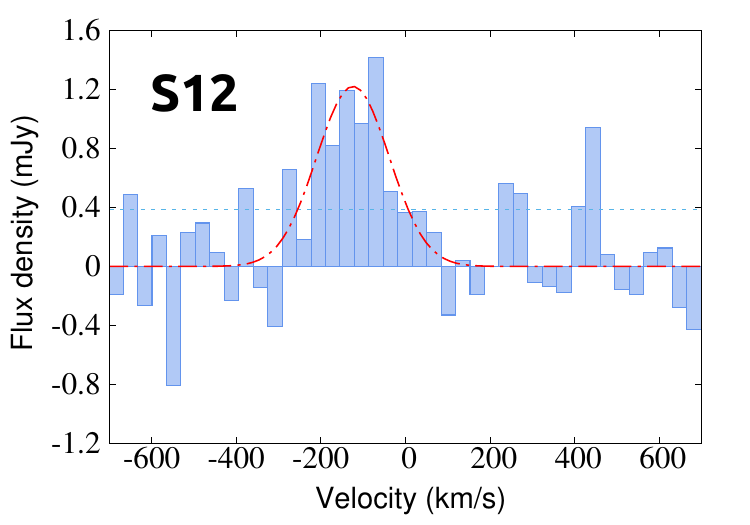}
\end{tabular}
\caption{ 
\hicm\ emission spectra of detected \hi\ sources listed in Table \ref{tab:HI}. All the spectra are extracted from the cube with 35 \kms\ channels (see Table \ref{tab:cubes}). 
The blue-dotted line marks  the rms noise level. 
The best fitted Gaussian functions to the spectra are shown with red-dot-dashed lines. 
For G3 with a double-horn-like  \hi\ spectrum  we also fitted a busy-function \citep[][]{Westmeier14-2014MNRAS.438.1176W} to the \hicm\ emission line,  presented by the  purple-dot-dashed line. This results to a $\rm \Delta v_{0.037}$ of -5 \kms, and a   FWHM$_{\rm HI}$ of 317 \kms, consistent with those obtained by fitting a Gaussian.
The \hicm\ emission line from S10 seemed appear to be like a double-horn emision line; but after carefully checking the \hicm\ emission in individual channels, we concluded the emission to originate from at least two individual components. We therefore fitted a double-Gaussian function to the \hi\ spectrum of this source to have a better understanding of its velocity spread. This is presented with the purple-dot-dashed line.  
\label{fig:specs}}
\end{figure*}

\begin{figure*}
\centering
\begin{tabular}{ccc} 
\includegraphics[width=0.33 \textwidth]{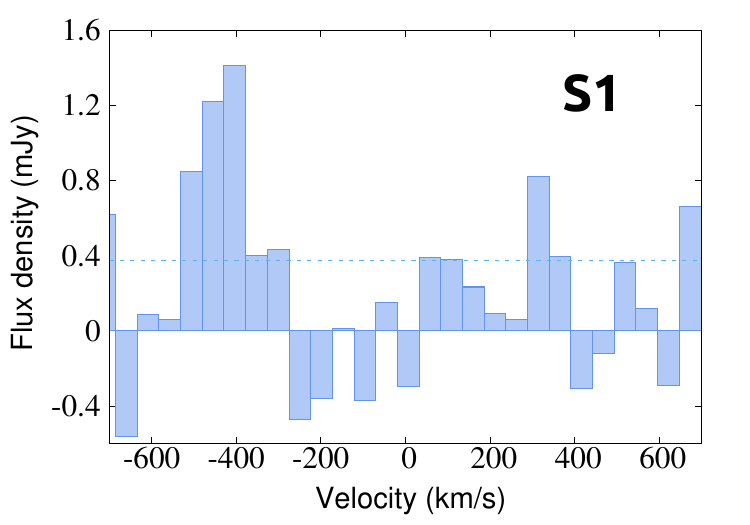}&
\includegraphics[width=0.33 \textwidth]{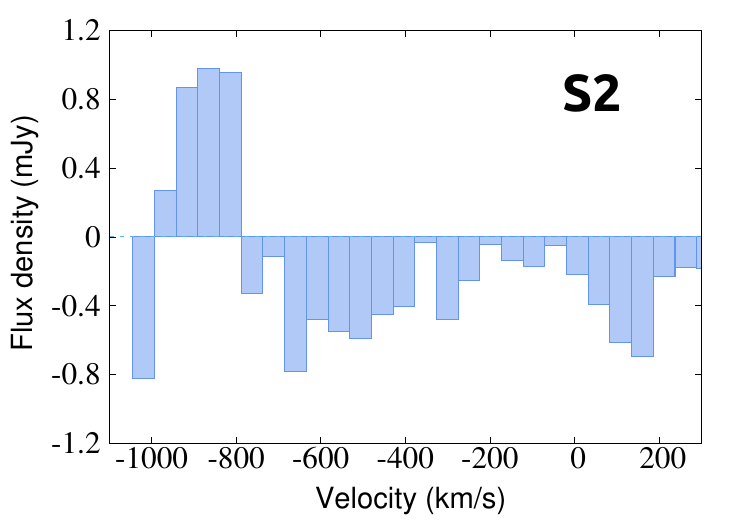}&
\includegraphics[width=0.33 \textwidth]{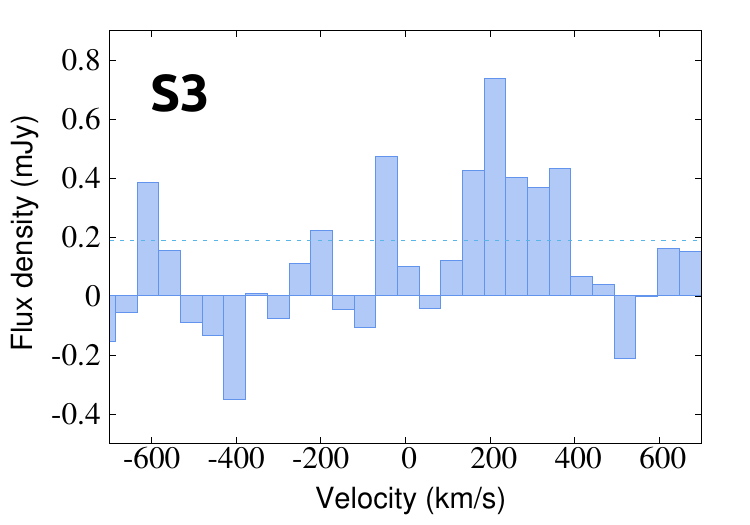}
\\
\includegraphics[width=0.33 \textwidth]{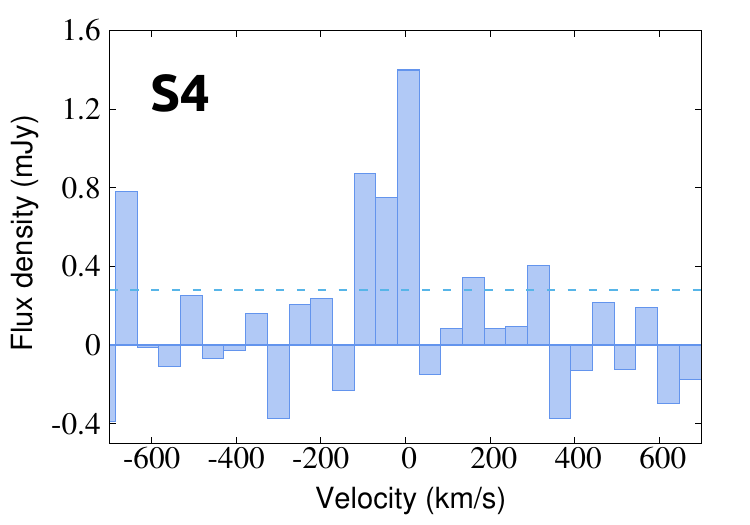}&
\includegraphics[width=0.33 \textwidth]{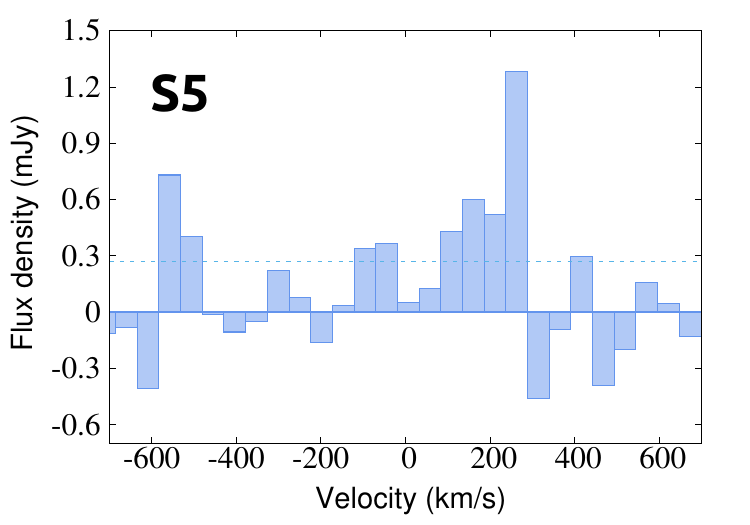}&
\includegraphics[width=0.33 \textwidth]{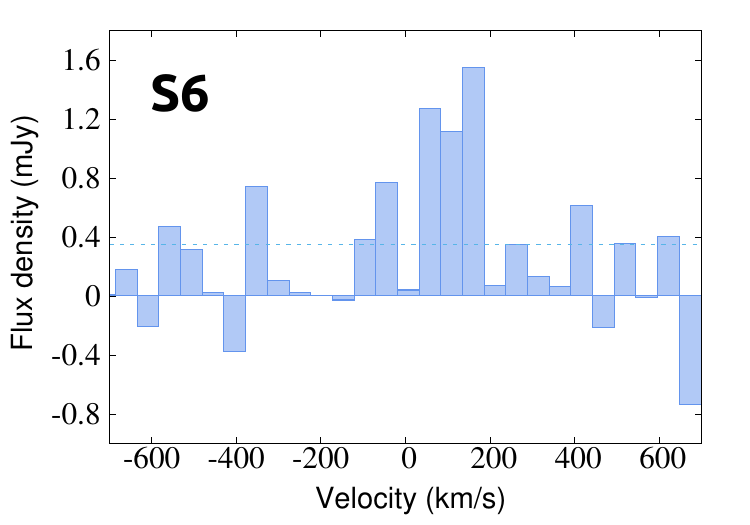}
\\
\includegraphics[width=0.33 \textwidth]{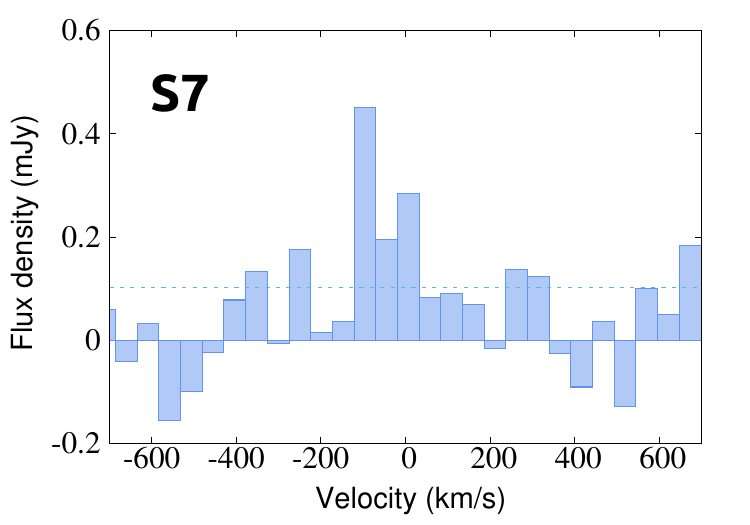}&
\includegraphics[width=0.33 \textwidth]{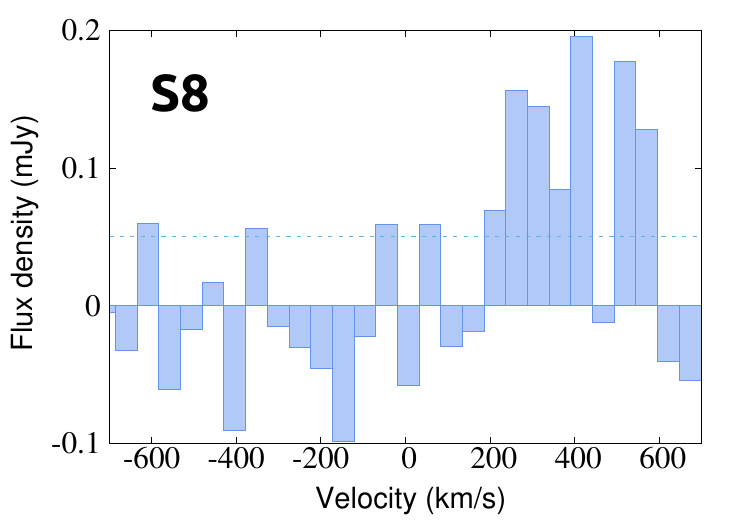}&
\includegraphics[width=0.33 \textwidth]{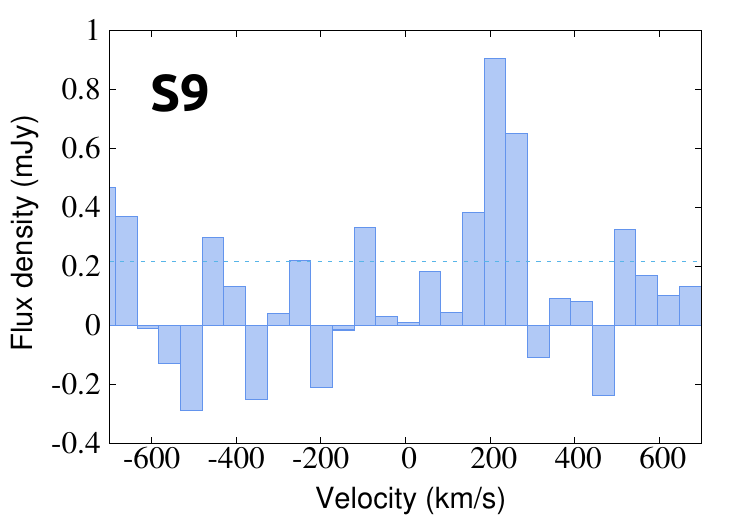}
\\
\includegraphics[width=0.33 \textwidth]{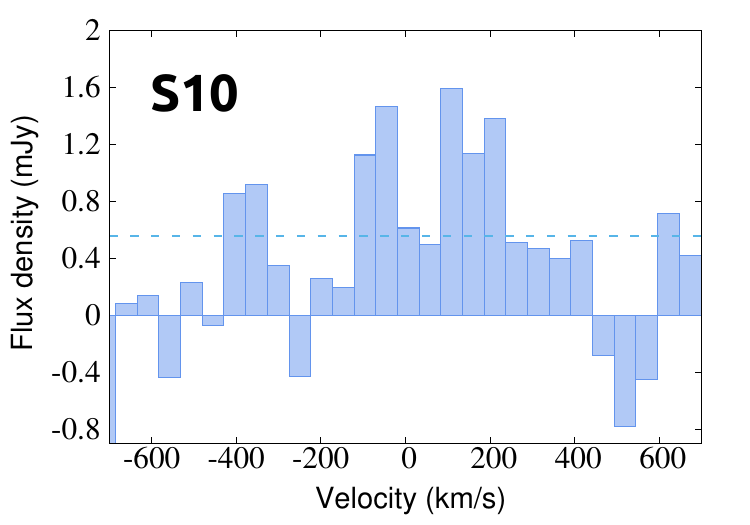}&
\includegraphics[width=0.33 \textwidth]{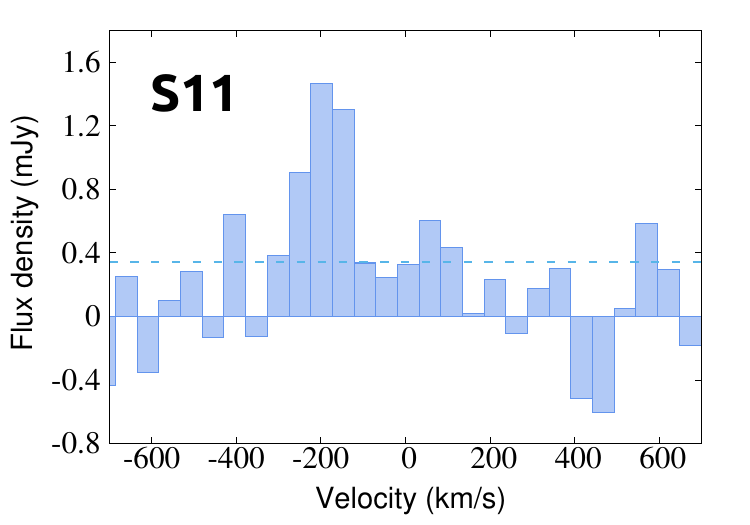}&
\includegraphics[width=0.33 \textwidth]{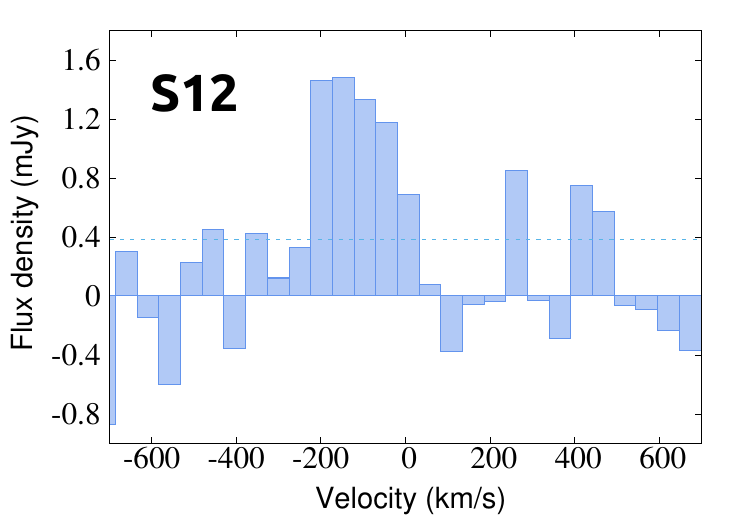}
\end{tabular}
\caption{ 
\hicm\ emission spectra of S1--S12 listed in Table \ref{tab:HI},  extracted from the cube with 50 \kms\ channels (see Table \ref{tab:cubes}), for comparison with the spectra presented in Figure \ref{fig:specs}. 
The blu-dotted line marks  the rms noise level. 
\label{fig:specs-50}}
\end{figure*}

Finally, we only keep those  detected \hi\ sources  which have signal-to-noise rations (SNRs) of 5 and above in both 35 and 50 \kms, cubes. 
We measured the \hi\ integrated flux density  by integrating over the adjacent channels with fluxes above the rms noise, and converted this to  a \hi\ mass \citep[][]{Meyer17-2017PASA...34...52M}. We  also fitted a Gaussian function to the \hicm\ emission line to measure the redshift and the FWHM of the emission. 
Our final list of detected \hi\ sources is provided in Table \ref{tab:HI}. 
The total intensity maps and the intensity-weighted velocity maps of \hicm\ emission line for the detected \hi\ sources are presented in Figure \ref{fig:maps}.  The extracted \hicm\ emission  spectra are shown  in Figures \ref{fig:specs} and \ref{fig:specs-50}.

\section{SED modellings}
\label{app:opt}

\begin{figure*}
\centering
\begin{tabular}{ccc} 
\includegraphics[width=0.33 \textwidth]{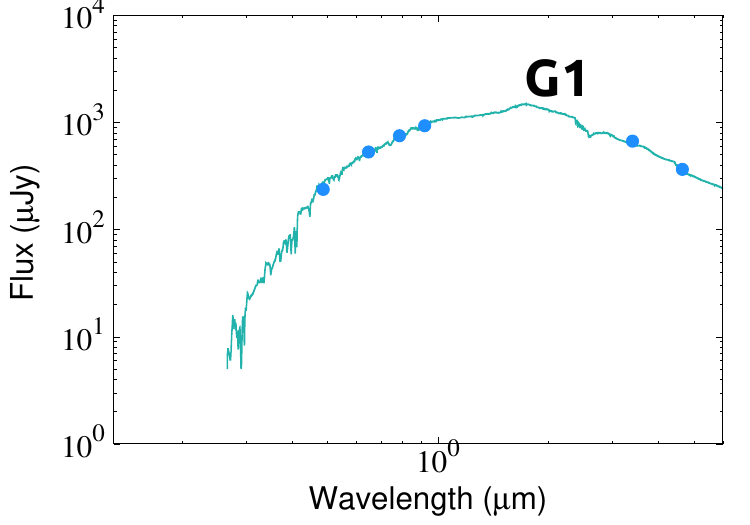}&
\includegraphics[width=0.33 \textwidth]{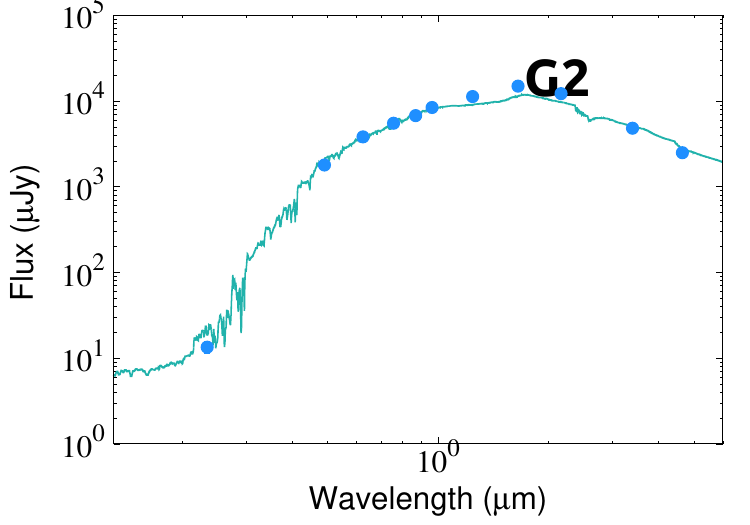}&
\includegraphics[width=0.33 \textwidth]{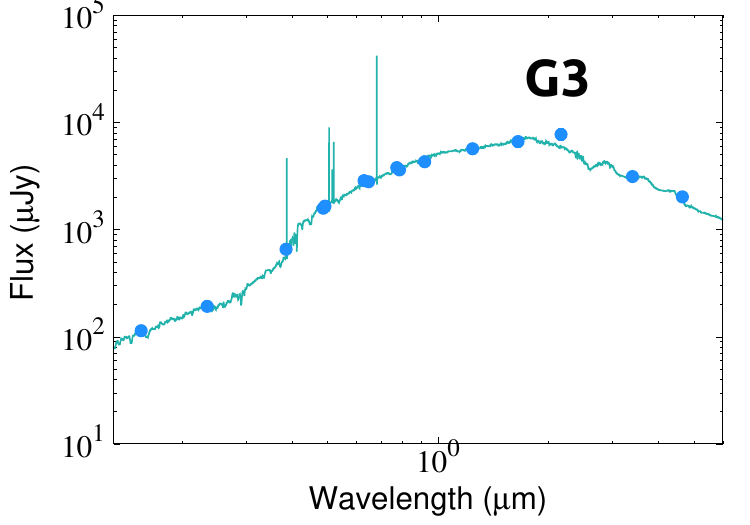}
\\
\includegraphics[width=0.33 \textwidth]{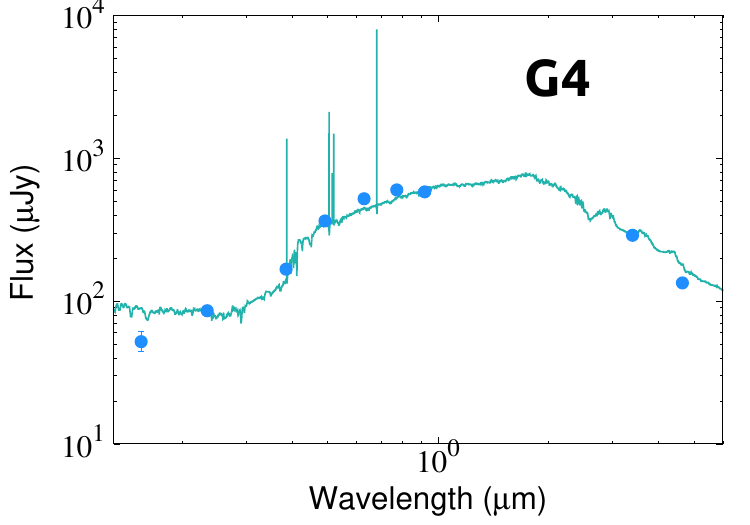}&
\includegraphics[width=0.33 \textwidth]{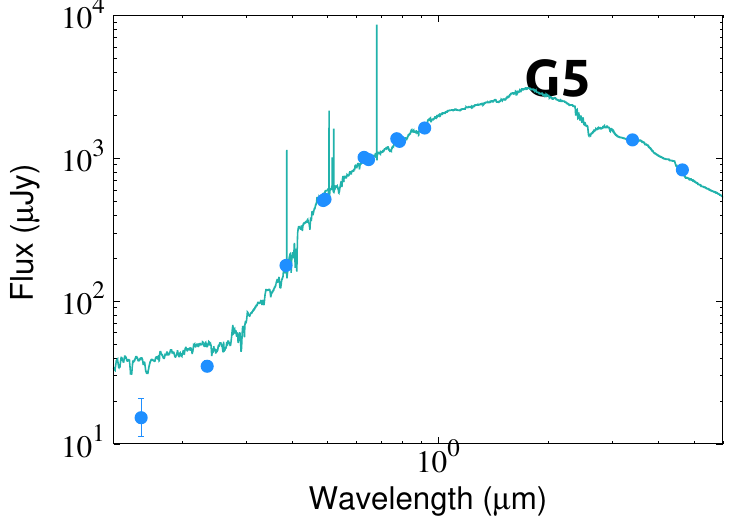}&
\includegraphics[width=0.33 \textwidth]{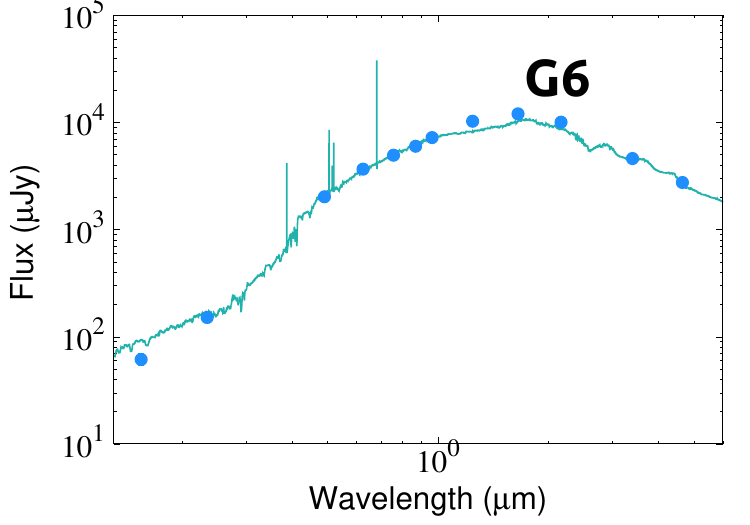}
\\
\includegraphics[width=0.33 \textwidth]{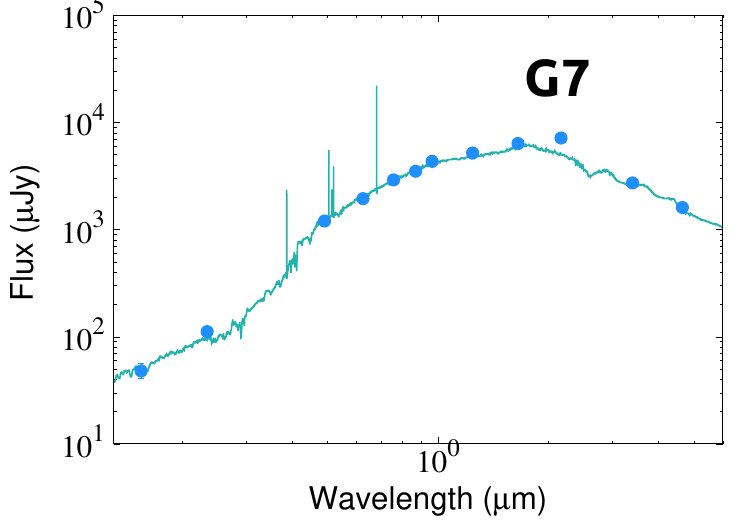}&
\includegraphics[width=0.33 \textwidth]{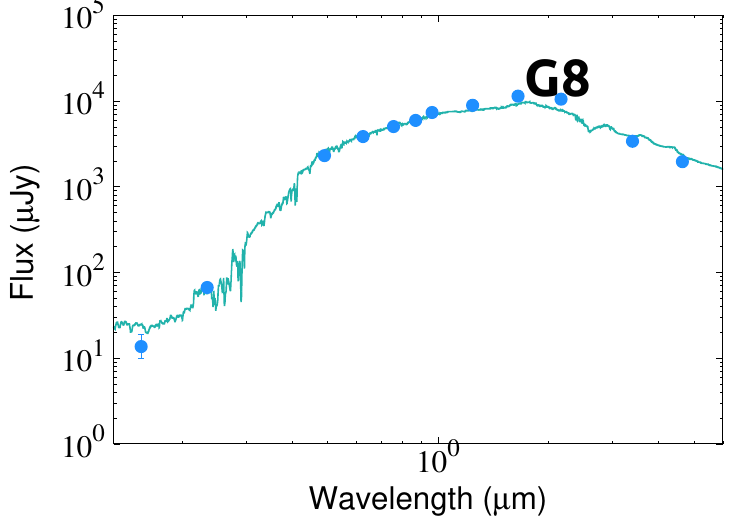}&
\end{tabular}
\caption{
The fitted Spectral Energy Distribution (SED) for the eight galaxies along the filament (G1-G8). The SEDs clearly demonstrate that G3-G7 are actively star-forming galaxies while G1, G2 and G8 (the three galaxies at the two ends of the filament) seem to be passive galaxies (with no emission lines). The physical properties of all galaxies, obtained from the best fitted SEDs, are listed in Table \ref{tab:g8}. 
\label{fig:seds-galaxies}}
\end{figure*}

The the optical magnitudes of the eight galaxies (G1--G8) are provide in Table \ref{tab:g8-mags} and the SED models of them are presented in Figure \ref{fig:seds-galaxies}. 
For the optical counterpart candidates  with $\gtrsim 1\%$ probability of having $z_{phot} \lesssim 0.1$ 
the derived properties are listed in Table \ref{tab:optical}.

\begin{table*}
\begin{center}
\caption{The optical AB magnitudes of the eight galaxies that reside along a narrow filament shown in  the right panel of Figure \ref{fig:field}.}
\label{tab:g8-mags}
\begin{tabular}{lcccccc}
\hline
ID &  GALEX-FUV & GALEX-NUV & Pan-STARRS-g & Pan-STARRS-r & Pan-STARRS-i & Pan-STARRS-z \\
\hline
G1&	          	... &	          	... &	            	...    &	            	...    &	            	...    &		            	...    \\	
G2&	          	... &	21.08$\pm$	0.17&	15.767 $\pm$    0.00154&	14.9462$\pm$	0.00076&	14.55  $\pm$    0.00068&		14.3274$\pm$	0.00096\\ 	
G3&	18.76$\pm$	0.12&	18.19$\pm$	0.03&	15.9313$\pm$	0.00228&	15.3308$\pm$	0.00138&	15.0497$\pm$	0.00141&		14.8951$\pm$	0.00224\\	
G4&	19.61$\pm$	0.18&	19.07$\pm$	0.05&	                ...    &                         ...   &                        ...    &	                        ...    \\	
G5&	20.94$\pm$	0.33&	20.04$\pm$	0.10&	17.0395$\pm$	0.01315&	16.5134$\pm$	0.00333&  	16.2298$\pm$	0.00291&		16.0221$\pm$	0.00459\\	
G6&	19.43$\pm$	0.15&	18.45$\pm$	0.04&	15.6338$\pm$	0.00145&	14.9911$\pm$	0.00103&	14.6652$\pm$	0.00081&		14.4563$\pm$	0.00174\\	
G7&	19.70$\pm$	0.17&	18.78$\pm$	0.05&	16.2019$\pm$	0.00394&	15.6776$\pm$	0.00145&	15.242 $\pm$    0.00150&		15.0405$\pm$	0.00179\\	
G8&	21.06$\pm$	0.35&	19.34$\pm$	0.07&	15.4931$\pm$	0.00087&	14.9373$\pm$	0.00071&	14.6468$\pm$	0.00059&		14.4661$\pm$	0.00090\\	
\hline
&&&&&&\\
ID & Pan-STARRS-y & 2MASS-J & 2MASS-H & 2MASS-Ks & ALLWISE-1 & ALLWISE2 \\
\hline
G1&	            	...    &	 16.70 $\pm$	0.08 &		16.56 $\pm$	0.10 &		16.61 $\pm$	0.11 &		17.0  $\pm$	0.038&		17.616$\pm$	0.05 \\	
G2&	14.0895$\pm$	0.00205&	 13.773$\pm$	0.044&		13.468$\pm$	0.058&		13.688$\pm$	0.082&		14.694$\pm$	0.023&		15.409$\pm$	0.024\\ 
G3&	14.6917$\pm$	0.00500&	 14.515$\pm$	0.096&		14.348$\pm$	0.127&		14.181$\pm$	0.142&		16.016$\pm$	0.026&		16.51 $\pm$     0.029\\	
G4&	                 ...   &	                  ...&		                ...  &	                        ...  &	        18.498$\pm$	0.053&		19.048$\pm$	0.128\\	
G5&	15.7913$\pm$	0.01054&	                  ...&		                ...  &	                        ...  &	        16.762$\pm$	0.032&		17.229$\pm$	0.045\\	
G6&	14.2541$\pm$	0.00254&	 13.873$\pm$	0.039&		13.701$\pm$	0.052&	  	13.896$\pm$	0.080&		14.743$\pm$	0.022&		15.302$\pm$	0.022\\	
G7&	14.8082$\pm$	0.00375&	 14.615$\pm$	0.069&		14.393$\pm$	0.089&		14.265$\pm$	0.100&		15.311$\pm$	0.023&		15.884$\pm$	0.025\\	
G8&	14.2366$\pm$	0.00162&	 14.028$\pm$	0.052&		13.76 $\pm$     0.050&		13.847$\pm$	0.078&		15.075$\pm$	0.033&		15.671$\pm$	0.035\\	
\hline
&&&&&&\\
ID & MegaCam-u & MegaCam-g & MegaCam-r & MegaCam-i &  &  \\
\hline
G3&	16.8593$\pm$	0.001&	 15.8529$\pm$	0.0004&		15.2587$\pm$	0.0004&		14.9554$\pm$	0.0006&		&		\\	
G4&	18.3428$\pm$	0.0027&	 17.4997$\pm$	0.0014&		17.1106$\pm$	0.0014&		16.955$\pm$	0.0029&		&		\\	
G5&	18.2766$\pm$	0.0019&	 17.1194$\pm$	0.0008&		16.3876$\pm$	0.0006&		16.0618$\pm$	0.001&		&		\\	
\hline
\end{tabular}   
\end{center}
\flushleft{
The CFHT/MegaCam magnitudes are derived as explained in Section \ref{sec:cfht}.  The urls for obtaining the other  magnitudes are listed below.\\
GALEX: https://galex.stsci.edu/GR6/?page=mastform\\
Pan-STARRS: https://outerspace.stsci.edu/display/PANSTARRS/ \\
2MASS: https://irsa.ipac.caltech.edu/cgi-bin/Gator/nph-scan?submit=Select\&projshort=2MASS \\
ALLWISE: https://irsa.ipac.caltech.edu/cgi-bin/Gator/nph-scan?submit=Select\&projshort=WISE \\
}
\end{table*}

\begin{table*}
\begin{center}
\caption{The properties of the possible optical counterparts of the detected \hi\ sources, marked with yellow circles in the middle panels of Figure \ref{fig:maps}. For the listed optical sources below the probability of having  a photometric redshift $\rm z_{photo}\leq 0.1$ obtained from  SED fitting  is $\geq$ 1\%. Hence we do not rule  $z=0.037$  out for these sources. 
}
\begin{tabular}{ccccccc}
\hline
\hi-ID & Opt-ID  &  $\rm P({z_{photo}\leq 0.1})$  &  $\rm log_{10}(M_{*}/M_{\odot})$  &  $\rm log_{10}(SFR/M_{\odot}\,yr^{-1})$  & E(B-V) & $\rm (M_{HI}/M_{*})_{LL}$  \\
\hline
\multirow{2}{*}{S1} &b  &  42\%     &      7.1$\pm$0.1    &     -7.5$\pm$0.6        & 0 & \multirow{2}{*}{89}\\
& d  &  1\%     &      6.2$\pm$0.4    &     -2.8$\pm$0.6        & 0.7 \\
\hline
S2 & a   & 65\%   &  9.10$\pm$0.03  &  -1.3$\pm$0.1  &  0  & 1\\
\hline
S3  & a  &  6\%     &       5.4$\pm$0.2   &     -3.3$\pm$0.3         & 0 & 2570\\
\hline
S4  & & & & & & 10000\\
\hline
\multirow{3}{*}{S5} & a & 1\% & 6.3$\pm$0.1 & -1.5$\pm$0.2 & 0.7 & \multirow{3}{*}{13/5370$^*$}\\ 
& b  &   3\%   &       5.2$\pm$0.3   &     -3.2$\pm$0.3         & 0 \\
& e  &   100\%    &       7.8$\pm$0.1   &     -2.0$\pm$0.1         & 0 \\
\hline
S6  & b  &   7\%    &       6.1$\pm$0.2   &     -2.9$\pm$0.4         & 0 & 1000\\
\hline
S7 & b  &  6\%     &      6.0$\pm$0.3    &     -2.8$\pm$0.5        & 0.3 & 316 \\
\hline
S8  & b  &  3\%     &      5.5$\pm$0.3    &     -2.9$\pm$0.4        & 0.4 & 912\\
\hline
S9  & a  &  1\%     &      6.5$\pm$0.3    &     -2.7$\pm$0.4        & 0.5 & 224\\
\hline
S10 & c  &  1\%     &      5.2$\pm$0.3    &     -3.0$\pm$0.3        & 0.2 & 15488\\
\hline
\multirow{4}{*}{S11} & a  & 5\%    &    5.7$\pm$0.2  &    -2.6$\pm$0.4 & 0.3 & \multirow{4}{*}{96} \\
& b  & 5\%    &    6.7$\pm$0.3  &    -2.6$\pm$0.7 & 0.7 \\
& c  & 4\%    &    6.4$\pm$0.3  &    -2.5$\pm$0.6 & 0.6 \\
& d  & 3\%    &    6.7$\pm$0.3  &    -2.6$\pm$0.7 & 0 \\
\hline
\multirow{3}{*}{S12} & g (93)   &  100\%    &      6.2$\pm$0.2    &     -1.7$\pm$0.2        & 0.6 & \multirow{3}{*}{28}\\
 & m (106)  &  11\%     &      7.4$\pm$0.1    &     -5.4$\pm$1.0        & 0.0 & \\
 & p (119)  &  49\%      &      7.3$\pm$0.1    &     -1.4$\pm$0.2        & 0.5 & \\
\hline
\hline
\end{tabular}   
\end{center}
\flushleft{
The first and second  columns are, respectively,   the identification names and numbers  of  the \hi\ sources and their possible optical counterparts.  The third column is the probaility of a  photomertic redshift $\rm z_{photo}\leq 0.1$ for the optical source, obtained from SED fitting. Columns 4 to 6 are, respectively, stellar mass, star formation rate, and E(B-V) obtained from SED fitting with assuming $z=0.037$. Column 7 is the lower limit on $\rm M_{HI}/M_{*}$ assuming that all the optical sources listed in the Table could be associated with the \hi\ sources. \\
$^*$ For this source we have provided the $\rm M_{HI}/M_{*}$  with including/excluding the optical source that is outside the \hi\ emission (marked by e in Figure \ref{fig:maps}). 
}
\label{tab:optical}
\end{table*}

\end{document}